\documentclass[%
reprint, 
preprintnumbers,
 amsmath,amssymb,
 aps,
 prl,
floatfix,
]{revtex4-1}

\usepackage{graphicx}
\usepackage{dcolumn}
\usepackage{booktabs}
\usepackage{bm}

\begin{document}

\graphicspath{{./figures}}


\title{
    Laser-Plasma Ion Beam Booster\\ 
    Based on Hollow-Channel Magnetic Vortex Acceleration
    }%

\author{Marco Garten}%
\author{Stepan S. Bulanov}%
\author{Sahel Hakimi}%
\author{Lieselotte Obst-Huebl}%
\author{Chad E. Mitchell}%
\author{Carl Schroeder}%
\author{Eric Esarey}%
\author{Cameron G. R. Geddes}%
\author{Jean-Luc Vay}%
\author{Axel Huebl}%
 \email{axelhuebl@lbl.gov} 
\affiliation{%
 Lawrence Berkeley National Laboratory\\
 1 Cyclotron Rd, Berkeley (CA), 94720, USA
}


\date{\today}

\begin{abstract}
Laser-driven ion acceleration provides ultra-short, high-charge, low-emittance beams, which are desirable for a wide range of high-impact applications.
Yet after decades of research, a significant increase in maximum ion energy is still needed.
This work introduces a quality-preserving staging concept for ultra-intense ion bunches that is seamlessly applicable from the non-relativistic plasma source to the relativistic regime.
Full 3D particle-in-cell simulations prove robustness and capture of a high-charge proton bunch, suitable for readily available and near-term laser facilities.
\end{abstract}

\maketitle




Laser-driven ion acceleration provides unique particle bunches of very high charge ($\gg\,100$\,pC), large current ($>$\,kA), ultra-low emittance ($\ll\,10$\,nm) on very compact acceleration footprints~\cite{mourou.rmp.2006,daido.rpp.2012,macchi.rmp.2013,Cowan.PRL.2004,Borghesi.PRL.2004,Nuernberg.RSI.2009}.
Such high-intensity ion beams can have significant impact on advancing both fundamental and applied research applications in physics, industry and society, ranging from next-generation hadron colliders or neutrino factories~\cite{abada_fcc-hh_2019,Geer.ProgPartNuclPhys.2007}, drivers for inertial fusion energy~\cite{roth.prl.2001,mackinnon.prl.2006}, radiotherapy~\cite{Favaudon2014,MontayGruel2017,Bin2022,SVBulanov2002,bulanov.pu.2014,LKarsch2017}, nuclear physics~\cite{Chen2019}, warm-dense matter research~\cite{borghesi.ppcf.2001,pelka.prl.2010}, secondary radiation generation~\cite{Ledingham.Science.2003,Yurevich.PhysPartNucl.2010} for material research and security applications, and possibly even radiation hardness of spacecraft~\cite{durante.rmp.2011,Miyazawa.iScience.2018}.
However after decades of research, the demonstrated maximum particle energy, between 60-100 MeV for protons~\cite{Wagner2016,Higginson2018,dover.lsa.2023,Rehwald2023}, still remains short of desired ranges for many of the mentioned pivotal applications.

With state-of-the-art Petawatt (PW) laser facilities~\cite{danson.hplse.2019,gonoskov.rmp.2022} that can deliver high-intensity laser pulses reaching $10^{23}$\,W/cm$^2$~\cite{yoon.optica.2021}, laser-ion acceleration remains an indirect process due to the high mass of ions.
The laser pulse interacts primarily with plasma electrons, creating a localized ($<\mu$m) charge separation that imposes a strong ($>$TV/m) electric field, orders of magnitude higher than achievable in established cyclotrons and radiofrequency accelerators, which then facilitates the acceleration.
Depending on the laser intensity and contrast~\cite{daido.rpp.2012,Higginson2018,Ziegler.SciRep.2021}, target density, thickness, and composition, a specific laser-ion acceleration mechanism might dominate~\cite{mourou.rmp.2006,daido.rpp.2012,macchi.rmp.2013,bulanov.pop.2016}.
In these mechanisms, the maximum ion energy scales with laser intensity on target, for some even linearly, while demanding high control of temporal laser contrast~\cite{bulanov.pop.2016}.
Yet, increasing laser power is a multi-decade-long undertaking, and significant advances initially reduce laser repetition rate and temporal laser contrast, increase facility cost and footprint~\cite{danson.hplse.2019,gonoskov.rmp.2022}.

In this work, we propose a solution to the longstanding challenge of achieving relativistic ion energies from a laser-plasma accelerator, the first quality-preserving, self-consistently modeled concept of ion staging.
This concept for staging boosts particle energies from the sub-relativistic over arbitrary energy levels to the relativistic regime.
Staging was already demonstrated in plasma acceleration of electrons~\cite{steinke.nature.2016,Kurz.NatComm.2021} and is a candidate for next generation electron-positron colliders for high-energy physics and light sources~\cite{geddes.af6.2022}.
In contrast to staged electron acceleration, where the initial laser-plasma source readily provides relativistic particles, ions need to be accelerated in multiple stages while undergoing significant velocity changes.
In this work, we design an ultra-compact stage with quasi-static fields of a specifically designed, ultra-compact laser-driven plasma element and show that repeated injection into stages of the same compact design boosts ion energy progressively from non-relativistic to relativistic values.

This letter explores staging via a laser-target setup previously used for for magnetic vortex acceleration (MVA)~\cite{SSBulanov2010,Park2019,Hakimi2022}.
Realizing MVA involves the interaction of an intense laser pulse with a near-critical density (NCD), $n_\mathrm{e} \approx n_\mathrm{cr}$, plasma target of several tens of micrometer thickness~\cite{Kuznetsov.PRR.2001,SVBulanov.PRR.2005,SVBulanov.PRL.2007}. 
Here $n_e$ and $n_{cr}=m_e\omega^2/4\pi e^2$ are plasma density and critical plasma density, $e$ and $m_e$ are electron charge and mass, and $\omega$ is the central laser pulse frequency.
The ponderomotive force of the laser pulse generates a plasma channel, first in electron density, then in ion density.
During its propagation through the channel, the laser pulse causes a strong electron current in the forward direction, which later starts to pinch, creating an electron and ion filament along the laser propagation axis.
As the laser, electrons and ions exit the target from the back side, strong electric ($\sim$\,TV/m) and magnetic fields ($\sim10^4$\,T) are generated there, arising from the forward-accelerated electron filament and cold return currents in the channel walls.
This produces a high-flux ion beam of ultra-short time structure and with nano-Coulombs of charge~\cite{SSBulanov2010,Park2019}. 

For MVA, the electric fields created at the rear side of the plasma channel both accelerate and focus the ions.
Ion acceleration from near-critical density plasmas has been studied in multiple experiments~\cite{Matsukado.PRL.2003,Willingale.PRL.2006,Yogo.PRE.2008,Fukuda.PRL.2009,Willingale.PRL.2009,Willingale.POP.2011,Rehwald2023}, and MVA was found to be robust against laser contrast and incidence angle variations in numerical modeling studies~\cite{Hakimi2022}.
The near-critical density regime of MVA, readily reached with high repetition rate targets~\cite{Goede.PRL.2017,Rehwald2023}, relaxes the requirements for microscopic fabrication precision needed.
Particularly attractive for energy boosting is that upon laser irradiation no potential barrier is created at the target front-side, which would decelerate particles~\cite{Jaeckel.CLEO.2009,Wang.PoP.2013,Kawata.APPC12.2014,Pfotenhauer2010}.

Existing theoretical work on boosting the energy of laser-driven ion beams can be classified into different categories:
First, energy boosting from a cascade of acceleration mechanisms, using a single laser pulse and composite target~\cite{Zhang.PoP.2007,Toncian2006,Kar.NatComm.2016,Ferguson.NJP.2023} or multiple targets~\cite{Liu.PoP.2018,An.HPLSE.2023}.
Second, energy boosting using multiple laser pulses on a single target~\cite{Bychenkov.PPR.2007,Yogo.SciRep.2017,Ferri.CommPhys.2019,Kim.PRR.2022}.
The first approach is limited by the available laser energy, the second is sensitive to control of the spatio-temporal laser-plasma interaction.
Third, proposing to avoid these limitations, energy boosting with multiple laser pulses and staged targets~\cite{Jaeckel.CLEO.2009,Wang.PoP.2017,Kawata.APPC12.2014,Ting.AAC.2017}, so far with limited exploration of performance with intense beams or preservation of important beam moments, besides gaining energy.
Consequently, no experimental studies exist so far.

Here, MVA is studied as a scalable mechanism for boosting an ultra-intense, high-charge beam that is implementable with high-repetition rate stages~\cite{Kim2016,Goede.PRL.2017,Rehwald2023}.
We investigate and propose a robust, quality-preserving variation of MVA as a plasma booster, which is insensitive to the accepted ion beam velocities, and systematically evaluate temporal and position acceptance tolerances.
Critical for staging into subsequent high--$\beta$ ($\beta=v/c$ is the normalized ion velocity) acceleration stages, our approach can conserve the high charge and the ultra-low emittance of laser-plasma generated beams.
The presented booster also accepts substantial energy spreads.

\begin{figure}[!htbp]
\includegraphics[width=\columnwidth]{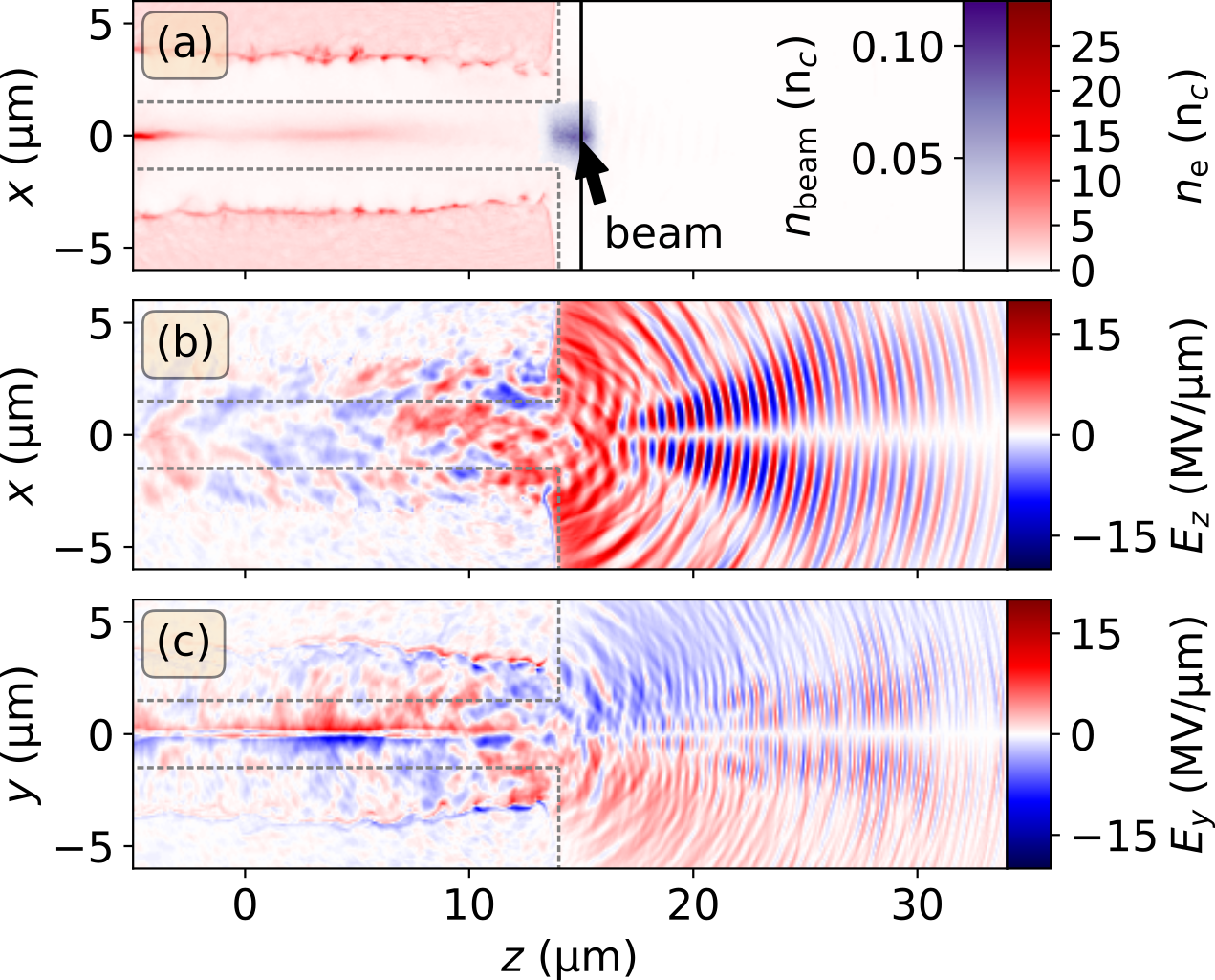}%
\caption{\label{fig:densities_fields}
    \textit{Fields of the booster stage, loaded with a 200\,pC proton bunch.} The stage center is at $z = 0\,\mathrm{\mu m}$.
    \textbf{(a)} Electron density $n_\mathrm{e}$ in red, reference plane at the accelerating bucket $z_\mathrm{ab} = 15\,\mathrm{\mu m}$ (black vertical line), and beam density $n_\mathrm{beam}$ in purple located behind the exit of the stage.
    \textbf{(b)} Accelerating electric field $E_z$, and \textbf{(c)} focusing electric field $E_y$, all shown as the beam reaches $z_\mathrm{ab}$.
    Dashed grey lines mark the outline of the initial electron density of the hollow channel.%
}%
\end{figure}

However, there is a significant distinction between a typical MVA target and the one proposed here, visualized in Figure~\ref{fig:densities_fields}.
In MVA, the accelerated ions originate from a central filament, which is formed from the background ions under the action of the pinched electron current. 
Those ions mainly come from a small region at the back of the target near the channel axis. 
Thus, in order to suppress additional low-energy ion injection from the booster stage itself, an NCD target with a pre-formed hollow region is chosen (dashed lines in Fig.~\ref{fig:densities_fields})~\cite{gong.prr.2022, Lezhnin.PRR.2022, zhu.prappl.2022,Tazes.HPLSE.2022}. 

Typically, MVA relies on the intense laser pulse self-channeling in an NCD plasma~\cite{SSBulanov2010, bulanov.prab.2015, Park2019}.
The balance of the ponderomotive push by the laser and the Coulomb attraction between the displaced electrons and remaining ions determine the radius of such a channel.
In the case of an NCD plasma with a pre-formed channel of radius $r_\mathrm{h}$ (for simplicity we simulate a density step function $n_\mathrm{e} (r < r_\mathrm{h}) = 0$) the radius of the self-generated channel $R_\mathrm{ch}$ and the amplitude of the laser field $a_\mathrm{ch}$ in the channel can be determined using the same line of reasoning:
\begin{eqnarray}
    \frac{\pi^2}{\lambda^2}\left(R_\mathrm{ch}^2-r_\mathrm{h}^2\right) &=& a_\mathrm{ch}(r_\mathrm{h})\frac{n_\mathrm{cr}}{n_\mathrm{e}},\\
    \frac{\pi^2 R_\mathrm{ch}^2}{\lambda^2}a_\mathrm{ch}^2 &=& \frac{2P}{K P_\mathrm{c}}. \label{eq:channel_radius}
\end{eqnarray}
Here $a_\mathrm{ch}(r_\mathrm{h})=a_\mathrm{ch}\left[J_0(\kappa r_\mathrm{h})-J_2(\kappa r_\mathrm{h})\right]$, which comes from the solution of the wave equation in the wave guide~\cite{SSBulanov2010}, $J_0$ and $J_2$ are the Bessel functions, and $\kappa=1.84/R_\mathrm{ch}$. $P_\mathrm{c}=8 \pi \epsilon_0 m_\mathrm{e}^2 c^5/e^2=17$ GW is a characteristic power for relativistic self-focusing, $\lambda$ and $P$ are the laser wavelength and power respectively, and $K=1/13.5$ is a geometric factor.
In the case $r_\mathrm{h}=0$, these two equations reduce to the results of Ref.~\cite{bulanov.prab.2015}.
For the parameters used below the radius of the channel is $R_\mathrm{ch}=2.1$ $\mu$m, which is only 5\% smaller than in the case of a homogeneous NCD plasma. 

As mentioned above, the accelerating and focusing fields at the back of the target are due to the strong electron current flowing through the channel, following the laser pulse.
The peak value of the magnetic field generated by such a current scales as $B_{ch}\sim n_e R_{ch}^3$. 
Thus, the main difference in field strength between homogeneous and hollow targets is caused by a factor $(1-r_\mathrm{h}^2/R_\mathrm{ch}^2)$ difference of electrons available in the channel.


For the study of the booster concept, 3D3V particle-in-cell simulations with the code WarpX~\cite{FedeliHuebl2022,WarpX} are performed, using parameters readily available at existing PW laser facilities like the BELLA iP2 beamline~\cite{ObstHueblSPIE2023}.
The laser pulse in all simulations discussed in this letter has 9.6\,J of energy on target (Gaussian, central laser mode), a normalized amplitude of $a_0 = 42$, a central wavelength of $815\,\mathrm{nm}$, a duration of $29.8\,\mathrm{f s}$, a beam waist of $2.12\,\mathrm{\mu m}$ and is linearly polarized (in $x$).

The booster stage targets are each $28\,\mathrm{\mu m}$ in length, with a density of $2\,n_\mathrm{cr}$ and pre-formed channel radius of $r_\mathrm{h}=1.5\,\mathrm{\mu m}$. 
The hollow channel prevents interactions between the ion beam and the plasma within $r_\mathrm{h}$, which would otherwise impact the beam quality.
The laser pulse waist is larger than the pre-formed channel, so as to drive a strong channel current at radius $R_\mathrm{ch}$ for acceleration at the stage rear.
The laser-target interaction pulls electrons off the channel wall and creates a forward-propagating electron filament on the channel axis, as in regular MVA (see Fig.~\ref{fig:densities_fields}).
When the laser pulse and the electron filament pass the target rear, MVA-typical accelerating and focusing fields are created, forming an ``accelerating bucket''.
The ion beam experiences optimal acceleration before the electron filament disperses and field strengths subside.
It is temporally phase matched to the peak field in space $z_\mathrm{ab}$ and time.


Before exploring the collective boost of a high-charge bunch, the properties of the hollow MVA stage in terms of longitudinal and transverse single-particle acceptance are studied.
Characterizing the proton phase space that can be energy-boosted, non-interacting protons are varied in initial position and momentum space and tracked through the electromagnetic fields of the self-consistently modeled laser-driven plasma stage.

\begin{figure}[ht]
\includegraphics[width=\columnwidth]{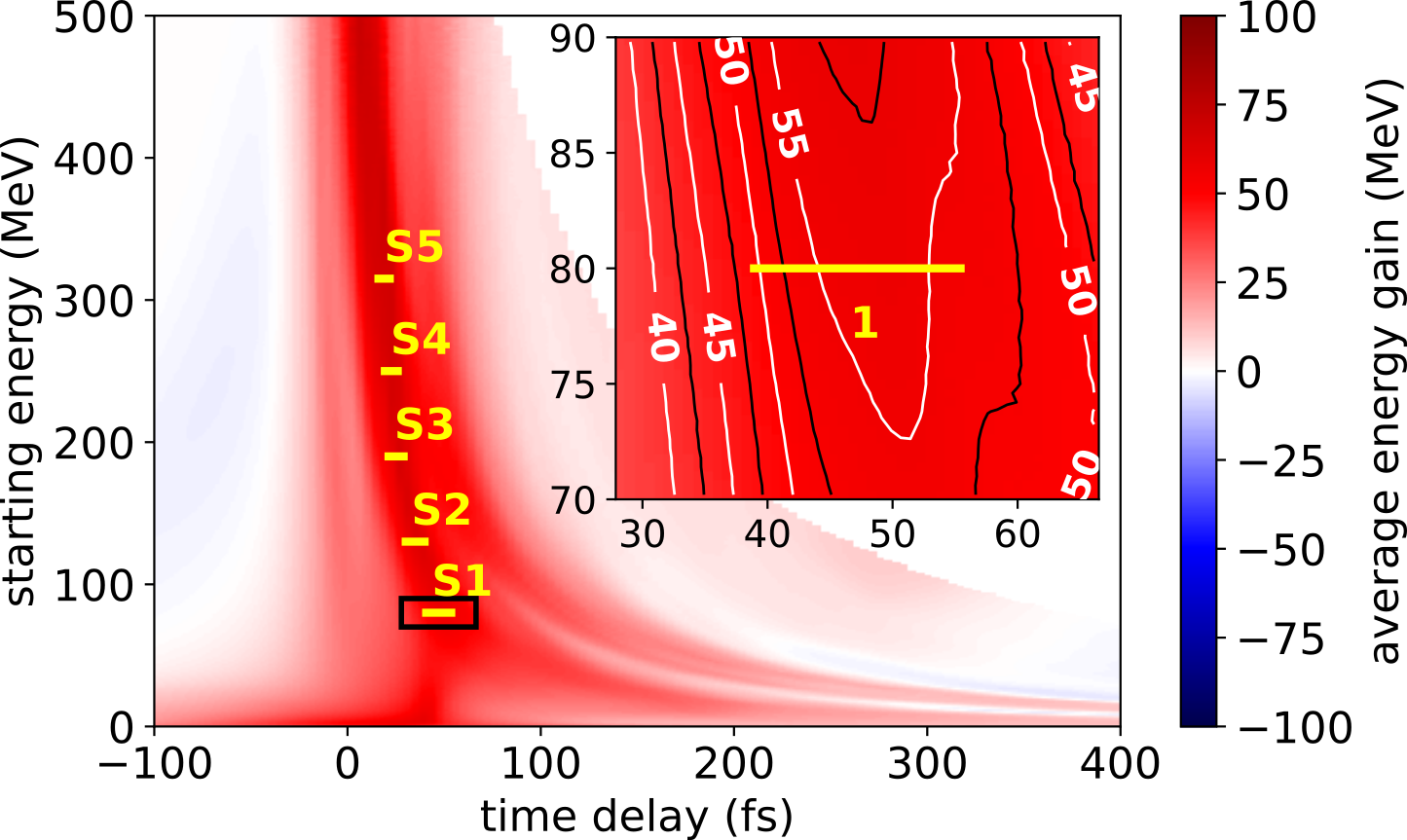}%
\caption{\label{fig:longitudinal_acceptance} 
    \textit{Longitudinal beam acceptance.}
    Average energy gain for tracked beam particles versus their starting energy and arrival time offset with respect to the laser maximum at $z_\mathrm{ab} = 15\,\mathrm{\mu m}$.
    The laser maximum is behind (before) the particle beam for negative (positive) times.
    Yellow horizontal lines mark design energies for five consecutive stages (S1--5) to boost a beam from $80\,\mathrm{MeV}$ to $380\,\mathrm{MeV}$ starting energy.
    The inset zooms into a region around stage S1 (black rectangle) where contour lines mark levels of constant energy gain.
}
\end{figure}

Figure~\ref{fig:longitudinal_acceptance} shows the longitudinal bunch acceptance, a measure for accepted bunch length and temporal jitter between bunch particles and laser pulse with respect to a reference plane at $z_\mathrm{ab}$.
Since the incoming ion beam is non-relativistic for early stages, it is timed to enter the plasma stage before the laser pulse and is overtaken by it during the course of the propagation.
Tracked particles were initially uniformly distributed within $0 \leq r_0 \leq 0.25\,\mathrm{\mu m}$ and without transverse momentum.

As long as the particle bunch arrives with or after the laser maximum in the accelerating bucket, for a large tolerance of over 200\,fs, an energy boost between $30 - 80\,\mathrm{MeV}$ is observed.
Particles that arrive outside this window are decelerated by less than $15\,\mathrm{MeV}$.
In Fig. \ref{fig:longitudinal_acceptance}, five possible consecutive stages are marked from $80$ to $380\,\mathrm{MeV}$.
The inset shows a zoomed detail for stage 1, with a ``flattened'' region of uniform acceleration for intense beams of up to 15\,fs in duration.

\newcolumntype{R}[1]{>{\raggedleft\let\newline\\\arraybackslash\hspace{0pt}}m{#1}}

\begin{figure}[ht]
    \centering
\includegraphics[width=\columnwidth]{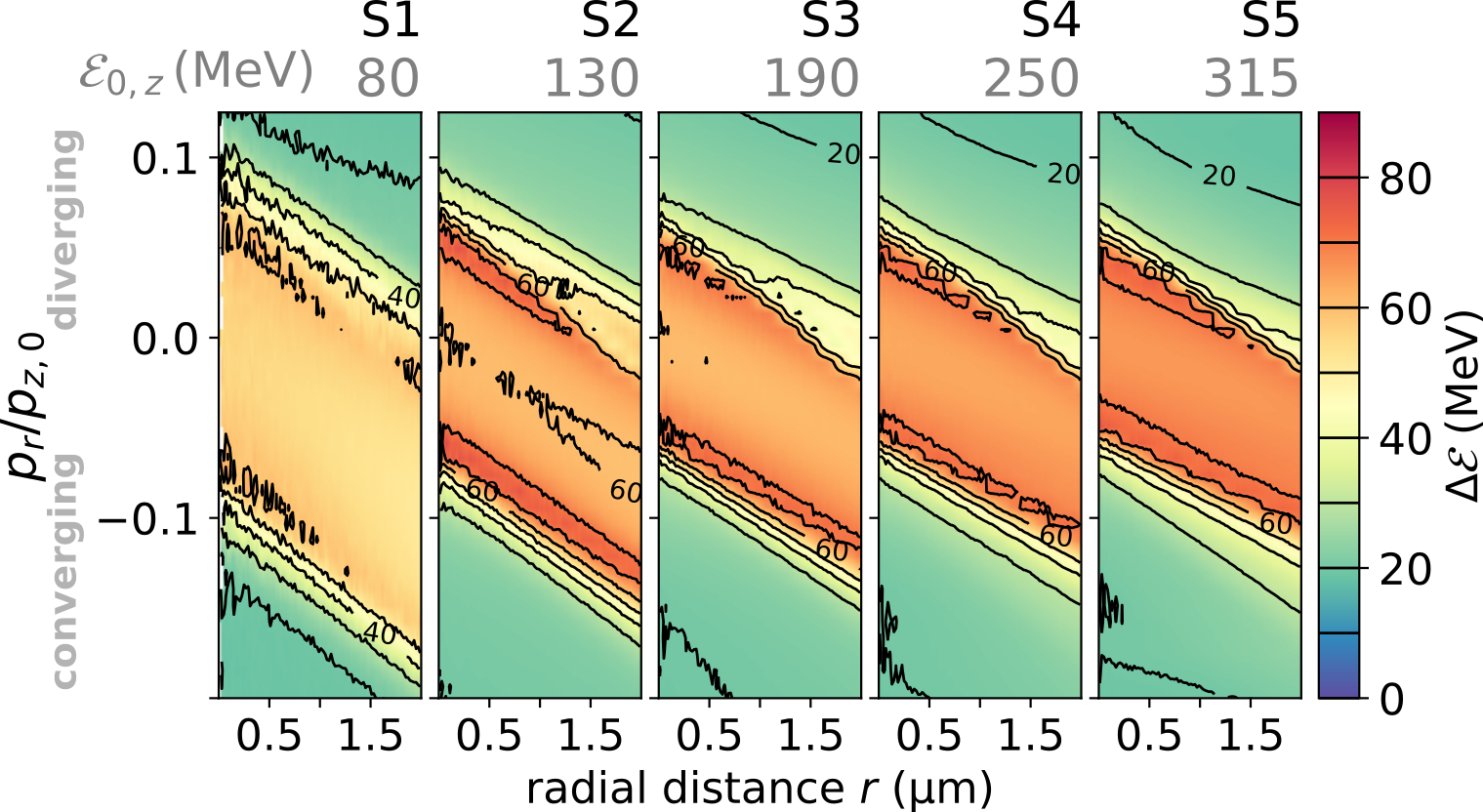}%
\newline
\centering
\begin{tabular}{r*{5}{R{1.08cm}}}
$\Delta \mathcal{E}$ & \textbf{S1} & \textbf{S2} & \textbf{S3} & \textbf{S4} & \textbf{S5} \\
\midrule
\textbf{30 MeV} & 29.0 & 33.0 & 35.2 & 39.5 & 44.3 \\
\textbf{40 MeV} & 25.9 & 29.2 & 30.0 & 32.3 & 34.9 \\
\textbf{50 MeV} & 22.7 & 26.0 & 26.2 & 28.8 & 31.5 \\
\textbf{60 MeV} &  \phantom{0}1.3 & 20.7 & 24.4 & 26.9 & 29.4 \\
\end{tabular}
\hspace{3em}
\caption{\label{fig:transverse_acceptance_fig_and_table}
    \textit{Transverse beam acceptance.}
    (top) Average energy gain $\Delta \mathcal{E}$ of tracked particles with respect to initial radial distance $r$ and initial normalized transverse momentum $p_r / p_{z,0}$.
    (bottom) Maximum accepted normalized transverse emittance in nm.
    Values are for minimum energy gains $\Delta \mathcal{E}$ of 30, 40, 50 and 60 MeV per stage, respectively.}
\end{figure}

Figure~\ref{fig:transverse_acceptance_fig_and_table} shows the correlation of transverse normalized momenta with radial position for beam acceptance into the booster stage.
For each stage S1 -- S5 in Fig.~\ref{fig:longitudinal_acceptance}, protons were tracked for a range of transverse initial conditions, and initial longitudinal momenta corresponding to their marked energies, respectively.

Every bunch initially measured $2\,\mathrm{\mu m}$ in length and $4\,\mathrm{\mu m}$ in diameter, was uniformly distributed and centered around a $z$-position corresponding to the delay expected to deliver the highest energy boost.
Transverse momenta were also uniformly distributed. 
A fraction of particles traversed the background plasma in the channel wall region to check that the transverse acceptance is not limited by the channel aperture, e.g., due to laser-plasma interaction dynamics.
Accommodating the cylindrical channel symmetry, we show the normalized radial momentum $p_r/p_z$ versus radial distance $r$, where $p_r = (p_x^2 + p_y^2)^{1/2}$.
For Fig.~\ref{fig:transverse_acceptance_fig_and_table}, particles had initial azimuthal momenta $p_\theta / mc$ smaller than $0.02$. 
Like before, the total energy gain is color-coded.
Spaced at every $10\,\mathrm{MeV}$, contour lines mark a constant energy gain between $0$ and $90\,\mathrm{MeV}$.
Slightly diverging particles on axis are still accepted by the stage.
With growing radial distance only focused, converging particles (negative normalized radial momentum) are still accepted.
Interestingly, for beams with initial energies larger than $130\,\mathrm{MeV}$, particles with a specific position--angle relation not aligned with the central propagation axis gain the highest energies in the stages S2--S5.

Calculated from the phase space regions contained in the respective contours in Fig.~\ref{fig:transverse_acceptance_fig_and_table}, the table lists the maximum accepted transverse emittances for each stage and selected minimum energy boosts:
For a beam with a Kapchinskij-Vladimirskij (KV) distribution~\cite{kvdist.cern.1959}, one can estimate this maximum accepted emittance by dividing the phase space area by $2 \pi$ (instead of $4 \pi$, due to projection of phase space to radial distance instead of transverse position) and multiplying by $p_{0,z}/mc$.
For boosts between $30$ and $50\,\mathrm{MeV}$, all stages accept a bunch with normalized emittance up to $23$--$44\,\mathrm{nm}$. 
The accepted emittance is sufficient for bunches of MVA sources ($<$\,20\,nm) and even increases acceptance for higher input energies. 



\begin{figure}[ht]
\includegraphics[width=\columnwidth]{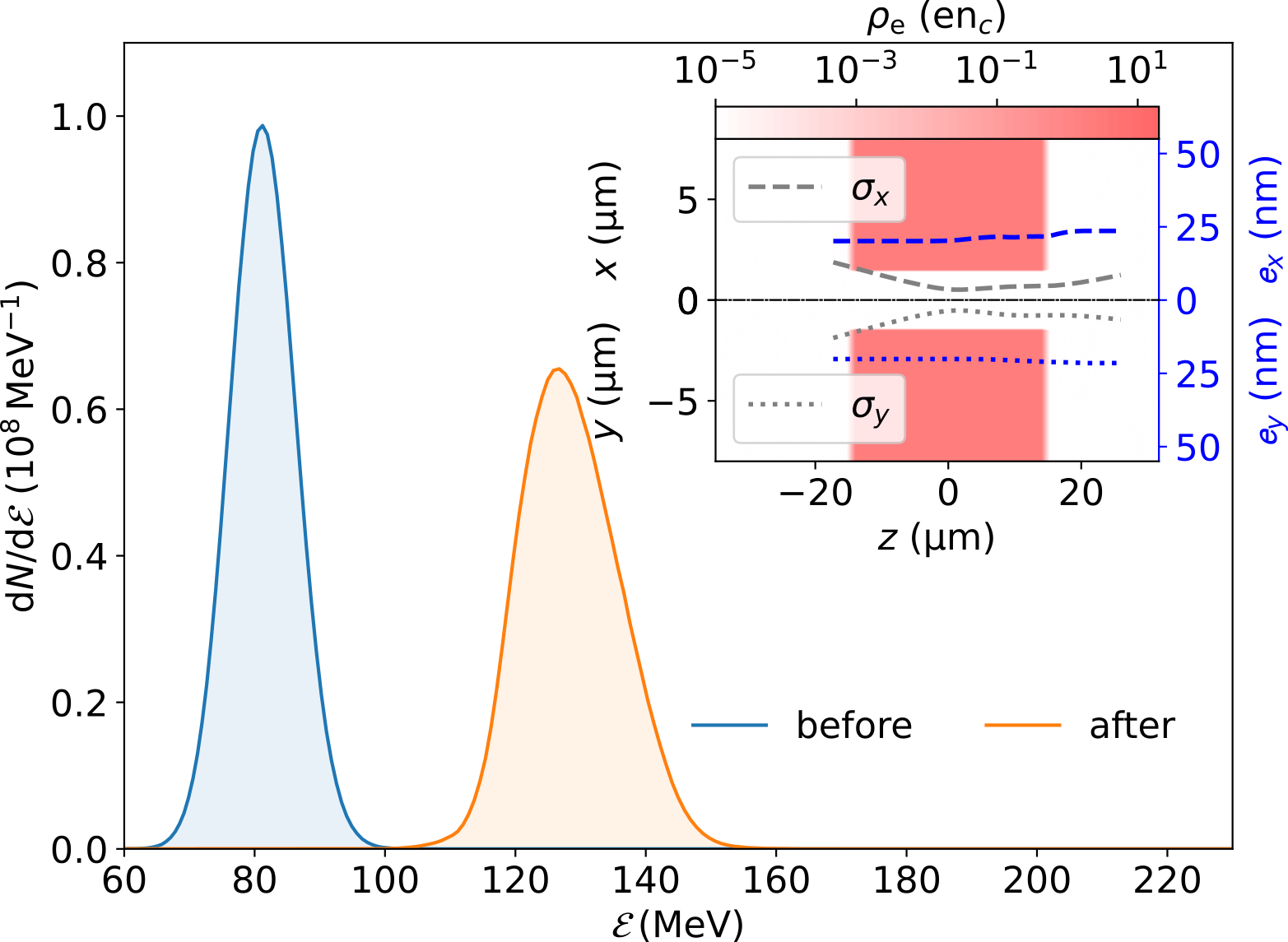}%
\caption{\label{fig:gain}
    \textit{Energy spectrum of a $200\,\mathrm{pC}$ proton beam} before (blue) and after (orange) the MVA booster element. 
    The inset shows the initial electron density distribution (red shaded area) of the stage in the $x$--$z$ and $y$--$z$ half-planes.
    Furthermore, grey lines show the evolution of transverse RMS beam sizes $\sigma_x$ (dashed) and $\sigma_y$ (dotted, mirrored downwards) along the longitudinal coordinate.
    Blue lines show the evolution of normalized transverse emittance $e_x$ (dashed) and $e_y$ (dotted), respectively.
}
\end{figure}

Including all collective effects of the energy-boosted beam, Fig. \ref{fig:gain} presents a fully self-consistent electromagnetic 3D WarpX simulation with an ultra-intense (35\,kA), high charge (200\,pC), narrowly focused proton bunch.
Informed by Figs.~\ref{fig:longitudinal_acceptance} and \ref{fig:transverse_acceptance_fig_and_table}, bunch parameters are a transverse KV-distribution, uniform in time, and a Gaussian, sub-relativistic  of 80$\pm 5$\,MeV (energy spread $\sigma_\mathcal{E}=5.1$\,\%).
In practice, the $200\,\mathrm{pC}$ of charge could be energy-selected from a wider spectrum of a present day PW laser ion source and ideally transported via an apochromatic beamline~\cite{Lindstroem2016,Brack2020,Wang2020}. 
It is assumed that the bunch has been phase-space rotated in a similar fashion as described by~\textcite{Busold2014}.

The inset of Fig.~\ref{fig:gain} presents the bunch envelope and emittance evolution.
Performed beam-loaded 3D simulations showed that emittance growth from space charge and subsequent channel-wall interaction occur in the second half of the channel ($z > 0$).
Mitigating this effect, the proton beam focal position was adjusted from $z_\mathrm{ab}$ at the target rear to $z_\mathrm{f} = 3\,\mu$m, exploiting that the beam is pinched in the strong laser-induced fields of the plasma channel (grey lines).

Fixing the transverse bunch size in focus ($2\sigma_{x,y}=1\,\mu$m) and choosing the transverse emittance to be between results from~\textcite{Hakimi2022} and the maximum accepted emittance from Fig.~\ref{fig:transverse_acceptance_fig_and_table}, namely $e_x = e_y = 20\,\mathrm{nm}$, results in Courant-Snyder (Twiss) parameters at the initialization plane at $\langle z \rangle_0 = -17.3\,\mathrm{\mu m}$: $\beta_\mathrm{x,0} = \beta_\mathrm{y,0} = 74.2\,\mathrm{\mu m}$, $\alpha_\mathrm{x,0} = \alpha_\mathrm{y,0} = 3.62$, and $\gamma_\mathrm{x,f} = \gamma_\mathrm{y,f} = 190\,000\,\mathrm{m^{-1}}$ via the beam envelope equations~\cite{CourantSnyder.AoP.1958} for a drift.

The central bunch energy is boosted in the stage by $50\,\mathrm{MeV}$, as predicted by tracking simulations in Fig.~\ref{fig:longitudinal_acceptance}.
No bunch charge is lost during the acceleration process.
The energy spread grew from 5 to $7\,\%$.
After the first booster stage, the emittance has grown by only $3.5\,\mathrm{nm}$.
These results are well within range for the above predicted coupling requirements to higher-energy stages and provide the opportunity for future research on design optimizations for energy-boosting to multiple stages.

In summary, a novel self-consistently modeled scheme is presented for boosting the energy of ultra-intense ion bunches of arbitrary $\beta$:
Staging can overcome the limits of achievable beam energy in a single laser-plasma interaction, which currently impedes the impact of laser-driven ion acceleration.
Central for a staging approach, 3D simulations confirm the conservation of charge, energy spread and emittance.
Additionally, the same laser-plasma stage can be used for a scalable range of input velocities, facilitating the coupling of multiple stages via the control of the temporal delay between laser pulse and boosted ion bunch.
Benefiting from the combined strong acceleration and focusing fields in magnetic vortex acceleration (10s of MV/$\mu$m), a hollow target is suitable as a robust plasma booster stage.
Source and booster laser-plasma parameter ranges are realistic for state-of-the-art and upcoming laser facilities around the world~\cite{ObstHueblSPIE2023,dipiazza2022multipetawatt}, motivating experimental validation and implementation for applications.

\newpage
\begin{acknowledgments}
We thank R. T. Sandberg, D. Terzani, R. Lehe and J. Qiang for helpful discussions.
Simulations used the open source particle-in-cell code WarpX in version \texttt{23.01}.
We acknowledge all WarpX contributors.
Simulation inputs, analysis scripts, source code and data are available in Refs.~\cite{WarpX,supplementary_material}.
This material is based upon work supported by the Defense Advanced Research Projects Agency via Northrop Grumman Corporation.
%
Partly supported by the U.S. DOE FES Postdoctoral Research Program, administered by ORISE under contract DE-SC0014664,
%
the U.S. DOE Office of Science Offices of ASCR, HEP and FES (incl. LaserNetUS) under Contract No. DE-AC02-05CH11231,
%
the
Scientific Discovery through Advanced Computing (SciDAC) program,
and the Exascale Computing Project (17-SC-20-SC).
%
This research used resources of the Oak Ridge Leadership Computing Facility at the ORNL (DE-AC05-00OR22725, ALCC program) and the National Energy Research Scientific Computing Center (DE-AC02-05CH11231, FES-ERCAP0024250).
\end{acknowledgments}



\bibliography{lpa_ion_booster}

\providecommand{\noopsort}[1]{}\providecommand{\singleletter}[1]{#1}%
\begin{thebibliography}{80}%
\makeatletter
\providecommand \@ifxundefined [1]{%
 \@ifx{#1\undefined}
}%
\providecommand \@ifnum [1]{%
 \ifnum #1\expandafter \@firstoftwo
 \else \expandafter \@secondoftwo
 \fi
}%
\providecommand \@ifx [1]{%
 \ifx #1\expandafter \@firstoftwo
 \else \expandafter \@secondoftwo
 \fi
}%
\providecommand \natexlab [1]{#1}%
\providecommand \enquote  [1]{``#1''}%
\providecommand \bibnamefont  [1]{#1}%
\providecommand \bibfnamefont [1]{#1}%
\providecommand \citenamefont [1]{#1}%
\providecommand \href@noop [0]{\@secondoftwo}%
\providecommand \href [0]{\begingroup \@sanitize@url \@href}%
\providecommand \@href[1]{\@@startlink{#1}\@@href}%
\providecommand \@@href[1]{\endgroup#1\@@endlink}%
\providecommand \@sanitize@url [0]{\catcode `\\12\catcode `\$12\catcode
  `\&12\catcode `\#12\catcode `\^12\catcode `\_12\catcode `\%12\relax}%
\providecommand \@@startlink[1]{}%
\providecommand \@@endlink[0]{}%
\providecommand \url  [0]{\begingroup\@sanitize@url \@url }%
\providecommand \@url [1]{\endgroup\@href {#1}{\urlprefix }}%
\providecommand \urlprefix  [0]{URL }%
\providecommand \Eprint [0]{\href }%
\providecommand \doibase [0]{http://dx.doi.org/}%
\providecommand \selectlanguage [0]{\@gobble}%
\providecommand \bibinfo  [0]{\@secondoftwo}%
\providecommand \bibfield  [0]{\@secondoftwo}%
\providecommand \translation [1]{[#1]}%
\providecommand \BibitemOpen [0]{}%
\providecommand \bibitemStop [0]{}%
\providecommand \bibitemNoStop [0]{.\EOS\space}%
\providecommand \EOS [0]{\spacefactor3000\relax}%
\providecommand \BibitemShut  [1]{\csname bibitem#1\endcsname}%
\let\auto@bib@innerbib\@empty
\bibitem [{\citenamefont {Mourou}\ \emph {et~al.}(2006)\citenamefont {Mourou},
  \citenamefont {Tajima},\ and\ \citenamefont {Bulanov}}]{mourou.rmp.2006}%
  \BibitemOpen
  \bibfield  {author} {\bibinfo {author} {\bibfnamefont {G.~A.}\ \bibnamefont
  {Mourou}}, \bibinfo {author} {\bibfnamefont {T.}~\bibnamefont {Tajima}}, \
  and\ \bibinfo {author} {\bibfnamefont {S.~V.}\ \bibnamefont {Bulanov}},\
  }\href@noop {} {\bibfield  {journal} {\bibinfo  {journal} {Rev. Mod. Phys.}\
  }\textbf {\bibinfo {volume} {78}},\ \bibinfo {pages} {309} (\bibinfo {year}
  {2006})}\BibitemShut {NoStop}%
\bibitem [{\citenamefont {Daido}\ \emph {et~al.}(2012)\citenamefont {Daido},
  \citenamefont {Nishiuchi},\ and\ \citenamefont {Pirozhkov}}]{daido.rpp.2012}%
  \BibitemOpen
  \bibfield  {author} {\bibinfo {author} {\bibfnamefont {H.}~\bibnamefont
  {Daido}}, \bibinfo {author} {\bibfnamefont {M.}~\bibnamefont {Nishiuchi}}, \
  and\ \bibinfo {author} {\bibfnamefont {A.~S.}\ \bibnamefont {Pirozhkov}},\
  }\href@noop {} {\bibfield  {journal} {\bibinfo  {journal} {Reports on
  Progress in Physics}\ }\textbf {\bibinfo {volume} {75}},\ \bibinfo {pages}
  {056401} (\bibinfo {year} {2012})}\BibitemShut {NoStop}%
\bibitem [{\citenamefont {Macchi}\ \emph {et~al.}(2013)\citenamefont {Macchi},
  \citenamefont {Borghesi},\ and\ \citenamefont {Passoni}}]{macchi.rmp.2013}%
  \BibitemOpen
  \bibfield  {author} {\bibinfo {author} {\bibfnamefont {A.}~\bibnamefont
  {Macchi}}, \bibinfo {author} {\bibfnamefont {M.}~\bibnamefont {Borghesi}}, \
  and\ \bibinfo {author} {\bibfnamefont {M.}~\bibnamefont {Passoni}},\
  }\href@noop {} {\bibfield  {journal} {\bibinfo  {journal} {Reviews of Modern
  Physics}\ }\textbf {\bibinfo {volume} {85}},\ \bibinfo {pages} {751}
  (\bibinfo {year} {2013})}\BibitemShut {NoStop}%
\bibitem [{\citenamefont {Cowan}\ \emph {et~al.}(2004)\citenamefont {Cowan},
  \citenamefont {Fuchs}, \citenamefont {Ruhl}, \citenamefont {Kemp},
  \citenamefont {Audebert}, \citenamefont {Roth}, \citenamefont {Stephens},
  \citenamefont {Barton}, \citenamefont {Blazevic}, \citenamefont {Brambrink},
  \citenamefont {Cobble}, \citenamefont {Fern\'andez}, \citenamefont
  {Gauthier}, \citenamefont {Geissel}, \citenamefont {Hegelich}, \citenamefont
  {Kaae}, \citenamefont {Karsch}, \citenamefont {Le~Sage}, \citenamefont
  {Letzring}, \citenamefont {Manclossi}, \citenamefont {Meyroneinc},
  \citenamefont {Newkirk}, \citenamefont {P\'epin},\ and\ \citenamefont
  {Renard-LeGalloudec}}]{Cowan.PRL.2004}%
  \BibitemOpen
  \bibfield  {author} {\bibinfo {author} {\bibfnamefont {T.~E.}\ \bibnamefont
  {Cowan}}, \bibinfo {author} {\bibfnamefont {J.}~\bibnamefont {Fuchs}},
  \bibinfo {author} {\bibfnamefont {H.}~\bibnamefont {Ruhl}}, \bibinfo {author}
  {\bibfnamefont {A.}~\bibnamefont {Kemp}}, \bibinfo {author} {\bibfnamefont
  {P.}~\bibnamefont {Audebert}}, \bibinfo {author} {\bibfnamefont
  {M.}~\bibnamefont {Roth}}, \bibinfo {author} {\bibfnamefont {R.}~\bibnamefont
  {Stephens}}, \bibinfo {author} {\bibfnamefont {I.}~\bibnamefont {Barton}},
  \bibinfo {author} {\bibfnamefont {A.}~\bibnamefont {Blazevic}}, \bibinfo
  {author} {\bibfnamefont {E.}~\bibnamefont {Brambrink}}, \bibinfo {author}
  {\bibfnamefont {J.}~\bibnamefont {Cobble}}, \bibinfo {author} {\bibfnamefont
  {J.}~\bibnamefont {Fern\'andez}}, \bibinfo {author} {\bibfnamefont {J.-C.}\
  \bibnamefont {Gauthier}}, \bibinfo {author} {\bibfnamefont {M.}~\bibnamefont
  {Geissel}}, \bibinfo {author} {\bibfnamefont {M.}~\bibnamefont {Hegelich}},
  \bibinfo {author} {\bibfnamefont {J.}~\bibnamefont {Kaae}}, \bibinfo {author}
  {\bibfnamefont {S.}~\bibnamefont {Karsch}}, \bibinfo {author} {\bibfnamefont
  {G.~P.}\ \bibnamefont {Le~Sage}}, \bibinfo {author} {\bibfnamefont
  {S.}~\bibnamefont {Letzring}}, \bibinfo {author} {\bibfnamefont
  {M.}~\bibnamefont {Manclossi}}, \bibinfo {author} {\bibfnamefont
  {S.}~\bibnamefont {Meyroneinc}}, \bibinfo {author} {\bibfnamefont
  {A.}~\bibnamefont {Newkirk}}, \bibinfo {author} {\bibfnamefont
  {H.}~\bibnamefont {P\'epin}}, \ and\ \bibinfo {author} {\bibfnamefont
  {N.}~\bibnamefont {Renard-LeGalloudec}},\ }\href {\doibase
  10.1103/PhysRevLett.92.204801} {\bibfield  {journal} {\bibinfo  {journal}
  {Phys. Rev. Lett.}\ }\textbf {\bibinfo {volume} {92}},\ \bibinfo {pages}
  {204801} (\bibinfo {year} {2004})}\BibitemShut {NoStop}%
\bibitem [{\citenamefont {Borghesi}\ \emph {et~al.}(2004)\citenamefont
  {Borghesi}, \citenamefont {Mackinnon}, \citenamefont {Campbell},
  \citenamefont {Hicks}, \citenamefont {Kar}, \citenamefont {Patel},
  \citenamefont {Price}, \citenamefont {Romagnani}, \citenamefont {Schiavi},\
  and\ \citenamefont {Willi}}]{Borghesi.PRL.2004}%
  \BibitemOpen
  \bibfield  {author} {\bibinfo {author} {\bibfnamefont {M.}~\bibnamefont
  {Borghesi}}, \bibinfo {author} {\bibfnamefont {A.~J.}\ \bibnamefont
  {Mackinnon}}, \bibinfo {author} {\bibfnamefont {D.~H.}\ \bibnamefont
  {Campbell}}, \bibinfo {author} {\bibfnamefont {D.~G.}\ \bibnamefont {Hicks}},
  \bibinfo {author} {\bibfnamefont {S.}~\bibnamefont {Kar}}, \bibinfo {author}
  {\bibfnamefont {P.~K.}\ \bibnamefont {Patel}}, \bibinfo {author}
  {\bibfnamefont {D.}~\bibnamefont {Price}}, \bibinfo {author} {\bibfnamefont
  {L.}~\bibnamefont {Romagnani}}, \bibinfo {author} {\bibfnamefont
  {A.}~\bibnamefont {Schiavi}}, \ and\ \bibinfo {author} {\bibfnamefont
  {O.}~\bibnamefont {Willi}},\ }\href {\doibase 10.1103/PhysRevLett.92.055003}
  {\bibfield  {journal} {\bibinfo  {journal} {Physical Review Letters}\
  }\textbf {\bibinfo {volume} {92}},\ \bibinfo {pages} {055003} (\bibinfo
  {year} {2004})}\BibitemShut {NoStop}%
\bibitem [{\citenamefont {Nürnberg}\ \emph {et~al.}(2009)\citenamefont
  {Nürnberg}, \citenamefont {Schollmeier}, \citenamefont {Brambrink},
  \citenamefont {Blažević}, \citenamefont {Carroll}, \citenamefont {Flippo},
  \citenamefont {Gautier}, \citenamefont {Geißel}, \citenamefont {Harres},
  \citenamefont {Hegelich}, \citenamefont {Lundh}, \citenamefont {Markey},
  \citenamefont {McKenna}, \citenamefont {Neely}, \citenamefont {Schreiber},\
  and\ \citenamefont {Roth}}]{Nuernberg.RSI.2009}%
  \BibitemOpen
  \bibfield  {author} {\bibinfo {author} {\bibfnamefont {F.}~\bibnamefont
  {Nürnberg}}, \bibinfo {author} {\bibfnamefont {M.}~\bibnamefont
  {Schollmeier}}, \bibinfo {author} {\bibfnamefont {E.}~\bibnamefont
  {Brambrink}}, \bibinfo {author} {\bibfnamefont {A.}~\bibnamefont
  {Blažević}}, \bibinfo {author} {\bibfnamefont {D.~C.}\ \bibnamefont
  {Carroll}}, \bibinfo {author} {\bibfnamefont {K.}~\bibnamefont {Flippo}},
  \bibinfo {author} {\bibfnamefont {D.~C.}\ \bibnamefont {Gautier}}, \bibinfo
  {author} {\bibfnamefont {M.}~\bibnamefont {Geißel}}, \bibinfo {author}
  {\bibfnamefont {K.}~\bibnamefont {Harres}}, \bibinfo {author} {\bibfnamefont
  {B.~M.}\ \bibnamefont {Hegelich}}, \bibinfo {author} {\bibfnamefont
  {O.}~\bibnamefont {Lundh}}, \bibinfo {author} {\bibfnamefont
  {K.}~\bibnamefont {Markey}}, \bibinfo {author} {\bibfnamefont
  {P.}~\bibnamefont {McKenna}}, \bibinfo {author} {\bibfnamefont
  {D.}~\bibnamefont {Neely}}, \bibinfo {author} {\bibfnamefont
  {J.}~\bibnamefont {Schreiber}}, \ and\ \bibinfo {author} {\bibfnamefont
  {M.}~\bibnamefont {Roth}},\ }\href {\doibase 10.1063/1.3086424} {\bibfield
  {journal} {\bibinfo  {journal} {Review of Scientific Instruments}\ }\textbf
  {\bibinfo {volume} {80}} (\bibinfo {year} {2009}),\
  10.1063/1.3086424}\BibitemShut {NoStop}%
\bibitem [{\citenamefont {Abada}(2019)}]{abada_fcc-hh_2019}%
  \BibitemOpen
  \bibfield  {author} {\bibinfo {author} {\bibfnamefont {A.~e. a. T. F.~C.}\
  \bibnamefont {Abada}},\ }\href {\doibase 10.1140/epjst/e2019-900087-0}
  {\bibfield  {journal} {\bibinfo  {journal} {The European Physical Journal
  Special Topics}\ }\textbf {\bibinfo {volume} {228}},\ \bibinfo {pages} {755}
  (\bibinfo {year} {2019})}\BibitemShut {NoStop}%
\bibitem [{\citenamefont {Geer}\ and\ \citenamefont
  {Zisman}(2007)}]{Geer.ProgPartNuclPhys.2007}%
  \BibitemOpen
  \bibfield  {author} {\bibinfo {author} {\bibfnamefont {S.}~\bibnamefont
  {Geer}}\ and\ \bibinfo {author} {\bibfnamefont {M.~S.}\ \bibnamefont
  {Zisman}},\ }\href {\doibase https://doi.org/10.1016/j.ppnp.2007.03.001}
  {\bibfield  {journal} {\bibinfo  {journal} {Progress in Particle and Nuclear
  Physics}\ }\textbf {\bibinfo {volume} {59}},\ \bibinfo {pages} {631}
  (\bibinfo {year} {2007})}\BibitemShut {NoStop}%
\bibitem [{\citenamefont {Roth}\ \emph {et~al.}(2001)\citenamefont {Roth},
  \citenamefont {Cowan}, \citenamefont {Key}, \citenamefont {Hatchett},
  \citenamefont {Brown}, \citenamefont {Fountain}, \citenamefont {Johnson},
  \citenamefont {Pennington}, \citenamefont {Snavely}, \citenamefont {Wilks},
  \citenamefont {Yasuike}, \citenamefont {Ruhl}, \citenamefont {Pegoraro},
  \citenamefont {Bulanov}, \citenamefont {Campbell}, \citenamefont {Perry},\
  and\ \citenamefont {Powell}}]{roth.prl.2001}%
  \BibitemOpen
  \bibfield  {author} {\bibinfo {author} {\bibfnamefont {M.}~\bibnamefont
  {Roth}}, \bibinfo {author} {\bibfnamefont {T.~E.}\ \bibnamefont {Cowan}},
  \bibinfo {author} {\bibfnamefont {M.~H.}\ \bibnamefont {Key}}, \bibinfo
  {author} {\bibfnamefont {S.~P.}\ \bibnamefont {Hatchett}}, \bibinfo {author}
  {\bibfnamefont {C.}~\bibnamefont {Brown}}, \bibinfo {author} {\bibfnamefont
  {W.}~\bibnamefont {Fountain}}, \bibinfo {author} {\bibfnamefont
  {J.}~\bibnamefont {Johnson}}, \bibinfo {author} {\bibfnamefont {D.~M.}\
  \bibnamefont {Pennington}}, \bibinfo {author} {\bibfnamefont {R.~A.}\
  \bibnamefont {Snavely}}, \bibinfo {author} {\bibfnamefont {S.~C.}\
  \bibnamefont {Wilks}}, \bibinfo {author} {\bibfnamefont {K.}~\bibnamefont
  {Yasuike}}, \bibinfo {author} {\bibfnamefont {H.}~\bibnamefont {Ruhl}},
  \bibinfo {author} {\bibfnamefont {F.}~\bibnamefont {Pegoraro}}, \bibinfo
  {author} {\bibfnamefont {S.~V.}\ \bibnamefont {Bulanov}}, \bibinfo {author}
  {\bibfnamefont {E.~M.}\ \bibnamefont {Campbell}}, \bibinfo {author}
  {\bibfnamefont {M.~D.}\ \bibnamefont {Perry}}, \ and\ \bibinfo {author}
  {\bibfnamefont {H.}~\bibnamefont {Powell}},\ }\href {\doibase
  10.1103/PhysRevLett.86.436} {\bibfield  {journal} {\bibinfo  {journal} {Phys.
  Rev. Lett.}\ }\textbf {\bibinfo {volume} {86}},\ \bibinfo {pages} {436}
  (\bibinfo {year} {2001})}\BibitemShut {NoStop}%
\bibitem [{\citenamefont {Mackinnon}\ \emph {et~al.}(2006)\citenamefont
  {Mackinnon}, \citenamefont {Patel}, \citenamefont {Borghesi}, \citenamefont
  {Clarke}, \citenamefont {Freeman}, \citenamefont {Habara}, \citenamefont
  {Hatchett}, \citenamefont {Hey}, \citenamefont {Hicks}, \citenamefont {Kar},
  \citenamefont {Key}, \citenamefont {King}, \citenamefont {Lancaster},
  \citenamefont {Neely}, \citenamefont {Nikkro}, \citenamefont {Norreys},
  \citenamefont {Notley}, \citenamefont {Phillips}, \citenamefont {Romagnani},
  \citenamefont {Snavely}, \citenamefont {Stephens},\ and\ \citenamefont
  {Town}}]{mackinnon.prl.2006}%
  \BibitemOpen
  \bibfield  {author} {\bibinfo {author} {\bibfnamefont {A.~J.}\ \bibnamefont
  {Mackinnon}}, \bibinfo {author} {\bibfnamefont {P.~K.}\ \bibnamefont
  {Patel}}, \bibinfo {author} {\bibfnamefont {M.}~\bibnamefont {Borghesi}},
  \bibinfo {author} {\bibfnamefont {R.~C.}\ \bibnamefont {Clarke}}, \bibinfo
  {author} {\bibfnamefont {R.~R.}\ \bibnamefont {Freeman}}, \bibinfo {author}
  {\bibfnamefont {H.}~\bibnamefont {Habara}}, \bibinfo {author} {\bibfnamefont
  {S.~P.}\ \bibnamefont {Hatchett}}, \bibinfo {author} {\bibfnamefont
  {D.}~\bibnamefont {Hey}}, \bibinfo {author} {\bibfnamefont {D.~G.}\
  \bibnamefont {Hicks}}, \bibinfo {author} {\bibfnamefont {S.}~\bibnamefont
  {Kar}}, \bibinfo {author} {\bibfnamefont {M.~H.}\ \bibnamefont {Key}},
  \bibinfo {author} {\bibfnamefont {J.~A.}\ \bibnamefont {King}}, \bibinfo
  {author} {\bibfnamefont {K.}~\bibnamefont {Lancaster}}, \bibinfo {author}
  {\bibfnamefont {D.}~\bibnamefont {Neely}}, \bibinfo {author} {\bibfnamefont
  {A.}~\bibnamefont {Nikkro}}, \bibinfo {author} {\bibfnamefont {P.~A.}\
  \bibnamefont {Norreys}}, \bibinfo {author} {\bibfnamefont {M.~M.}\
  \bibnamefont {Notley}}, \bibinfo {author} {\bibfnamefont {T.~W.}\
  \bibnamefont {Phillips}}, \bibinfo {author} {\bibfnamefont {L.}~\bibnamefont
  {Romagnani}}, \bibinfo {author} {\bibfnamefont {R.~A.}\ \bibnamefont
  {Snavely}}, \bibinfo {author} {\bibfnamefont {R.~B.}\ \bibnamefont
  {Stephens}}, \ and\ \bibinfo {author} {\bibfnamefont {R.~P.~J.}\ \bibnamefont
  {Town}},\ }\href {\doibase 10.1103/PhysRevLett.97.045001} {\bibfield
  {journal} {\bibinfo  {journal} {Phys. Rev. Lett.}\ }\textbf {\bibinfo
  {volume} {97}},\ \bibinfo {pages} {045001} (\bibinfo {year}
  {2006})}\BibitemShut {NoStop}%
\bibitem [{\citenamefont {Favaudon}\ \emph {et~al.}(2014)\citenamefont
  {Favaudon}, \citenamefont {Caplier}, \citenamefont {Monceau}, \citenamefont
  {Pouzoulet}, \citenamefont {Sayarath}, \citenamefont {Fouillade},
  \citenamefont {Poupon}, \citenamefont {Brito}, \citenamefont {Hupé},
  \citenamefont {Bourhis}, \citenamefont {Hall}, \citenamefont {Fontaine},\
  and\ \citenamefont {Vozenin}}]{Favaudon2014}%
  \BibitemOpen
  \bibfield  {author} {\bibinfo {author} {\bibfnamefont {V.}~\bibnamefont
  {Favaudon}}, \bibinfo {author} {\bibfnamefont {L.}~\bibnamefont {Caplier}},
  \bibinfo {author} {\bibfnamefont {V.}~\bibnamefont {Monceau}}, \bibinfo
  {author} {\bibfnamefont {F.}~\bibnamefont {Pouzoulet}}, \bibinfo {author}
  {\bibfnamefont {M.}~\bibnamefont {Sayarath}}, \bibinfo {author}
  {\bibfnamefont {C.}~\bibnamefont {Fouillade}}, \bibinfo {author}
  {\bibfnamefont {M.-F.}\ \bibnamefont {Poupon}}, \bibinfo {author}
  {\bibfnamefont {I.}~\bibnamefont {Brito}}, \bibinfo {author} {\bibfnamefont
  {P.}~\bibnamefont {Hupé}}, \bibinfo {author} {\bibfnamefont
  {J.}~\bibnamefont {Bourhis}}, \bibinfo {author} {\bibfnamefont
  {J.}~\bibnamefont {Hall}}, \bibinfo {author} {\bibfnamefont {J.-J.}\
  \bibnamefont {Fontaine}}, \ and\ \bibinfo {author} {\bibfnamefont {M.-C.}\
  \bibnamefont {Vozenin}},\ }\href {\doibase 10.1126/scitranslmed.3008973}
  {\bibfield  {journal} {\bibinfo  {journal} {Science Translational Medicine}\
  }\textbf {\bibinfo {volume} {6}},\ \bibinfo {pages} {245ra93} (\bibinfo
  {year} {2014})}\BibitemShut {NoStop}%
\bibitem [{\citenamefont {Montay-Gruel}\ \emph {et~al.}(2017)\citenamefont
  {Montay-Gruel}, \citenamefont {Petersson}, \citenamefont {Jaccard},
  \citenamefont {Boivin}, \citenamefont {Germond}, \citenamefont {Petit},
  \citenamefont {Doenlen}, \citenamefont {Favaudon}, \citenamefont {Bochud},
  \citenamefont {Bailat}, \citenamefont {Bourhis},\ and\ \citenamefont
  {Vozenin}}]{MontayGruel2017}%
  \BibitemOpen
  \bibfield  {author} {\bibinfo {author} {\bibfnamefont {P.}~\bibnamefont
  {Montay-Gruel}}, \bibinfo {author} {\bibfnamefont {K.}~\bibnamefont
  {Petersson}}, \bibinfo {author} {\bibfnamefont {M.}~\bibnamefont {Jaccard}},
  \bibinfo {author} {\bibfnamefont {G.}~\bibnamefont {Boivin}}, \bibinfo
  {author} {\bibfnamefont {J.-F.}\ \bibnamefont {Germond}}, \bibinfo {author}
  {\bibfnamefont {B.}~\bibnamefont {Petit}}, \bibinfo {author} {\bibfnamefont
  {R.}~\bibnamefont {Doenlen}}, \bibinfo {author} {\bibfnamefont
  {V.}~\bibnamefont {Favaudon}}, \bibinfo {author} {\bibfnamefont
  {F.}~\bibnamefont {Bochud}}, \bibinfo {author} {\bibfnamefont
  {C.}~\bibnamefont {Bailat}}, \bibinfo {author} {\bibfnamefont
  {J.}~\bibnamefont {Bourhis}}, \ and\ \bibinfo {author} {\bibfnamefont
  {M.-C.}\ \bibnamefont {Vozenin}},\ }\href {\doibase
  10.1016/j.radonc.2017.05.003} {\bibfield  {journal} {\bibinfo  {journal}
  {Radiotherapy and Oncology}\ }\textbf {\bibinfo {volume} {124}},\ \bibinfo
  {pages} {365} (\bibinfo {year} {2017})},\ \bibinfo {note} {15th International
  Wolfsberg Meeting 2017}\BibitemShut {NoStop}%
\bibitem [{\citenamefont {Bin}\ \emph {et~al.}(2022)\citenamefont {Bin},
  \citenamefont {Obst-Huebl}, \citenamefont {Mao}, \citenamefont {Nakamura},
  \citenamefont {Geulig}, \citenamefont {Chang}, \citenamefont {Ji},
  \citenamefont {He}, \citenamefont {De~Chant}, \citenamefont {Kober},
  \citenamefont {Gonsalves}, \citenamefont {Bulanov}, \citenamefont {Celniker},
  \citenamefont {Schroeder}, \citenamefont {Geddes}, \citenamefont {Esarey},
  \citenamefont {Simmons}, \citenamefont {Schenkel}, \citenamefont {Blakely},
  \citenamefont {Steinke},\ and\ \citenamefont {Snijders}}]{Bin2022}%
  \BibitemOpen
  \bibfield  {author} {\bibinfo {author} {\bibfnamefont {J.}~\bibnamefont
  {Bin}}, \bibinfo {author} {\bibfnamefont {L.}~\bibnamefont {Obst-Huebl}},
  \bibinfo {author} {\bibfnamefont {J.-H.}\ \bibnamefont {Mao}}, \bibinfo
  {author} {\bibfnamefont {K.}~\bibnamefont {Nakamura}}, \bibinfo {author}
  {\bibfnamefont {L.~D.}\ \bibnamefont {Geulig}}, \bibinfo {author}
  {\bibfnamefont {H.}~\bibnamefont {Chang}}, \bibinfo {author} {\bibfnamefont
  {Q.}~\bibnamefont {Ji}}, \bibinfo {author} {\bibfnamefont {L.}~\bibnamefont
  {He}}, \bibinfo {author} {\bibfnamefont {J.}~\bibnamefont {De~Chant}},
  \bibinfo {author} {\bibfnamefont {Z.}~\bibnamefont {Kober}}, \bibinfo
  {author} {\bibfnamefont {A.~J.}\ \bibnamefont {Gonsalves}}, \bibinfo {author}
  {\bibfnamefont {S.}~\bibnamefont {Bulanov}}, \bibinfo {author} {\bibfnamefont
  {S.~E.}\ \bibnamefont {Celniker}}, \bibinfo {author} {\bibfnamefont {C.~B.}\
  \bibnamefont {Schroeder}}, \bibinfo {author} {\bibfnamefont {C.~G.~R.}\
  \bibnamefont {Geddes}}, \bibinfo {author} {\bibfnamefont {E.}~\bibnamefont
  {Esarey}}, \bibinfo {author} {\bibfnamefont {B.~A.}\ \bibnamefont {Simmons}},
  \bibinfo {author} {\bibfnamefont {T.}~\bibnamefont {Schenkel}}, \bibinfo
  {author} {\bibfnamefont {E.~A.}\ \bibnamefont {Blakely}}, \bibinfo {author}
  {\bibfnamefont {S.}~\bibnamefont {Steinke}}, \ and\ \bibinfo {author}
  {\bibfnamefont {A.~M.}\ \bibnamefont {Snijders}},\ }\href {\doibase
  10.1038/s41598-022-05181-3} {\bibfield  {journal} {\bibinfo  {journal}
  {Scientific Reports}\ }\textbf {\bibinfo {volume} {12}} (\bibinfo {year}
  {2022}),\ 10.1038/s41598-022-05181-3}\BibitemShut {NoStop}%
\bibitem [{\citenamefont {Bulanov}\ and\ \citenamefont
  {Khoroshkov}(2002)}]{SVBulanov2002}%
  \BibitemOpen
  \bibfield  {author} {\bibinfo {author} {\bibfnamefont {S.~V.}\ \bibnamefont
  {Bulanov}}\ and\ \bibinfo {author} {\bibfnamefont {V.~S.}\ \bibnamefont
  {Khoroshkov}},\ }\href {\doibase 10.1134/1.1478534} {\bibfield  {journal}
  {\bibinfo  {journal} {Plasma Physics Reports}\ }\textbf {\bibinfo {volume}
  {28}},\ \bibinfo {pages} {453} (\bibinfo {year} {2002})}\BibitemShut
  {NoStop}%
\bibitem [{\citenamefont {Bulanov}\ \emph {et~al.}(2014)\citenamefont
  {Bulanov}, \citenamefont {Wilkens}, \citenamefont {Esirkepov}, \citenamefont
  {Korn}, \citenamefont {Kraft}, \citenamefont {Kraft}, \citenamefont {Molls},\
  and\ \citenamefont {Khoroshkov}}]{bulanov.pu.2014}%
  \BibitemOpen
  \bibfield  {author} {\bibinfo {author} {\bibfnamefont {S.~V.}\ \bibnamefont
  {Bulanov}}, \bibinfo {author} {\bibfnamefont {J.~J.}\ \bibnamefont
  {Wilkens}}, \bibinfo {author} {\bibfnamefont {T.~Z.}\ \bibnamefont
  {Esirkepov}}, \bibinfo {author} {\bibfnamefont {G.}~\bibnamefont {Korn}},
  \bibinfo {author} {\bibfnamefont {G.}~\bibnamefont {Kraft}}, \bibinfo
  {author} {\bibfnamefont {S.~D.}\ \bibnamefont {Kraft}}, \bibinfo {author}
  {\bibfnamefont {M.}~\bibnamefont {Molls}}, \ and\ \bibinfo {author}
  {\bibfnamefont {V.~S.}\ \bibnamefont {Khoroshkov}},\ }\href {\doibase
  10.3367/ufne.0184.201412a.1265} {\bibfield  {journal} {\bibinfo  {journal}
  {Physics-Uspekhi}\ }\textbf {\bibinfo {volume} {57}},\ \bibinfo {pages}
  {1149} (\bibinfo {year} {2014})}\BibitemShut {NoStop}%
\bibitem [{\citenamefont {Karsch}\ \emph {et~al.}(2017)\citenamefont {Karsch},
  \citenamefont {Beyreuther}, \citenamefont {Enghardt}, \citenamefont {Gotz},
  \citenamefont {Masood}, \citenamefont {Schramm}, \citenamefont {Zeil},\ and\
  \citenamefont {Pawelke}}]{LKarsch2017}%
  \BibitemOpen
  \bibfield  {author} {\bibinfo {author} {\bibfnamefont {L.}~\bibnamefont
  {Karsch}}, \bibinfo {author} {\bibfnamefont {E.}~\bibnamefont {Beyreuther}},
  \bibinfo {author} {\bibfnamefont {W.}~\bibnamefont {Enghardt}}, \bibinfo
  {author} {\bibfnamefont {M.}~\bibnamefont {Gotz}}, \bibinfo {author}
  {\bibfnamefont {U.}~\bibnamefont {Masood}}, \bibinfo {author} {\bibfnamefont
  {U.}~\bibnamefont {Schramm}}, \bibinfo {author} {\bibfnamefont
  {K.}~\bibnamefont {Zeil}}, \ and\ \bibinfo {author} {\bibfnamefont
  {J.}~\bibnamefont {Pawelke}},\ }\href {\doibase
  10.1080/0284186X.2017.1355111} {\bibfield  {journal} {\bibinfo  {journal}
  {Acta Oncologica}\ }\textbf {\bibinfo {volume} {56}},\ \bibinfo {pages}
  {1359} (\bibinfo {year} {2017})},\ \bibinfo {note} {pMID: 28828925},\ \Eprint
  {http://arxiv.org/abs/https://doi.org/10.1080/0284186X.2017.1355111}
  {https://doi.org/10.1080/0284186X.2017.1355111} \BibitemShut {NoStop}%
\bibitem [{\citenamefont {Borghesi}\ \emph {et~al.}(2001)\citenamefont
  {Borghesi}, \citenamefont {Schiavi}, \citenamefont {Campbell}, \citenamefont
  {Haines}, \citenamefont {Willi}, \citenamefont {MacKinnon}, \citenamefont
  {Gizzi}, \citenamefont {Galimberti}, \citenamefont {Clarke},\ and\
  \citenamefont {Ruhl}}]{borghesi.ppcf.2001}%
  \BibitemOpen
  \bibfield  {author} {\bibinfo {author} {\bibfnamefont {M.}~\bibnamefont
  {Borghesi}}, \bibinfo {author} {\bibfnamefont {A.}~\bibnamefont {Schiavi}},
  \bibinfo {author} {\bibfnamefont {D.~H.}\ \bibnamefont {Campbell}}, \bibinfo
  {author} {\bibfnamefont {M.~G.}\ \bibnamefont {Haines}}, \bibinfo {author}
  {\bibfnamefont {O.}~\bibnamefont {Willi}}, \bibinfo {author} {\bibfnamefont
  {A.~J.}\ \bibnamefont {MacKinnon}}, \bibinfo {author} {\bibfnamefont {L.~A.}\
  \bibnamefont {Gizzi}}, \bibinfo {author} {\bibfnamefont {M.}~\bibnamefont
  {Galimberti}}, \bibinfo {author} {\bibfnamefont {R.~J.}\ \bibnamefont
  {Clarke}}, \ and\ \bibinfo {author} {\bibfnamefont {H.}~\bibnamefont
  {Ruhl}},\ }\href {\doibase 10.1088/0741-3335/43/12A/320} {\bibfield
  {journal} {\bibinfo  {journal} {Plasma Physics and Controlled Fusion}\
  }\textbf {\bibinfo {volume} {43}},\ \bibinfo {pages} {A267} (\bibinfo {year}
  {2001})}\BibitemShut {NoStop}%
\bibitem [{\citenamefont {Pelka}\ \emph {et~al.}(2010)\citenamefont {Pelka},
  \citenamefont {Gregori}, \citenamefont {Gericke}, \citenamefont {Vorberger},
  \citenamefont {Glenzer}, \citenamefont {G\"unther}, \citenamefont {Harres},
  \citenamefont {Heathcote}, \citenamefont {Kritcher}, \citenamefont {Kugland},
  \citenamefont {Li}, \citenamefont {Makita}, \citenamefont {Mithen},
  \citenamefont {Neely}, \citenamefont {Niemann}, \citenamefont {Otten},
  \citenamefont {Riley}, \citenamefont {Schaumann}, \citenamefont
  {Schollmeier}, \citenamefont {Tauschwitz},\ and\ \citenamefont
  {Roth}}]{pelka.prl.2010}%
  \BibitemOpen
  \bibfield  {author} {\bibinfo {author} {\bibfnamefont {A.}~\bibnamefont
  {Pelka}}, \bibinfo {author} {\bibfnamefont {G.}~\bibnamefont {Gregori}},
  \bibinfo {author} {\bibfnamefont {D.~O.}\ \bibnamefont {Gericke}}, \bibinfo
  {author} {\bibfnamefont {J.}~\bibnamefont {Vorberger}}, \bibinfo {author}
  {\bibfnamefont {S.~H.}\ \bibnamefont {Glenzer}}, \bibinfo {author}
  {\bibfnamefont {M.~M.}\ \bibnamefont {G\"unther}}, \bibinfo {author}
  {\bibfnamefont {K.}~\bibnamefont {Harres}}, \bibinfo {author} {\bibfnamefont
  {R.}~\bibnamefont {Heathcote}}, \bibinfo {author} {\bibfnamefont {A.~L.}\
  \bibnamefont {Kritcher}}, \bibinfo {author} {\bibfnamefont {N.~L.}\
  \bibnamefont {Kugland}}, \bibinfo {author} {\bibfnamefont {B.}~\bibnamefont
  {Li}}, \bibinfo {author} {\bibfnamefont {M.}~\bibnamefont {Makita}}, \bibinfo
  {author} {\bibfnamefont {J.}~\bibnamefont {Mithen}}, \bibinfo {author}
  {\bibfnamefont {D.}~\bibnamefont {Neely}}, \bibinfo {author} {\bibfnamefont
  {C.}~\bibnamefont {Niemann}}, \bibinfo {author} {\bibfnamefont
  {A.}~\bibnamefont {Otten}}, \bibinfo {author} {\bibfnamefont
  {D.}~\bibnamefont {Riley}}, \bibinfo {author} {\bibfnamefont
  {G.}~\bibnamefont {Schaumann}}, \bibinfo {author} {\bibfnamefont
  {M.}~\bibnamefont {Schollmeier}}, \bibinfo {author} {\bibfnamefont
  {A.}~\bibnamefont {Tauschwitz}}, \ and\ \bibinfo {author} {\bibfnamefont
  {M.}~\bibnamefont {Roth}},\ }\href {\doibase 10.1103/PhysRevLett.105.265701}
  {\bibfield  {journal} {\bibinfo  {journal} {Phys. Rev. Lett.}\ }\textbf
  {\bibinfo {volume} {105}},\ \bibinfo {pages} {265701} (\bibinfo {year}
  {2010})}\BibitemShut {NoStop}%
\bibitem [{\citenamefont {Ledingham}\ \emph {et~al.}(2003)\citenamefont
  {Ledingham}, \citenamefont {McKenna},\ and\ \citenamefont
  {Singhal}}]{Ledingham.Science.2003}%
  \BibitemOpen
  \bibfield  {author} {\bibinfo {author} {\bibfnamefont {K.~W.~D.}\
  \bibnamefont {Ledingham}}, \bibinfo {author} {\bibfnamefont {P.}~\bibnamefont
  {McKenna}}, \ and\ \bibinfo {author} {\bibfnamefont {R.~P.}\ \bibnamefont
  {Singhal}},\ }\href {\doibase 10.1126/science.1080552} {\bibfield  {journal}
  {\bibinfo  {journal} {Science}\ }\textbf {\bibinfo {volume} {300}},\ \bibinfo
  {pages} {1107} (\bibinfo {year} {2003})},\ \bibinfo {note} {publisher:
  American Association for the Advancement of Science}\BibitemShut {NoStop}%
\bibitem [{\citenamefont {Yurevich}(2010)}]{Yurevich.PhysPartNucl.2010}%
  \BibitemOpen
  \bibfield  {author} {\bibinfo {author} {\bibfnamefont {V.~I.}\ \bibnamefont
  {Yurevich}},\ }\href {\doibase 10.1134/S1063779610050060} {\bibfield
  {journal} {\bibinfo  {journal} {Physics of Particles and Nuclei}\ }\textbf
  {\bibinfo {volume} {41}},\ \bibinfo {pages} {778} (\bibinfo {year}
  {2010})}\BibitemShut {NoStop}%
\bibitem [{\citenamefont {Durante}\ and\ \citenamefont
  {Cucinotta}(2011)}]{durante.rmp.2011}%
  \BibitemOpen
  \bibfield  {author} {\bibinfo {author} {\bibfnamefont {M.}~\bibnamefont
  {Durante}}\ and\ \bibinfo {author} {\bibfnamefont {F.~A.}\ \bibnamefont
  {Cucinotta}},\ }\href {\doibase 10.1103/RevModPhys.83.1245} {\bibfield
  {journal} {\bibinfo  {journal} {Rev. Mod. Phys.}\ }\textbf {\bibinfo {volume}
  {83}},\ \bibinfo {pages} {1245} (\bibinfo {year} {2011})}\BibitemShut
  {NoStop}%
\bibitem [{\citenamefont {Miyazawa}\ \emph {et~al.}(2018)\citenamefont
  {Miyazawa}, \citenamefont {Ikegami}, \citenamefont {Chen}, \citenamefont
  {Ohshima}, \citenamefont {Imaizumi}, \citenamefont {Hirose},\ and\
  \citenamefont {Miyasaka}}]{Miyazawa.iScience.2018}%
  \BibitemOpen
  \bibfield  {author} {\bibinfo {author} {\bibfnamefont {Y.}~\bibnamefont
  {Miyazawa}}, \bibinfo {author} {\bibfnamefont {M.}~\bibnamefont {Ikegami}},
  \bibinfo {author} {\bibfnamefont {H.-W.}\ \bibnamefont {Chen}}, \bibinfo
  {author} {\bibfnamefont {T.}~\bibnamefont {Ohshima}}, \bibinfo {author}
  {\bibfnamefont {M.}~\bibnamefont {Imaizumi}}, \bibinfo {author}
  {\bibfnamefont {K.}~\bibnamefont {Hirose}}, \ and\ \bibinfo {author}
  {\bibfnamefont {T.}~\bibnamefont {Miyasaka}},\ }\href {\doibase
  https://doi.org/10.1016/j.isci.2018.03.020} {\bibfield  {journal} {\bibinfo
  {journal} {iScience}\ }\textbf {\bibinfo {volume} {2}},\ \bibinfo {pages}
  {148} (\bibinfo {year} {2018})}\BibitemShut {NoStop}%
\bibitem [{\citenamefont {Wagner}\ \emph {et~al.}(2016)\citenamefont {Wagner},
  \citenamefont {Deppert}, \citenamefont {Brabetz}, \citenamefont {Fiala},
  \citenamefont {Kleinschmidt}, \citenamefont {Poth}, \citenamefont {Schanz},
  \citenamefont {Tebartz}, \citenamefont {Zielbauer}, \citenamefont {Roth},
  \citenamefont {St\"ohlker},\ and\ \citenamefont {Bagnoud}}]{Wagner2016}%
  \BibitemOpen
  \bibfield  {author} {\bibinfo {author} {\bibfnamefont {F.}~\bibnamefont
  {Wagner}}, \bibinfo {author} {\bibfnamefont {O.}~\bibnamefont {Deppert}},
  \bibinfo {author} {\bibfnamefont {C.}~\bibnamefont {Brabetz}}, \bibinfo
  {author} {\bibfnamefont {P.}~\bibnamefont {Fiala}}, \bibinfo {author}
  {\bibfnamefont {A.}~\bibnamefont {Kleinschmidt}}, \bibinfo {author}
  {\bibfnamefont {P.}~\bibnamefont {Poth}}, \bibinfo {author} {\bibfnamefont
  {V.~A.}\ \bibnamefont {Schanz}}, \bibinfo {author} {\bibfnamefont
  {A.}~\bibnamefont {Tebartz}}, \bibinfo {author} {\bibfnamefont
  {B.}~\bibnamefont {Zielbauer}}, \bibinfo {author} {\bibfnamefont
  {M.}~\bibnamefont {Roth}}, \bibinfo {author} {\bibfnamefont {T.}~\bibnamefont
  {St\"ohlker}}, \ and\ \bibinfo {author} {\bibfnamefont {V.}~\bibnamefont
  {Bagnoud}},\ }\href {\doibase 10.1103/PhysRevLett.116.205002} {\bibfield
  {journal} {\bibinfo  {journal} {Phys. Rev. Lett.}\ }\textbf {\bibinfo
  {volume} {116}},\ \bibinfo {pages} {205002} (\bibinfo {year}
  {2016})}\BibitemShut {NoStop}%
\bibitem [{\citenamefont {Higginson}\ \emph {et~al.}(2018)\citenamefont
  {Higginson}, \citenamefont {Gray}, \citenamefont {King}, \citenamefont
  {Dance}, \citenamefont {Williamson}, \citenamefont {Butler}, \citenamefont
  {Wilson}, \citenamefont {Capdessus}, \citenamefont {Armstrong}, \citenamefont
  {Green}, \citenamefont {Hawkes}, \citenamefont {Martin}, \citenamefont {Wei},
  \citenamefont {Mirfayzi}, \citenamefont {Yuan}, \citenamefont {Kar},
  \citenamefont {Borghesi}, \citenamefont {Clarke}, \citenamefont {Neely},\
  and\ \citenamefont {McKenna}}]{Higginson2018}%
  \BibitemOpen
  \bibfield  {author} {\bibinfo {author} {\bibfnamefont {A.}~\bibnamefont
  {Higginson}}, \bibinfo {author} {\bibfnamefont {R.~J.}\ \bibnamefont {Gray}},
  \bibinfo {author} {\bibfnamefont {M.}~\bibnamefont {King}}, \bibinfo {author}
  {\bibfnamefont {R.~J.}\ \bibnamefont {Dance}}, \bibinfo {author}
  {\bibfnamefont {S.~D.~R.}\ \bibnamefont {Williamson}}, \bibinfo {author}
  {\bibfnamefont {N.~M.~H.}\ \bibnamefont {Butler}}, \bibinfo {author}
  {\bibfnamefont {R.}~\bibnamefont {Wilson}}, \bibinfo {author} {\bibfnamefont
  {R.}~\bibnamefont {Capdessus}}, \bibinfo {author} {\bibfnamefont
  {C.}~\bibnamefont {Armstrong}}, \bibinfo {author} {\bibfnamefont {J.~S.}\
  \bibnamefont {Green}}, \bibinfo {author} {\bibfnamefont {S.~J.}\ \bibnamefont
  {Hawkes}}, \bibinfo {author} {\bibfnamefont {P.}~\bibnamefont {Martin}},
  \bibinfo {author} {\bibfnamefont {W.~Q.}\ \bibnamefont {Wei}}, \bibinfo
  {author} {\bibfnamefont {S.~R.}\ \bibnamefont {Mirfayzi}}, \bibinfo {author}
  {\bibfnamefont {X.~H.}\ \bibnamefont {Yuan}}, \bibinfo {author}
  {\bibfnamefont {S.}~\bibnamefont {Kar}}, \bibinfo {author} {\bibfnamefont
  {M.}~\bibnamefont {Borghesi}}, \bibinfo {author} {\bibfnamefont {R.~J.}\
  \bibnamefont {Clarke}}, \bibinfo {author} {\bibfnamefont {D.}~\bibnamefont
  {Neely}}, \ and\ \bibinfo {author} {\bibfnamefont {P.}~\bibnamefont
  {McKenna}},\ }\href {\doibase 10.1038/s41467-018-03063-9} {\bibfield
  {journal} {\bibinfo  {journal} {Nature Communications}\ }\textbf {\bibinfo
  {volume} {9}} (\bibinfo {year} {2018}),\
  10.1038/s41467-018-03063-9}\BibitemShut {NoStop}%
\bibitem [{\citenamefont {Dover}\ \emph {et~al.}(2023)\citenamefont {Dover},
  \citenamefont {Ziegler}, \citenamefont {Assenbaum}, \citenamefont {Bernert},
  \citenamefont {Bock}, \citenamefont {Brack}, \citenamefont {Cowan},
  \citenamefont {Ditter}, \citenamefont {Garten}, \citenamefont {Gaus},
  \citenamefont {Goethel}, \citenamefont {Hicks}, \citenamefont {Kiriyama},
  \citenamefont {Kluge}, \citenamefont {Koga}, \citenamefont {Kon},
  \citenamefont {Kondo}, \citenamefont {Kraft}, \citenamefont {Kroll},
  \citenamefont {Lowe}, \citenamefont {Metzkes-Ng}, \citenamefont {Miyatake},
  \citenamefont {Najmudin}, \citenamefont {P{\"u}schel}, \citenamefont
  {Rehwald}, \citenamefont {Reimold}, \citenamefont {Sakaki}, \citenamefont
  {Schlenvoigt}, \citenamefont {Shiokawa}, \citenamefont {Umlandt},
  \citenamefont {Schramm}, \citenamefont {Zeil},\ and\ \citenamefont
  {Nishiuchi}}]{dover.lsa.2023}%
  \BibitemOpen
  \bibfield  {author} {\bibinfo {author} {\bibfnamefont {N.~P.}\ \bibnamefont
  {Dover}}, \bibinfo {author} {\bibfnamefont {T.}~\bibnamefont {Ziegler}},
  \bibinfo {author} {\bibfnamefont {S.}~\bibnamefont {Assenbaum}}, \bibinfo
  {author} {\bibfnamefont {C.}~\bibnamefont {Bernert}}, \bibinfo {author}
  {\bibfnamefont {S.}~\bibnamefont {Bock}}, \bibinfo {author} {\bibfnamefont
  {F.-E.}\ \bibnamefont {Brack}}, \bibinfo {author} {\bibfnamefont {T.~E.}\
  \bibnamefont {Cowan}}, \bibinfo {author} {\bibfnamefont {E.~J.}\ \bibnamefont
  {Ditter}}, \bibinfo {author} {\bibfnamefont {M.}~\bibnamefont {Garten}},
  \bibinfo {author} {\bibfnamefont {L.}~\bibnamefont {Gaus}}, \bibinfo {author}
  {\bibfnamefont {I.}~\bibnamefont {Goethel}}, \bibinfo {author} {\bibfnamefont
  {G.~S.}\ \bibnamefont {Hicks}}, \bibinfo {author} {\bibfnamefont
  {H.}~\bibnamefont {Kiriyama}}, \bibinfo {author} {\bibfnamefont
  {T.}~\bibnamefont {Kluge}}, \bibinfo {author} {\bibfnamefont {J.~K.}\
  \bibnamefont {Koga}}, \bibinfo {author} {\bibfnamefont {A.}~\bibnamefont
  {Kon}}, \bibinfo {author} {\bibfnamefont {K.}~\bibnamefont {Kondo}}, \bibinfo
  {author} {\bibfnamefont {S.}~\bibnamefont {Kraft}}, \bibinfo {author}
  {\bibfnamefont {F.}~\bibnamefont {Kroll}}, \bibinfo {author} {\bibfnamefont
  {H.~F.}\ \bibnamefont {Lowe}}, \bibinfo {author} {\bibfnamefont
  {J.}~\bibnamefont {Metzkes-Ng}}, \bibinfo {author} {\bibfnamefont
  {T.}~\bibnamefont {Miyatake}}, \bibinfo {author} {\bibfnamefont
  {Z.}~\bibnamefont {Najmudin}}, \bibinfo {author} {\bibfnamefont
  {T.}~\bibnamefont {P{\"u}schel}}, \bibinfo {author} {\bibfnamefont
  {M.}~\bibnamefont {Rehwald}}, \bibinfo {author} {\bibfnamefont
  {M.}~\bibnamefont {Reimold}}, \bibinfo {author} {\bibfnamefont
  {H.}~\bibnamefont {Sakaki}}, \bibinfo {author} {\bibfnamefont {H.-P.}\
  \bibnamefont {Schlenvoigt}}, \bibinfo {author} {\bibfnamefont
  {K.}~\bibnamefont {Shiokawa}}, \bibinfo {author} {\bibfnamefont {M.~E.~P.}\
  \bibnamefont {Umlandt}}, \bibinfo {author} {\bibfnamefont {U.}~\bibnamefont
  {Schramm}}, \bibinfo {author} {\bibfnamefont {K.}~\bibnamefont {Zeil}}, \
  and\ \bibinfo {author} {\bibfnamefont {M.}~\bibnamefont {Nishiuchi}},\ }\href
  {\doibase 10.1038/s41377-023-01083-9} {\bibfield  {journal} {\bibinfo
  {journal} {Light: Science \& Applications}\ }\textbf {\bibinfo {volume}
  {12}},\ \bibinfo {pages} {71} (\bibinfo {year} {2023})}\BibitemShut {NoStop}%
\bibitem [{\citenamefont {Rehwald}\ \emph {et~al.}(2023)\citenamefont
  {Rehwald}, \citenamefont {Assenbaum}, \citenamefont {Bernert}, \citenamefont
  {Brack}, \citenamefont {Bussmann}, \citenamefont {Cowan}, \citenamefont
  {Curry}, \citenamefont {Fiuza}, \citenamefont {Garten}, \citenamefont {Gaus},
  \citenamefont {Gauthier}, \citenamefont {Göde}, \citenamefont {Göthel},
  \citenamefont {Glenzer}, \citenamefont {Huang}, \citenamefont {Huebl},
  \citenamefont {Kim}, \citenamefont {Kluge}, \citenamefont {Kraft},
  \citenamefont {Kroll}, \citenamefont {Metzkes-Ng}, \citenamefont
  {Miethlinger}, \citenamefont {Loeser}, \citenamefont {Obst-Huebl},
  \citenamefont {Reimold}, \citenamefont {Schlenvoigt}, \citenamefont
  {Schoenwaelder}, \citenamefont {Schramm}, \citenamefont {Siebold},
  \citenamefont {Treffert}, \citenamefont {Yang}, \citenamefont {Ziegler},\
  and\ \citenamefont {Zeil}}]{Rehwald2023}%
  \BibitemOpen
  \bibfield  {author} {\bibinfo {author} {\bibfnamefont {M.}~\bibnamefont
  {Rehwald}}, \bibinfo {author} {\bibfnamefont {S.}~\bibnamefont {Assenbaum}},
  \bibinfo {author} {\bibfnamefont {C.}~\bibnamefont {Bernert}}, \bibinfo
  {author} {\bibfnamefont {F.-E.}\ \bibnamefont {Brack}}, \bibinfo {author}
  {\bibfnamefont {M.}~\bibnamefont {Bussmann}}, \bibinfo {author}
  {\bibfnamefont {T.~E.}\ \bibnamefont {Cowan}}, \bibinfo {author}
  {\bibfnamefont {C.~B.}\ \bibnamefont {Curry}}, \bibinfo {author}
  {\bibfnamefont {F.}~\bibnamefont {Fiuza}}, \bibinfo {author} {\bibfnamefont
  {M.}~\bibnamefont {Garten}}, \bibinfo {author} {\bibfnamefont
  {L.}~\bibnamefont {Gaus}}, \bibinfo {author} {\bibfnamefont {M.}~\bibnamefont
  {Gauthier}}, \bibinfo {author} {\bibfnamefont {S.}~\bibnamefont {Göde}},
  \bibinfo {author} {\bibfnamefont {I.}~\bibnamefont {Göthel}}, \bibinfo
  {author} {\bibfnamefont {S.~H.}\ \bibnamefont {Glenzer}}, \bibinfo {author}
  {\bibfnamefont {L.}~\bibnamefont {Huang}}, \bibinfo {author} {\bibfnamefont
  {A.}~\bibnamefont {Huebl}}, \bibinfo {author} {\bibfnamefont {J.~B.}\
  \bibnamefont {Kim}}, \bibinfo {author} {\bibfnamefont {T.}~\bibnamefont
  {Kluge}}, \bibinfo {author} {\bibfnamefont {S.}~\bibnamefont {Kraft}},
  \bibinfo {author} {\bibfnamefont {F.}~\bibnamefont {Kroll}}, \bibinfo
  {author} {\bibfnamefont {J.}~\bibnamefont {Metzkes-Ng}}, \bibinfo {author}
  {\bibfnamefont {T.}~\bibnamefont {Miethlinger}}, \bibinfo {author}
  {\bibfnamefont {M.}~\bibnamefont {Loeser}}, \bibinfo {author} {\bibfnamefont
  {L.}~\bibnamefont {Obst-Huebl}}, \bibinfo {author} {\bibfnamefont
  {M.}~\bibnamefont {Reimold}}, \bibinfo {author} {\bibfnamefont {H.-P.}\
  \bibnamefont {Schlenvoigt}}, \bibinfo {author} {\bibfnamefont
  {C.}~\bibnamefont {Schoenwaelder}}, \bibinfo {author} {\bibfnamefont
  {U.}~\bibnamefont {Schramm}}, \bibinfo {author} {\bibfnamefont
  {M.}~\bibnamefont {Siebold}}, \bibinfo {author} {\bibfnamefont
  {F.}~\bibnamefont {Treffert}}, \bibinfo {author} {\bibfnamefont
  {L.}~\bibnamefont {Yang}}, \bibinfo {author} {\bibfnamefont {T.}~\bibnamefont
  {Ziegler}}, \ and\ \bibinfo {author} {\bibfnamefont {K.}~\bibnamefont
  {Zeil}},\ }\href {\doibase 10.1038/s41467-023-39739-0} {\bibfield  {journal}
  {\bibinfo  {journal} {Nature Communications}\ }\textbf {\bibinfo {volume}
  {14}},\ \bibinfo {pages} {4009} (\bibinfo {year} {2023})}\BibitemShut
  {NoStop}%
\bibitem [{\citenamefont {Danson}\ \emph {et~al.}(2019)\citenamefont {Danson},
  \citenamefont {Haefner}, \citenamefont {Bromage}, \citenamefont {Butcher},
  \citenamefont {Chanteloup}, \citenamefont {Chowdhury}, \citenamefont
  {Galvanauskas}, \citenamefont {Gizzi}, \citenamefont {Hein}, \citenamefont
  {Hillier} \emph {et~al.}}]{danson.hplse.2019}%
  \BibitemOpen
  \bibfield  {author} {\bibinfo {author} {\bibfnamefont {C.~N.}\ \bibnamefont
  {Danson}}, \bibinfo {author} {\bibfnamefont {C.}~\bibnamefont {Haefner}},
  \bibinfo {author} {\bibfnamefont {J.}~\bibnamefont {Bromage}}, \bibinfo
  {author} {\bibfnamefont {T.}~\bibnamefont {Butcher}}, \bibinfo {author}
  {\bibfnamefont {J.-C.~F.}\ \bibnamefont {Chanteloup}}, \bibinfo {author}
  {\bibfnamefont {E.~A.}\ \bibnamefont {Chowdhury}}, \bibinfo {author}
  {\bibfnamefont {A.}~\bibnamefont {Galvanauskas}}, \bibinfo {author}
  {\bibfnamefont {L.~A.}\ \bibnamefont {Gizzi}}, \bibinfo {author}
  {\bibfnamefont {J.}~\bibnamefont {Hein}}, \bibinfo {author} {\bibfnamefont
  {D.~I.}\ \bibnamefont {Hillier}},  \emph {et~al.},\ }\href@noop {} {\bibfield
   {journal} {\bibinfo  {journal} {High Power Laser Science and Engineering}\
  }\textbf {\bibinfo {volume} {7}} (\bibinfo {year} {2019})}\BibitemShut
  {NoStop}%
\bibitem [{\citenamefont {Gonoskov}\ \emph {et~al.}(2022)\citenamefont
  {Gonoskov}, \citenamefont {Blackburn}, \citenamefont {Marklund},\ and\
  \citenamefont {Bulanov}}]{gonoskov.rmp.2022}%
  \BibitemOpen
  \bibfield  {author} {\bibinfo {author} {\bibfnamefont {A.}~\bibnamefont
  {Gonoskov}}, \bibinfo {author} {\bibfnamefont {T.~G.}\ \bibnamefont
  {Blackburn}}, \bibinfo {author} {\bibfnamefont {M.}~\bibnamefont {Marklund}},
  \ and\ \bibinfo {author} {\bibfnamefont {S.~S.}\ \bibnamefont {Bulanov}},\
  }\href {\doibase 10.1103/RevModPhys.94.045001} {\bibfield  {journal}
  {\bibinfo  {journal} {Rev. Mod. Phys.}\ }\textbf {\bibinfo {volume} {94}},\
  \bibinfo {pages} {045001} (\bibinfo {year} {2022})}\BibitemShut {NoStop}%
\bibitem [{\citenamefont {Yoon}\ \emph {et~al.}(2021)\citenamefont {Yoon},
  \citenamefont {Kim}, \citenamefont {Choi}, \citenamefont {Sung},
  \citenamefont {Lee}, \citenamefont {Lee},\ and\ \citenamefont
  {Nam}}]{yoon.optica.2021}%
  \BibitemOpen
  \bibfield  {author} {\bibinfo {author} {\bibfnamefont {J.~W.}\ \bibnamefont
  {Yoon}}, \bibinfo {author} {\bibfnamefont {Y.~G.}\ \bibnamefont {Kim}},
  \bibinfo {author} {\bibfnamefont {I.}~\bibnamefont {Choi}}, \bibinfo {author}
  {\bibfnamefont {J.~H.}\ \bibnamefont {Sung}}, \bibinfo {author}
  {\bibfnamefont {H.~W.}\ \bibnamefont {Lee}}, \bibinfo {author} {\bibfnamefont
  {S.~K.}\ \bibnamefont {Lee}}, \ and\ \bibinfo {author} {\bibfnamefont
  {C.~H.}\ \bibnamefont {Nam}},\ }\href {\doibase 10.1364/optica.420520}
  {\bibfield  {journal} {\bibinfo  {journal} {Optica}\ }\textbf {\bibinfo
  {volume} {8}},\ \bibinfo {pages} {630} (\bibinfo {year} {2021})}\BibitemShut
  {NoStop}%
\bibitem [{\citenamefont {Ziegler}\ \emph {et~al.}(2021)\citenamefont
  {Ziegler}, \citenamefont {Albach}, \citenamefont {Bernert}, \citenamefont
  {Bock}, \citenamefont {Brack}, \citenamefont {Cowan}, \citenamefont {Dover},
  \citenamefont {Garten}, \citenamefont {Gaus}, \citenamefont {Gebhardt},
  \citenamefont {Goethel}, \citenamefont {Helbig}, \citenamefont {Irman},
  \citenamefont {Kiriyama}, \citenamefont {Kluge}, \citenamefont {Kon},
  \citenamefont {Kraft}, \citenamefont {Kroll}, \citenamefont {Loeser},
  \citenamefont {Metzkes-Ng}, \citenamefont {Nishiuchi}, \citenamefont
  {Obst-Huebl}, \citenamefont {Püschel}, \citenamefont {Rehwald},
  \citenamefont {Schlenvoigt}, \citenamefont {Schramm},\ and\ \citenamefont
  {Zeil}}]{Ziegler.SciRep.2021}%
  \BibitemOpen
  \bibfield  {author} {\bibinfo {author} {\bibfnamefont {T.}~\bibnamefont
  {Ziegler}}, \bibinfo {author} {\bibfnamefont {D.}~\bibnamefont {Albach}},
  \bibinfo {author} {\bibfnamefont {C.}~\bibnamefont {Bernert}}, \bibinfo
  {author} {\bibfnamefont {S.}~\bibnamefont {Bock}}, \bibinfo {author}
  {\bibfnamefont {F.~E.}\ \bibnamefont {Brack}}, \bibinfo {author}
  {\bibfnamefont {T.~E.}\ \bibnamefont {Cowan}}, \bibinfo {author}
  {\bibfnamefont {N.~P.}\ \bibnamefont {Dover}}, \bibinfo {author}
  {\bibfnamefont {M.}~\bibnamefont {Garten}}, \bibinfo {author} {\bibfnamefont
  {L.}~\bibnamefont {Gaus}}, \bibinfo {author} {\bibfnamefont {R.}~\bibnamefont
  {Gebhardt}}, \bibinfo {author} {\bibfnamefont {I.}~\bibnamefont {Goethel}},
  \bibinfo {author} {\bibfnamefont {U.}~\bibnamefont {Helbig}}, \bibinfo
  {author} {\bibfnamefont {A.}~\bibnamefont {Irman}}, \bibinfo {author}
  {\bibfnamefont {H.}~\bibnamefont {Kiriyama}}, \bibinfo {author}
  {\bibfnamefont {T.}~\bibnamefont {Kluge}}, \bibinfo {author} {\bibfnamefont
  {A.}~\bibnamefont {Kon}}, \bibinfo {author} {\bibfnamefont {S.}~\bibnamefont
  {Kraft}}, \bibinfo {author} {\bibfnamefont {F.}~\bibnamefont {Kroll}},
  \bibinfo {author} {\bibfnamefont {M.}~\bibnamefont {Loeser}}, \bibinfo
  {author} {\bibfnamefont {J.}~\bibnamefont {Metzkes-Ng}}, \bibinfo {author}
  {\bibfnamefont {M.}~\bibnamefont {Nishiuchi}}, \bibinfo {author}
  {\bibfnamefont {L.}~\bibnamefont {Obst-Huebl}}, \bibinfo {author}
  {\bibfnamefont {T.}~\bibnamefont {Püschel}}, \bibinfo {author}
  {\bibfnamefont {M.}~\bibnamefont {Rehwald}}, \bibinfo {author} {\bibfnamefont
  {H.~P.}\ \bibnamefont {Schlenvoigt}}, \bibinfo {author} {\bibfnamefont
  {U.}~\bibnamefont {Schramm}}, \ and\ \bibinfo {author} {\bibfnamefont
  {K.}~\bibnamefont {Zeil}},\ }\href {\doibase 10.1038/s41598-021-86547-x}
  {\bibfield  {journal} {\bibinfo  {journal} {Scientific Reports}\ }\textbf
  {\bibinfo {volume} {11}},\ \bibinfo {pages} {7338} (\bibinfo {year}
  {2021})}\BibitemShut {NoStop}%
\bibitem [{\citenamefont {Bulanov}\ \emph {et~al.}(2016)\citenamefont
  {Bulanov}, \citenamefont {Esarey}, \citenamefont {Schroeder}, \citenamefont
  {Bulanov}, \citenamefont {Esirkepov}, \citenamefont {Kando}, \citenamefont
  {Pegoraro},\ and\ \citenamefont {Leemans}}]{bulanov.pop.2016}%
  \BibitemOpen
  \bibfield  {author} {\bibinfo {author} {\bibfnamefont {S.~S.}\ \bibnamefont
  {Bulanov}}, \bibinfo {author} {\bibfnamefont {E.}~\bibnamefont {Esarey}},
  \bibinfo {author} {\bibfnamefont {C.~B.}\ \bibnamefont {Schroeder}}, \bibinfo
  {author} {\bibfnamefont {S.~V.}\ \bibnamefont {Bulanov}}, \bibinfo {author}
  {\bibfnamefont {T.~Z.}\ \bibnamefont {Esirkepov}}, \bibinfo {author}
  {\bibfnamefont {M.}~\bibnamefont {Kando}}, \bibinfo {author} {\bibfnamefont
  {F.}~\bibnamefont {Pegoraro}}, \ and\ \bibinfo {author} {\bibfnamefont
  {W.~P.}\ \bibnamefont {Leemans}},\ }\href {\doibase 10.1063/1.4946025}
  {\bibfield  {journal} {\bibinfo  {journal} {Physics of Plasmas}\ }\textbf
  {\bibinfo {volume} {23}},\ \bibinfo {pages} {056703} (\bibinfo {year}
  {2016})}\BibitemShut {NoStop}%
\bibitem [{\citenamefont {Steinke}\ \emph {et~al.}(2016)\citenamefont
  {Steinke}, \citenamefont {van Tilborg}, \citenamefont {Benedetti},
  \citenamefont {Geddes}, \citenamefont {Schroeder}, \citenamefont {Daniels},
  \citenamefont {Swanson}, \citenamefont {Gonsalves}, \citenamefont {Nakamura},
  \citenamefont {Matlis}, \citenamefont {Shaw}, \citenamefont {Esarey},\ and\
  \citenamefont {Leemans}}]{steinke.nature.2016}%
  \BibitemOpen
  \bibfield  {author} {\bibinfo {author} {\bibfnamefont {S.}~\bibnamefont
  {Steinke}}, \bibinfo {author} {\bibfnamefont {J.}~\bibnamefont {van
  Tilborg}}, \bibinfo {author} {\bibfnamefont {C.}~\bibnamefont {Benedetti}},
  \bibinfo {author} {\bibfnamefont {C.~G.~R.}\ \bibnamefont {Geddes}}, \bibinfo
  {author} {\bibfnamefont {C.~B.}\ \bibnamefont {Schroeder}}, \bibinfo {author}
  {\bibfnamefont {J.}~\bibnamefont {Daniels}}, \bibinfo {author} {\bibfnamefont
  {K.~K.}\ \bibnamefont {Swanson}}, \bibinfo {author} {\bibfnamefont {A.~J.}\
  \bibnamefont {Gonsalves}}, \bibinfo {author} {\bibfnamefont {K.}~\bibnamefont
  {Nakamura}}, \bibinfo {author} {\bibfnamefont {N.~H.}\ \bibnamefont
  {Matlis}}, \bibinfo {author} {\bibfnamefont {B.~H.}\ \bibnamefont {Shaw}},
  \bibinfo {author} {\bibfnamefont {E.}~\bibnamefont {Esarey}}, \ and\ \bibinfo
  {author} {\bibfnamefont {W.~P.}\ \bibnamefont {Leemans}},\ }\href {\doibase
  10.1038/nature16525} {\bibfield  {journal} {\bibinfo  {journal} {Nature}\
  }\textbf {\bibinfo {volume} {530}},\ \bibinfo {pages} {190} (\bibinfo {year}
  {2016})}\BibitemShut {NoStop}%
\bibitem [{\citenamefont {Kurz}\ \emph {et~al.}(2021)\citenamefont {Kurz},
  \citenamefont {Heinemann}, \citenamefont {Gilljohann}, \citenamefont {Chang},
  \citenamefont {Couperus~Cabadağ}, \citenamefont {Debus}, \citenamefont
  {Kononenko}, \citenamefont {Pausch}, \citenamefont {Schöbel}, \citenamefont
  {Assmann}, \citenamefont {Bussmann}, \citenamefont {Ding}, \citenamefont
  {Götzfried}, \citenamefont {Köhler}, \citenamefont {Raj}, \citenamefont
  {Schindler}, \citenamefont {Steiniger}, \citenamefont {Zarini}, \citenamefont
  {Corde}, \citenamefont {Döpp}, \citenamefont {Hidding}, \citenamefont
  {Karsch}, \citenamefont {Schramm}, \citenamefont {Martinez de~la Ossa},\ and\
  \citenamefont {Irman}}]{Kurz.NatComm.2021}%
  \BibitemOpen
  \bibfield  {author} {\bibinfo {author} {\bibfnamefont {T.}~\bibnamefont
  {Kurz}}, \bibinfo {author} {\bibfnamefont {T.}~\bibnamefont {Heinemann}},
  \bibinfo {author} {\bibfnamefont {M.~F.}\ \bibnamefont {Gilljohann}},
  \bibinfo {author} {\bibfnamefont {Y.~Y.}\ \bibnamefont {Chang}}, \bibinfo
  {author} {\bibfnamefont {J.~P.}\ \bibnamefont {Couperus~Cabadağ}}, \bibinfo
  {author} {\bibfnamefont {A.}~\bibnamefont {Debus}}, \bibinfo {author}
  {\bibfnamefont {O.}~\bibnamefont {Kononenko}}, \bibinfo {author}
  {\bibfnamefont {R.}~\bibnamefont {Pausch}}, \bibinfo {author} {\bibfnamefont
  {S.}~\bibnamefont {Schöbel}}, \bibinfo {author} {\bibfnamefont {R.~W.}\
  \bibnamefont {Assmann}}, \bibinfo {author} {\bibfnamefont {M.}~\bibnamefont
  {Bussmann}}, \bibinfo {author} {\bibfnamefont {H.}~\bibnamefont {Ding}},
  \bibinfo {author} {\bibfnamefont {J.}~\bibnamefont {Götzfried}}, \bibinfo
  {author} {\bibfnamefont {A.}~\bibnamefont {Köhler}}, \bibinfo {author}
  {\bibfnamefont {G.}~\bibnamefont {Raj}}, \bibinfo {author} {\bibfnamefont
  {S.}~\bibnamefont {Schindler}}, \bibinfo {author} {\bibfnamefont
  {K.}~\bibnamefont {Steiniger}}, \bibinfo {author} {\bibfnamefont
  {O.}~\bibnamefont {Zarini}}, \bibinfo {author} {\bibfnamefont
  {S.}~\bibnamefont {Corde}}, \bibinfo {author} {\bibfnamefont
  {A.}~\bibnamefont {Döpp}}, \bibinfo {author} {\bibfnamefont
  {B.}~\bibnamefont {Hidding}}, \bibinfo {author} {\bibfnamefont
  {S.}~\bibnamefont {Karsch}}, \bibinfo {author} {\bibfnamefont
  {U.}~\bibnamefont {Schramm}}, \bibinfo {author} {\bibfnamefont
  {A.}~\bibnamefont {Martinez de~la Ossa}}, \ and\ \bibinfo {author}
  {\bibfnamefont {A.}~\bibnamefont {Irman}},\ }\href {\doibase
  10.1038/s41467-021-23000-7} {\bibfield  {journal} {\bibinfo  {journal}
  {Nature Communications}\ }\textbf {\bibinfo {volume} {12}},\ \bibinfo {pages}
  {2895} (\bibinfo {year} {2021})},\ \bibinfo {note} {arXiv:
  1909.06676}\BibitemShut {NoStop}%
\bibitem [{\citenamefont {Geddes}\ \emph {et~al.}(2022)\citenamefont {Geddes},
  \citenamefont {Hogan}, \citenamefont {Musumeci},\ and\ \citenamefont
  {Assmann}}]{geddes.af6.2022}%
  \BibitemOpen
  \bibfield  {author} {\bibinfo {author} {\bibfnamefont {C.}~\bibnamefont
  {Geddes}}, \bibinfo {author} {\bibfnamefont {M.}~\bibnamefont {Hogan}},
  \bibinfo {author} {\bibfnamefont {P.}~\bibnamefont {Musumeci}}, \ and\
  \bibinfo {author} {\bibfnamefont {R.}~\bibnamefont {Assmann}},\ }\href
  {\doibase 10.48550/ARXIV.2208.13279} {\enquote {\bibinfo {title} {Report of
  snowmass 21 accelerator frontier topical group 6 on advanced accelerators},}\
  } (\bibinfo {year} {2022})\BibitemShut {NoStop}%
\bibitem [{\citenamefont {Bulanov}\ \emph {et~al.}(2010)\citenamefont
  {Bulanov}, \citenamefont {Bychenkov}, \citenamefont {Chvykov}, \citenamefont
  {Kalinchenko}, \citenamefont {Litzenberg}, \citenamefont {Matsuoka},
  \citenamefont {Thomas}, \citenamefont {Willingale}, \citenamefont {Yanovsky},
  \citenamefont {Krushelnick},\ and\ \citenamefont
  {Maksimchuk}}]{SSBulanov2010}%
  \BibitemOpen
  \bibfield  {author} {\bibinfo {author} {\bibfnamefont {S.~S.}\ \bibnamefont
  {Bulanov}}, \bibinfo {author} {\bibfnamefont {V.~Y.}\ \bibnamefont
  {Bychenkov}}, \bibinfo {author} {\bibfnamefont {V.}~\bibnamefont {Chvykov}},
  \bibinfo {author} {\bibfnamefont {G.}~\bibnamefont {Kalinchenko}}, \bibinfo
  {author} {\bibfnamefont {D.~W.}\ \bibnamefont {Litzenberg}}, \bibinfo
  {author} {\bibfnamefont {T.}~\bibnamefont {Matsuoka}}, \bibinfo {author}
  {\bibfnamefont {A.~G.~R.}\ \bibnamefont {Thomas}}, \bibinfo {author}
  {\bibfnamefont {L.}~\bibnamefont {Willingale}}, \bibinfo {author}
  {\bibfnamefont {V.}~\bibnamefont {Yanovsky}}, \bibinfo {author}
  {\bibfnamefont {K.}~\bibnamefont {Krushelnick}}, \ and\ \bibinfo {author}
  {\bibfnamefont {A.}~\bibnamefont {Maksimchuk}},\ }\href {\doibase
  10.1063/1.3372840} {\bibfield  {journal} {\bibinfo  {journal} {Physics of
  Plasmas}\ }\textbf {\bibinfo {volume} {17}},\ \bibinfo {pages} {043105}
  (\bibinfo {year} {2010})}\BibitemShut {NoStop}%
\bibitem [{\citenamefont {Park}\ \emph {et~al.}(2019)\citenamefont {Park},
  \citenamefont {Bulanov}, \citenamefont {Bin}, \citenamefont {Ji},
  \citenamefont {Steinke}, \citenamefont {Vay}, \citenamefont {Geddes},
  \citenamefont {Schroeder}, \citenamefont {Leemans}, \citenamefont
  {Schenkel},\ and\ \citenamefont {Esarey}}]{Park2019}%
  \BibitemOpen
  \bibfield  {author} {\bibinfo {author} {\bibfnamefont {J.}~\bibnamefont
  {Park}}, \bibinfo {author} {\bibfnamefont {S.~S.}\ \bibnamefont {Bulanov}},
  \bibinfo {author} {\bibfnamefont {J.}~\bibnamefont {Bin}}, \bibinfo {author}
  {\bibfnamefont {Q.}~\bibnamefont {Ji}}, \bibinfo {author} {\bibfnamefont
  {S.}~\bibnamefont {Steinke}}, \bibinfo {author} {\bibfnamefont {J.-L.}\
  \bibnamefont {Vay}}, \bibinfo {author} {\bibfnamefont {C.~G.~R.}\
  \bibnamefont {Geddes}}, \bibinfo {author} {\bibfnamefont {C.~B.}\
  \bibnamefont {Schroeder}}, \bibinfo {author} {\bibfnamefont {W.~P.}\
  \bibnamefont {Leemans}}, \bibinfo {author} {\bibfnamefont {T.}~\bibnamefont
  {Schenkel}}, \ and\ \bibinfo {author} {\bibfnamefont {E.}~\bibnamefont
  {Esarey}},\ }\href {\doibase 10.1063/1.5094045} {\bibfield  {journal}
  {\bibinfo  {journal} {Physics of Plasmas}\ }\textbf {\bibinfo {volume}
  {26}},\ \bibinfo {pages} {103108} (\bibinfo {year} {2019})}\BibitemShut
  {NoStop}%
\bibitem [{\citenamefont {Kuznetsov}\ \emph {et~al.}(2001)\citenamefont
  {Kuznetsov}, \citenamefont {Esirkepov}, \citenamefont {Kamenets},\ and\
  \citenamefont {Bulanov}}]{Kuznetsov.PRR.2001}%
  \BibitemOpen
  \bibfield  {author} {\bibinfo {author} {\bibfnamefont {A.~V.}\ \bibnamefont
  {Kuznetsov}}, \bibinfo {author} {\bibfnamefont {T.~Z.}\ \bibnamefont
  {Esirkepov}}, \bibinfo {author} {\bibfnamefont {F.~F.}\ \bibnamefont
  {Kamenets}}, \ and\ \bibinfo {author} {\bibfnamefont {S.~V.}\ \bibnamefont
  {Bulanov}},\ }\href {\doibase 10.1134/1.1354219} {\bibfield  {journal}
  {\bibinfo  {journal} {Plasma Physics Reports}\ }\textbf {\bibinfo {volume}
  {27}},\ \bibinfo {pages} {211} (\bibinfo {year} {2001})}\BibitemShut
  {NoStop}%
\bibitem [{\citenamefont {Bulanov}(2005)}]{SVBulanov.PRR.2005}%
  \BibitemOpen
  \bibfield  {author} {\bibinfo {author} {\bibfnamefont {S.~V.}\ \bibnamefont
  {Bulanov}},\ }\href {\doibase 10.1134/1.1925787} {\bibfield  {journal}
  {\bibinfo  {journal} {Plasma Physics Reports}\ }\textbf {\bibinfo {volume}
  {31}},\ \bibinfo {pages} {369} (\bibinfo {year} {2005})}\BibitemShut
  {NoStop}%
\bibitem [{\citenamefont {Bulanov}\ and\ \citenamefont
  {Esirkepov}(2007)}]{SVBulanov.PRL.2007}%
  \BibitemOpen
  \bibfield  {author} {\bibinfo {author} {\bibfnamefont {S.~V.}\ \bibnamefont
  {Bulanov}}\ and\ \bibinfo {author} {\bibfnamefont {T.~Z.}\ \bibnamefont
  {Esirkepov}},\ }\href {\doibase 10.1103/PhysRevLett.98.049503} {\bibfield
  {journal} {\bibinfo  {journal} {Physical Review Letters}\ }\textbf {\bibinfo
  {volume} {98}},\ \bibinfo {pages} {049503} (\bibinfo {year}
  {2007})}\BibitemShut {NoStop}%
\bibitem [{\citenamefont {Matsukado}\ \emph {et~al.}(2003)\citenamefont
  {Matsukado}, \citenamefont {Esirkepov}, \citenamefont {Daido}, \citenamefont
  {Utsumi}, \citenamefont {Li}, \citenamefont {Fukumi}, \citenamefont
  {Hayashi}, \citenamefont {Orimo}, \citenamefont {Nishiuchi}, \citenamefont
  {Bulanov}, \citenamefont {Tajima}, \citenamefont {Noda}, \citenamefont
  {Iwashita}, \citenamefont {Shirai}, \citenamefont {Takeuchi}, \citenamefont
  {Nakamura}, \citenamefont {Yamazaki}, \citenamefont {Ikegami}, \citenamefont
  {Mihara}, \citenamefont {Morita}, \citenamefont {Uesaka}, \citenamefont
  {Yoshii}, \citenamefont {Watanabe}, \citenamefont {Kinoshita}, \citenamefont
  {Hosokai}, \citenamefont {Zhidkov}, \citenamefont {Ogata}, \citenamefont
  {Wada},\ and\ \citenamefont {Kubota}}]{Matsukado.PRL.2003}%
  \BibitemOpen
  \bibfield  {author} {\bibinfo {author} {\bibfnamefont {K.}~\bibnamefont
  {Matsukado}}, \bibinfo {author} {\bibfnamefont {T.}~\bibnamefont
  {Esirkepov}}, \bibinfo {author} {\bibfnamefont {H.}~\bibnamefont {Daido}},
  \bibinfo {author} {\bibfnamefont {T.}~\bibnamefont {Utsumi}}, \bibinfo
  {author} {\bibfnamefont {Z.}~\bibnamefont {Li}}, \bibinfo {author}
  {\bibfnamefont {A.}~\bibnamefont {Fukumi}}, \bibinfo {author} {\bibfnamefont
  {Y.}~\bibnamefont {Hayashi}}, \bibinfo {author} {\bibfnamefont
  {S.}~\bibnamefont {Orimo}}, \bibinfo {author} {\bibfnamefont
  {M.}~\bibnamefont {Nishiuchi}}, \bibinfo {author} {\bibfnamefont {S.~V.}\
  \bibnamefont {Bulanov}}, \bibinfo {author} {\bibfnamefont {T.}~\bibnamefont
  {Tajima}}, \bibinfo {author} {\bibfnamefont {A.}~\bibnamefont {Noda}},
  \bibinfo {author} {\bibfnamefont {Y.}~\bibnamefont {Iwashita}}, \bibinfo
  {author} {\bibfnamefont {T.}~\bibnamefont {Shirai}}, \bibinfo {author}
  {\bibfnamefont {T.}~\bibnamefont {Takeuchi}}, \bibinfo {author}
  {\bibfnamefont {S.}~\bibnamefont {Nakamura}}, \bibinfo {author}
  {\bibfnamefont {A.}~\bibnamefont {Yamazaki}}, \bibinfo {author}
  {\bibfnamefont {M.}~\bibnamefont {Ikegami}}, \bibinfo {author} {\bibfnamefont
  {T.}~\bibnamefont {Mihara}}, \bibinfo {author} {\bibfnamefont
  {A.}~\bibnamefont {Morita}}, \bibinfo {author} {\bibfnamefont
  {M.}~\bibnamefont {Uesaka}}, \bibinfo {author} {\bibfnamefont
  {K.}~\bibnamefont {Yoshii}}, \bibinfo {author} {\bibfnamefont
  {T.}~\bibnamefont {Watanabe}}, \bibinfo {author} {\bibfnamefont
  {K.}~\bibnamefont {Kinoshita}}, \bibinfo {author} {\bibfnamefont
  {T.}~\bibnamefont {Hosokai}}, \bibinfo {author} {\bibfnamefont
  {A.}~\bibnamefont {Zhidkov}}, \bibinfo {author} {\bibfnamefont
  {A.}~\bibnamefont {Ogata}}, \bibinfo {author} {\bibfnamefont
  {Y.}~\bibnamefont {Wada}}, \ and\ \bibinfo {author} {\bibfnamefont
  {T.}~\bibnamefont {Kubota}},\ }\href {\doibase 10.1103/PhysRevLett.91.215001}
  {\bibfield  {journal} {\bibinfo  {journal} {Phys. Rev. Lett.}\ }\textbf
  {\bibinfo {volume} {91}},\ \bibinfo {pages} {215001} (\bibinfo {year}
  {2003})}\BibitemShut {NoStop}%
\bibitem [{\citenamefont {Willingale}\ \emph {et~al.}(2006)\citenamefont
  {Willingale}, \citenamefont {Mangles}, \citenamefont {Nilson}, \citenamefont
  {Clarke}, \citenamefont {Dangor}, \citenamefont {Kaluza}, \citenamefont
  {Karsch}, \citenamefont {Lancaster}, \citenamefont {Mori}, \citenamefont
  {Najmudin}, \citenamefont {Schreiber}, \citenamefont {Thomas}, \citenamefont
  {Wei},\ and\ \citenamefont {Krushelnick}}]{Willingale.PRL.2006}%
  \BibitemOpen
  \bibfield  {author} {\bibinfo {author} {\bibfnamefont {L.}~\bibnamefont
  {Willingale}}, \bibinfo {author} {\bibfnamefont {S.~P.~D.}\ \bibnamefont
  {Mangles}}, \bibinfo {author} {\bibfnamefont {P.~M.}\ \bibnamefont {Nilson}},
  \bibinfo {author} {\bibfnamefont {R.~J.}\ \bibnamefont {Clarke}}, \bibinfo
  {author} {\bibfnamefont {A.~E.}\ \bibnamefont {Dangor}}, \bibinfo {author}
  {\bibfnamefont {M.~C.}\ \bibnamefont {Kaluza}}, \bibinfo {author}
  {\bibfnamefont {S.}~\bibnamefont {Karsch}}, \bibinfo {author} {\bibfnamefont
  {K.~L.}\ \bibnamefont {Lancaster}}, \bibinfo {author} {\bibfnamefont {W.~B.}\
  \bibnamefont {Mori}}, \bibinfo {author} {\bibfnamefont {Z.}~\bibnamefont
  {Najmudin}}, \bibinfo {author} {\bibfnamefont {J.}~\bibnamefont {Schreiber}},
  \bibinfo {author} {\bibfnamefont {A.~G.~R.}\ \bibnamefont {Thomas}}, \bibinfo
  {author} {\bibfnamefont {M.~S.}\ \bibnamefont {Wei}}, \ and\ \bibinfo
  {author} {\bibfnamefont {K.}~\bibnamefont {Krushelnick}},\ }\href {\doibase
  10.1103/PhysRevLett.96.245002} {\bibfield  {journal} {\bibinfo  {journal}
  {Phys. Rev. Lett.}\ }\textbf {\bibinfo {volume} {96}},\ \bibinfo {pages}
  {245002} (\bibinfo {year} {2006})}\BibitemShut {NoStop}%
\bibitem [{\citenamefont {Yogo}\ \emph {et~al.}(2008)\citenamefont {Yogo},
  \citenamefont {Daido}, \citenamefont {Bulanov}, \citenamefont {Nemoto},
  \citenamefont {Oishi}, \citenamefont {Nayuki}, \citenamefont {Fujii},
  \citenamefont {Ogura}, \citenamefont {Orimo}, \citenamefont {Sagisaka},
  \citenamefont {Ma}, \citenamefont {Esirkepov}, \citenamefont {Mori},
  \citenamefont {Nishiuchi}, \citenamefont {Pirozhkov}, \citenamefont
  {Nakamura}, \citenamefont {Noda}, \citenamefont {Nagatomo}, \citenamefont
  {Kimura},\ and\ \citenamefont {Tajima}}]{Yogo.PRE.2008}%
  \BibitemOpen
  \bibfield  {author} {\bibinfo {author} {\bibfnamefont {A.}~\bibnamefont
  {Yogo}}, \bibinfo {author} {\bibfnamefont {H.}~\bibnamefont {Daido}},
  \bibinfo {author} {\bibfnamefont {S.~V.}\ \bibnamefont {Bulanov}}, \bibinfo
  {author} {\bibfnamefont {K.}~\bibnamefont {Nemoto}}, \bibinfo {author}
  {\bibfnamefont {Y.}~\bibnamefont {Oishi}}, \bibinfo {author} {\bibfnamefont
  {T.}~\bibnamefont {Nayuki}}, \bibinfo {author} {\bibfnamefont
  {T.}~\bibnamefont {Fujii}}, \bibinfo {author} {\bibfnamefont
  {K.}~\bibnamefont {Ogura}}, \bibinfo {author} {\bibfnamefont
  {S.}~\bibnamefont {Orimo}}, \bibinfo {author} {\bibfnamefont
  {A.}~\bibnamefont {Sagisaka}}, \bibinfo {author} {\bibfnamefont {J.-L.}\
  \bibnamefont {Ma}}, \bibinfo {author} {\bibfnamefont {T.~Z.}\ \bibnamefont
  {Esirkepov}}, \bibinfo {author} {\bibfnamefont {M.}~\bibnamefont {Mori}},
  \bibinfo {author} {\bibfnamefont {M.}~\bibnamefont {Nishiuchi}}, \bibinfo
  {author} {\bibfnamefont {A.~S.}\ \bibnamefont {Pirozhkov}}, \bibinfo {author}
  {\bibfnamefont {S.}~\bibnamefont {Nakamura}}, \bibinfo {author}
  {\bibfnamefont {A.}~\bibnamefont {Noda}}, \bibinfo {author} {\bibfnamefont
  {H.}~\bibnamefont {Nagatomo}}, \bibinfo {author} {\bibfnamefont
  {T.}~\bibnamefont {Kimura}}, \ and\ \bibinfo {author} {\bibfnamefont
  {T.}~\bibnamefont {Tajima}},\ }\href {\doibase 10.1103/PhysRevE.77.016401}
  {\bibfield  {journal} {\bibinfo  {journal} {Physical Review E}\ }\textbf
  {\bibinfo {volume} {77}},\ \bibinfo {pages} {016401} (\bibinfo {year}
  {2008})}\BibitemShut {NoStop}%
\bibitem [{\citenamefont {Fukuda}\ \emph {et~al.}(2009)\citenamefont {Fukuda},
  \citenamefont {Faenov}, \citenamefont {Tampo}, \citenamefont {Pikuz},
  \citenamefont {Nakamura}, \citenamefont {Kando}, \citenamefont {Hayashi},
  \citenamefont {Yogo}, \citenamefont {Sakaki}, \citenamefont {Kameshima},
  \citenamefont {Pirozhkov}, \citenamefont {Ogura}, \citenamefont {Mori},
  \citenamefont {Esirkepov}, \citenamefont {Koga}, \citenamefont {Boldarev},
  \citenamefont {Gasilov}, \citenamefont {Magunov}, \citenamefont {Yamauchi},
  \citenamefont {Kodama}, \citenamefont {Bolton}, \citenamefont {Kato},
  \citenamefont {Tajima}, \citenamefont {Daido},\ and\ \citenamefont
  {Bulanov}}]{Fukuda.PRL.2009}%
  \BibitemOpen
  \bibfield  {author} {\bibinfo {author} {\bibfnamefont {Y.}~\bibnamefont
  {Fukuda}}, \bibinfo {author} {\bibfnamefont {A.~Y.}\ \bibnamefont {Faenov}},
  \bibinfo {author} {\bibfnamefont {M.}~\bibnamefont {Tampo}}, \bibinfo
  {author} {\bibfnamefont {T.~A.}\ \bibnamefont {Pikuz}}, \bibinfo {author}
  {\bibfnamefont {T.}~\bibnamefont {Nakamura}}, \bibinfo {author}
  {\bibfnamefont {M.}~\bibnamefont {Kando}}, \bibinfo {author} {\bibfnamefont
  {Y.}~\bibnamefont {Hayashi}}, \bibinfo {author} {\bibfnamefont
  {A.}~\bibnamefont {Yogo}}, \bibinfo {author} {\bibfnamefont {H.}~\bibnamefont
  {Sakaki}}, \bibinfo {author} {\bibfnamefont {T.}~\bibnamefont {Kameshima}},
  \bibinfo {author} {\bibfnamefont {A.~S.}\ \bibnamefont {Pirozhkov}}, \bibinfo
  {author} {\bibfnamefont {K.}~\bibnamefont {Ogura}}, \bibinfo {author}
  {\bibfnamefont {M.}~\bibnamefont {Mori}}, \bibinfo {author} {\bibfnamefont
  {T.~Z.}\ \bibnamefont {Esirkepov}}, \bibinfo {author} {\bibfnamefont
  {J.}~\bibnamefont {Koga}}, \bibinfo {author} {\bibfnamefont {A.~S.}\
  \bibnamefont {Boldarev}}, \bibinfo {author} {\bibfnamefont {V.~A.}\
  \bibnamefont {Gasilov}}, \bibinfo {author} {\bibfnamefont {A.~I.}\
  \bibnamefont {Magunov}}, \bibinfo {author} {\bibfnamefont {T.}~\bibnamefont
  {Yamauchi}}, \bibinfo {author} {\bibfnamefont {R.}~\bibnamefont {Kodama}},
  \bibinfo {author} {\bibfnamefont {P.~R.}\ \bibnamefont {Bolton}}, \bibinfo
  {author} {\bibfnamefont {Y.}~\bibnamefont {Kato}}, \bibinfo {author}
  {\bibfnamefont {T.}~\bibnamefont {Tajima}}, \bibinfo {author} {\bibfnamefont
  {H.}~\bibnamefont {Daido}}, \ and\ \bibinfo {author} {\bibfnamefont {S.~V.}\
  \bibnamefont {Bulanov}},\ }\href {\doibase 10.1103/PhysRevLett.103.165002}
  {\bibfield  {journal} {\bibinfo  {journal} {Phys. Rev. Lett.}\ }\textbf
  {\bibinfo {volume} {103}},\ \bibinfo {pages} {165002} (\bibinfo {year}
  {2009})}\BibitemShut {NoStop}%
\bibitem [{\citenamefont {Willingale}\ \emph {et~al.}(2009)\citenamefont
  {Willingale}, \citenamefont {Nagel}, \citenamefont {Thomas}, \citenamefont
  {Bellei}, \citenamefont {Clarke}, \citenamefont {Dangor}, \citenamefont
  {Heathcote}, \citenamefont {Kaluza}, \citenamefont {Kamperidis},
  \citenamefont {Kneip}, \citenamefont {Krushelnick}, \citenamefont {Lopes},
  \citenamefont {Mangles}, \citenamefont {Nazarov}, \citenamefont {Nilson},\
  and\ \citenamefont {Najmudin}}]{Willingale.PRL.2009}%
  \BibitemOpen
  \bibfield  {author} {\bibinfo {author} {\bibfnamefont {L.}~\bibnamefont
  {Willingale}}, \bibinfo {author} {\bibfnamefont {S.~R.}\ \bibnamefont
  {Nagel}}, \bibinfo {author} {\bibfnamefont {A.~G.~R.}\ \bibnamefont
  {Thomas}}, \bibinfo {author} {\bibfnamefont {C.}~\bibnamefont {Bellei}},
  \bibinfo {author} {\bibfnamefont {R.~J.}\ \bibnamefont {Clarke}}, \bibinfo
  {author} {\bibfnamefont {A.~E.}\ \bibnamefont {Dangor}}, \bibinfo {author}
  {\bibfnamefont {R.}~\bibnamefont {Heathcote}}, \bibinfo {author}
  {\bibfnamefont {M.~C.}\ \bibnamefont {Kaluza}}, \bibinfo {author}
  {\bibfnamefont {C.}~\bibnamefont {Kamperidis}}, \bibinfo {author}
  {\bibfnamefont {S.}~\bibnamefont {Kneip}}, \bibinfo {author} {\bibfnamefont
  {K.}~\bibnamefont {Krushelnick}}, \bibinfo {author} {\bibfnamefont
  {N.}~\bibnamefont {Lopes}}, \bibinfo {author} {\bibfnamefont {S.~P.~D.}\
  \bibnamefont {Mangles}}, \bibinfo {author} {\bibfnamefont {W.}~\bibnamefont
  {Nazarov}}, \bibinfo {author} {\bibfnamefont {P.~M.}\ \bibnamefont {Nilson}},
  \ and\ \bibinfo {author} {\bibfnamefont {Z.}~\bibnamefont {Najmudin}},\
  }\href {\doibase 10.1103/PhysRevLett.102.125002} {\bibfield  {journal}
  {\bibinfo  {journal} {Physical Review Letters}\ }\textbf {\bibinfo {volume}
  {102}},\ \bibinfo {pages} {125002} (\bibinfo {year} {2009})}\BibitemShut
  {NoStop}%
\bibitem [{\citenamefont {Willingale}\ \emph {et~al.}(2011)\citenamefont
  {Willingale}, \citenamefont {Nilson}, \citenamefont {Thomas}, \citenamefont
  {Bulanov}, \citenamefont {Maksimchuk}, \citenamefont {Nazarov}, \citenamefont
  {Sangster}, \citenamefont {Stoeckl},\ and\ \citenamefont
  {Krushelnick}}]{Willingale.POP.2011}%
  \BibitemOpen
  \bibfield  {author} {\bibinfo {author} {\bibfnamefont {L.}~\bibnamefont
  {Willingale}}, \bibinfo {author} {\bibfnamefont {P.~M.}\ \bibnamefont
  {Nilson}}, \bibinfo {author} {\bibfnamefont {A.~G.~R.}\ \bibnamefont
  {Thomas}}, \bibinfo {author} {\bibfnamefont {S.~S.}\ \bibnamefont {Bulanov}},
  \bibinfo {author} {\bibfnamefont {A.}~\bibnamefont {Maksimchuk}}, \bibinfo
  {author} {\bibfnamefont {W.}~\bibnamefont {Nazarov}}, \bibinfo {author}
  {\bibfnamefont {T.~C.}\ \bibnamefont {Sangster}}, \bibinfo {author}
  {\bibfnamefont {C.}~\bibnamefont {Stoeckl}}, \ and\ \bibinfo {author}
  {\bibfnamefont {K.}~\bibnamefont {Krushelnick}},\ }\href {\doibase
  10.1063/1.3563438} {\bibfield  {journal} {\bibinfo  {journal} {Physics of
  Plasmas}\ }\textbf {\bibinfo {volume} {18}},\ \bibinfo {pages} {056706}
  (\bibinfo {year} {2011})}\BibitemShut {NoStop}%
\bibitem [{\citenamefont {Hakimi}\ \emph {et~al.}(2022)\citenamefont {Hakimi},
  \citenamefont {Obst-Huebl}, \citenamefont {Huebl}, \citenamefont {Nakamura},
  \citenamefont {Bulanov}, \citenamefont {Steinke}, \citenamefont {Leemans},
  \citenamefont {Kober}, \citenamefont {Ostermayr}, \citenamefont {Schenkel},
  \citenamefont {Gonsalves}, \citenamefont {Vay}, \citenamefont {van Tilborg},
  \citenamefont {Toth}, \citenamefont {Schroeder}, \citenamefont {Esarey},\
  and\ \citenamefont {Geddes}}]{Hakimi2022}%
  \BibitemOpen
  \bibfield  {author} {\bibinfo {author} {\bibfnamefont {S.}~\bibnamefont
  {Hakimi}}, \bibinfo {author} {\bibfnamefont {L.}~\bibnamefont {Obst-Huebl}},
  \bibinfo {author} {\bibfnamefont {A.}~\bibnamefont {Huebl}}, \bibinfo
  {author} {\bibfnamefont {K.}~\bibnamefont {Nakamura}}, \bibinfo {author}
  {\bibfnamefont {S.~S.}\ \bibnamefont {Bulanov}}, \bibinfo {author}
  {\bibfnamefont {S.}~\bibnamefont {Steinke}}, \bibinfo {author} {\bibfnamefont
  {W.~P.}\ \bibnamefont {Leemans}}, \bibinfo {author} {\bibfnamefont
  {Z.}~\bibnamefont {Kober}}, \bibinfo {author} {\bibfnamefont {T.~M.}\
  \bibnamefont {Ostermayr}}, \bibinfo {author} {\bibfnamefont {T.}~\bibnamefont
  {Schenkel}}, \bibinfo {author} {\bibfnamefont {A.~J.}\ \bibnamefont
  {Gonsalves}}, \bibinfo {author} {\bibfnamefont {J.-L.}\ \bibnamefont {Vay}},
  \bibinfo {author} {\bibfnamefont {J.}~\bibnamefont {van Tilborg}}, \bibinfo
  {author} {\bibfnamefont {C.}~\bibnamefont {Toth}}, \bibinfo {author}
  {\bibfnamefont {C.~B.}\ \bibnamefont {Schroeder}}, \bibinfo {author}
  {\bibfnamefont {E.}~\bibnamefont {Esarey}}, \ and\ \bibinfo {author}
  {\bibfnamefont {C.~G.~R.}\ \bibnamefont {Geddes}},\ }\href {\doibase
  10.1063/5.0089331} {\bibfield  {journal} {\bibinfo  {journal} {Physics of
  Plasmas}\ }\textbf {\bibinfo {volume} {29}},\ \bibinfo {pages} {083102}
  (\bibinfo {year} {2022})}\BibitemShut {NoStop}%
\bibitem [{\citenamefont {G\"ode}\ \emph {et~al.}(2017)\citenamefont {G\"ode},
  \citenamefont {R\"odel}, \citenamefont {Zeil}, \citenamefont {Mishra},
  \citenamefont {Gauthier}, \citenamefont {Brack}, \citenamefont {Kluge},
  \citenamefont {MacDonald}, \citenamefont {Metzkes}, \citenamefont {Obst},
  \citenamefont {Rehwald}, \citenamefont {Ruyer}, \citenamefont {Schlenvoigt},
  \citenamefont {Schumaker}, \citenamefont {Sommer}, \citenamefont {Cowan},
  \citenamefont {Schramm}, \citenamefont {Glenzer},\ and\ \citenamefont
  {Fiuza}}]{Goede.PRL.2017}%
  \BibitemOpen
  \bibfield  {author} {\bibinfo {author} {\bibfnamefont {S.}~\bibnamefont
  {G\"ode}}, \bibinfo {author} {\bibfnamefont {C.}~\bibnamefont {R\"odel}},
  \bibinfo {author} {\bibfnamefont {K.}~\bibnamefont {Zeil}}, \bibinfo {author}
  {\bibfnamefont {R.}~\bibnamefont {Mishra}}, \bibinfo {author} {\bibfnamefont
  {M.}~\bibnamefont {Gauthier}}, \bibinfo {author} {\bibfnamefont {F.-E.}\
  \bibnamefont {Brack}}, \bibinfo {author} {\bibfnamefont {T.}~\bibnamefont
  {Kluge}}, \bibinfo {author} {\bibfnamefont {M.~J.}\ \bibnamefont
  {MacDonald}}, \bibinfo {author} {\bibfnamefont {J.}~\bibnamefont {Metzkes}},
  \bibinfo {author} {\bibfnamefont {L.}~\bibnamefont {Obst}}, \bibinfo {author}
  {\bibfnamefont {M.}~\bibnamefont {Rehwald}}, \bibinfo {author} {\bibfnamefont
  {C.}~\bibnamefont {Ruyer}}, \bibinfo {author} {\bibfnamefont {H.-P.}\
  \bibnamefont {Schlenvoigt}}, \bibinfo {author} {\bibfnamefont
  {W.}~\bibnamefont {Schumaker}}, \bibinfo {author} {\bibfnamefont
  {P.}~\bibnamefont {Sommer}}, \bibinfo {author} {\bibfnamefont {T.~E.}\
  \bibnamefont {Cowan}}, \bibinfo {author} {\bibfnamefont {U.}~\bibnamefont
  {Schramm}}, \bibinfo {author} {\bibfnamefont {S.}~\bibnamefont {Glenzer}}, \
  and\ \bibinfo {author} {\bibfnamefont {F.}~\bibnamefont {Fiuza}},\ }\href
  {\doibase 10.1103/PhysRevLett.118.194801} {\bibfield  {journal} {\bibinfo
  {journal} {Phys. Rev. Lett.}\ }\textbf {\bibinfo {volume} {118}},\ \bibinfo
  {pages} {194801} (\bibinfo {year} {2017})}\BibitemShut {NoStop}%
\bibitem [{\citenamefont {Jäckel}\ \emph {et~al.}(2009)\citenamefont
  {Jäckel}, \citenamefont {Pfotenhauer}, \citenamefont {Polz}, \citenamefont
  {Steinke}, \citenamefont {Schlenvoigt}, \citenamefont {Heymann},
  \citenamefont {Robinson},\ and\ \citenamefont {Kaluza}}]{Jaeckel.CLEO.2009}%
  \BibitemOpen
  \bibfield  {author} {\bibinfo {author} {\bibfnamefont {O.}~\bibnamefont
  {Jäckel}}, \bibinfo {author} {\bibfnamefont {S.~M.}\ \bibnamefont
  {Pfotenhauer}}, \bibinfo {author} {\bibfnamefont {J.}~\bibnamefont {Polz}},
  \bibinfo {author} {\bibfnamefont {S.}~\bibnamefont {Steinke}}, \bibinfo
  {author} {\bibfnamefont {H.~P.}\ \bibnamefont {Schlenvoigt}}, \bibinfo
  {author} {\bibfnamefont {J.}~\bibnamefont {Heymann}}, \bibinfo {author}
  {\bibfnamefont {A.~P.~L.}\ \bibnamefont {Robinson}}, \ and\ \bibinfo {author}
  {\bibfnamefont {M.~C.}\ \bibnamefont {Kaluza}}\ }(\bibinfo  {publisher}
  {OSA},\ \bibinfo {year} {2009})\ p.\ \bibinfo {pages} {JFB1}\BibitemShut
  {NoStop}%
\bibitem [{\citenamefont {Wang}\ \emph {et~al.}(2013)\citenamefont {Wang},
  \citenamefont {Shen}, \citenamefont {Zhang}, \citenamefont {Wang},
  \citenamefont {Xu}, \citenamefont {Zhao}, \citenamefont {Yu}, \citenamefont
  {Yi}, \citenamefont {Shi}, \citenamefont {Zhang}, \citenamefont {Xu},\ and\
  \citenamefont {Xu}}]{Wang.PoP.2013}%
  \BibitemOpen
  \bibfield  {author} {\bibinfo {author} {\bibfnamefont {W.~P.}\ \bibnamefont
  {Wang}}, \bibinfo {author} {\bibfnamefont {B.~F.}\ \bibnamefont {Shen}},
  \bibinfo {author} {\bibfnamefont {X.~M.}\ \bibnamefont {Zhang}}, \bibinfo
  {author} {\bibfnamefont {X.~F.}\ \bibnamefont {Wang}}, \bibinfo {author}
  {\bibfnamefont {J.~C.}\ \bibnamefont {Xu}}, \bibinfo {author} {\bibfnamefont
  {X.~Y.}\ \bibnamefont {Zhao}}, \bibinfo {author} {\bibfnamefont {Y.~H.}\
  \bibnamefont {Yu}}, \bibinfo {author} {\bibfnamefont {L.~Q.}\ \bibnamefont
  {Yi}}, \bibinfo {author} {\bibfnamefont {Y.}~\bibnamefont {Shi}}, \bibinfo
  {author} {\bibfnamefont {L.~G.}\ \bibnamefont {Zhang}}, \bibinfo {author}
  {\bibfnamefont {T.~J.}\ \bibnamefont {Xu}}, \ and\ \bibinfo {author}
  {\bibfnamefont {Z.~Z.}\ \bibnamefont {Xu}},\ }\href {\doibase
  10.1063/1.4831943} {\bibfield  {journal} {\bibinfo  {journal} {Physics of
  Plasmas}\ }\textbf {\bibinfo {volume} {20}} (\bibinfo {year} {2013}),\
  10.1063/1.4831943}\BibitemShut {NoStop}%
\bibitem [{\citenamefont {Kawata}\ \emph {et~al.}(2014)\citenamefont {Kawata},
  \citenamefont {Sato}, \citenamefont {Izumiyama}, \citenamefont {Nagashima},
  \citenamefont {Takano}, \citenamefont {Barada}, \citenamefont {Ma},
  \citenamefont {Wang}, \citenamefont {Kong}, \citenamefont {Wang},\ and\
  \citenamefont {Gu}}]{Kawata.APPC12.2014}%
  \BibitemOpen
  \bibfield  {author} {\bibinfo {author} {\bibfnamefont {S.}~\bibnamefont
  {Kawata}}, \bibinfo {author} {\bibfnamefont {D.}~\bibnamefont {Sato}},
  \bibinfo {author} {\bibfnamefont {T.}~\bibnamefont {Izumiyama}}, \bibinfo
  {author} {\bibfnamefont {T.}~\bibnamefont {Nagashima}}, \bibinfo {author}
  {\bibfnamefont {M.}~\bibnamefont {Takano}}, \bibinfo {author} {\bibfnamefont
  {D.}~\bibnamefont {Barada}}, \bibinfo {author} {\bibfnamefont {Y.~Y.}\
  \bibnamefont {Ma}}, \bibinfo {author} {\bibfnamefont {W.~M.}\ \bibnamefont
  {Wang}}, \bibinfo {author} {\bibfnamefont {Q.}~\bibnamefont {Kong}}, \bibinfo
  {author} {\bibfnamefont {P.~X.}\ \bibnamefont {Wang}}, \ and\ \bibinfo
  {author} {\bibfnamefont {Y.~J.}\ \bibnamefont {Gu}}\ }(\bibinfo  {publisher}
  {Journal of the Physical Society of Japan},\ \bibinfo {year}
  {2014})\BibitemShut {NoStop}%
\bibitem [{\citenamefont {Pfotenhauer}\ \emph {et~al.}(2010)\citenamefont
  {Pfotenhauer}, \citenamefont {Jäckel}, \citenamefont {Polz}, \citenamefont
  {Steinke}, \citenamefont {Schlenvoigt}, \citenamefont {Heymann},
  \citenamefont {Robinson},\ and\ \citenamefont {Kaluza}}]{Pfotenhauer2010}%
  \BibitemOpen
  \bibfield  {author} {\bibinfo {author} {\bibfnamefont {S.~M.}\ \bibnamefont
  {Pfotenhauer}}, \bibinfo {author} {\bibfnamefont {O.}~\bibnamefont
  {Jäckel}}, \bibinfo {author} {\bibfnamefont {J.}~\bibnamefont {Polz}},
  \bibinfo {author} {\bibfnamefont {S.}~\bibnamefont {Steinke}}, \bibinfo
  {author} {\bibfnamefont {H.-P.}\ \bibnamefont {Schlenvoigt}}, \bibinfo
  {author} {\bibfnamefont {J.}~\bibnamefont {Heymann}}, \bibinfo {author}
  {\bibfnamefont {A.~P.~L.}\ \bibnamefont {Robinson}}, \ and\ \bibinfo {author}
  {\bibfnamefont {M.~C.}\ \bibnamefont {Kaluza}},\ }\href {\doibase
  10.1088/1367-2630/12/10/103009} {\bibfield  {journal} {\bibinfo  {journal}
  {New Journal of Physics}\ }\textbf {\bibinfo {volume} {12}},\ \bibinfo
  {pages} {103009} (\bibinfo {year} {2010})}\BibitemShut {NoStop}%
\bibitem [{\citenamefont {Zhang}\ \emph {et~al.}(2007)\citenamefont {Zhang},
  \citenamefont {Shen}, \citenamefont {Li}, \citenamefont {Jin},\ and\
  \citenamefont {Wang}}]{Zhang.PoP.2007}%
  \BibitemOpen
  \bibfield  {author} {\bibinfo {author} {\bibfnamefont {X.}~\bibnamefont
  {Zhang}}, \bibinfo {author} {\bibfnamefont {B.}~\bibnamefont {Shen}},
  \bibinfo {author} {\bibfnamefont {X.}~\bibnamefont {Li}}, \bibinfo {author}
  {\bibfnamefont {Z.}~\bibnamefont {Jin}}, \ and\ \bibinfo {author}
  {\bibfnamefont {F.}~\bibnamefont {Wang}},\ }\href {\doibase
  10.1063/1.2746810} {\bibfield  {journal} {\bibinfo  {journal} {Physics of
  Plasmas}\ }\textbf {\bibinfo {volume} {14}},\ \bibinfo {pages} {073101}
  (\bibinfo {year} {2007})}\BibitemShut {NoStop}%
\bibitem [{\citenamefont {Toncian}\ \emph {et~al.}(2006)\citenamefont
  {Toncian}, \citenamefont {Borghesi}, \citenamefont {Fuchs}, \citenamefont
  {d'Humieres}, \citenamefont {Antici}, \citenamefont {Audebert}, \citenamefont
  {Brambrink}, \citenamefont {Cecchetti}, \citenamefont {Pipahl}, \citenamefont
  {Romagnani},\ and\ \citenamefont {Willi}}]{Toncian2006}%
  \BibitemOpen
  \bibfield  {author} {\bibinfo {author} {\bibfnamefont {T.}~\bibnamefont
  {Toncian}}, \bibinfo {author} {\bibfnamefont {M.}~\bibnamefont {Borghesi}},
  \bibinfo {author} {\bibfnamefont {J.}~\bibnamefont {Fuchs}}, \bibinfo
  {author} {\bibfnamefont {E.}~\bibnamefont {d'Humieres}}, \bibinfo {author}
  {\bibfnamefont {P.}~\bibnamefont {Antici}}, \bibinfo {author} {\bibfnamefont
  {P.}~\bibnamefont {Audebert}}, \bibinfo {author} {\bibfnamefont
  {E.}~\bibnamefont {Brambrink}}, \bibinfo {author} {\bibfnamefont {C.~A.}\
  \bibnamefont {Cecchetti}}, \bibinfo {author} {\bibfnamefont {A.}~\bibnamefont
  {Pipahl}}, \bibinfo {author} {\bibfnamefont {L.}~\bibnamefont {Romagnani}}, \
  and\ \bibinfo {author} {\bibfnamefont {O.}~\bibnamefont {Willi}},\ }\href
  {\doibase 10.1126/science.1124412} {\bibfield  {journal} {\bibinfo  {journal}
  {Science}\ }\textbf {\bibinfo {volume} {312}},\ \bibinfo {pages} {410}
  (\bibinfo {year} {2006})}\BibitemShut {NoStop}%
\bibitem [{\citenamefont {Kar}\ \emph {et~al.}(2016)\citenamefont {Kar},
  \citenamefont {Ahmed}, \citenamefont {Prasad}, \citenamefont {Cerchez},
  \citenamefont {Brauckmann}, \citenamefont {Aurand}, \citenamefont {Cantono},
  \citenamefont {Hadjisolomou}, \citenamefont {Lewis}, \citenamefont {Macchi},
  \citenamefont {Nersisyan}, \citenamefont {Robinson}, \citenamefont {Schroer},
  \citenamefont {Swantusch}, \citenamefont {Zepf}, \citenamefont {Willi},\ and\
  \citenamefont {Borghesi}}]{Kar.NatComm.2016}%
  \BibitemOpen
  \bibfield  {author} {\bibinfo {author} {\bibfnamefont {S.}~\bibnamefont
  {Kar}}, \bibinfo {author} {\bibfnamefont {H.}~\bibnamefont {Ahmed}}, \bibinfo
  {author} {\bibfnamefont {R.}~\bibnamefont {Prasad}}, \bibinfo {author}
  {\bibfnamefont {M.}~\bibnamefont {Cerchez}}, \bibinfo {author} {\bibfnamefont
  {S.}~\bibnamefont {Brauckmann}}, \bibinfo {author} {\bibfnamefont
  {B.}~\bibnamefont {Aurand}}, \bibinfo {author} {\bibfnamefont
  {G.}~\bibnamefont {Cantono}}, \bibinfo {author} {\bibfnamefont
  {P.}~\bibnamefont {Hadjisolomou}}, \bibinfo {author} {\bibfnamefont
  {C.~L.~S.}\ \bibnamefont {Lewis}}, \bibinfo {author} {\bibfnamefont
  {A.}~\bibnamefont {Macchi}}, \bibinfo {author} {\bibfnamefont
  {G.}~\bibnamefont {Nersisyan}}, \bibinfo {author} {\bibfnamefont {A.~P.~L.}\
  \bibnamefont {Robinson}}, \bibinfo {author} {\bibfnamefont {A.~M.}\
  \bibnamefont {Schroer}}, \bibinfo {author} {\bibfnamefont {M.}~\bibnamefont
  {Swantusch}}, \bibinfo {author} {\bibfnamefont {M.}~\bibnamefont {Zepf}},
  \bibinfo {author} {\bibfnamefont {O.}~\bibnamefont {Willi}}, \ and\ \bibinfo
  {author} {\bibfnamefont {M.}~\bibnamefont {Borghesi}},\ }\href {\doibase
  10.1038/ncomms10792} {\bibfield  {journal} {\bibinfo  {journal} {Nature
  Communications}\ }\textbf {\bibinfo {volume} {7}},\ \bibinfo {pages} {10792}
  (\bibinfo {year} {2016})}\BibitemShut {NoStop}%
\bibitem [{\citenamefont {Ferguson}\ \emph {et~al.}(2023)\citenamefont
  {Ferguson}, \citenamefont {Martin}, \citenamefont {Ahmed}, \citenamefont
  {Aktan}, \citenamefont {Alanazi}, \citenamefont {Cerchez}, \citenamefont
  {Doria}, \citenamefont {Green}, \citenamefont {Greenwood}, \citenamefont
  {Odlozilik}, \citenamefont {Willi}, \citenamefont {Borghesi},\ and\
  \citenamefont {Kar}}]{Ferguson.NJP.2023}%
  \BibitemOpen
  \bibfield  {author} {\bibinfo {author} {\bibfnamefont {S.}~\bibnamefont
  {Ferguson}}, \bibinfo {author} {\bibfnamefont {P.}~\bibnamefont {Martin}},
  \bibinfo {author} {\bibfnamefont {H.}~\bibnamefont {Ahmed}}, \bibinfo
  {author} {\bibfnamefont {E.}~\bibnamefont {Aktan}}, \bibinfo {author}
  {\bibfnamefont {M.}~\bibnamefont {Alanazi}}, \bibinfo {author} {\bibfnamefont
  {M.}~\bibnamefont {Cerchez}}, \bibinfo {author} {\bibfnamefont
  {D.}~\bibnamefont {Doria}}, \bibinfo {author} {\bibfnamefont {J.~S.}\
  \bibnamefont {Green}}, \bibinfo {author} {\bibfnamefont {B.}~\bibnamefont
  {Greenwood}}, \bibinfo {author} {\bibfnamefont {B.}~\bibnamefont
  {Odlozilik}}, \bibinfo {author} {\bibfnamefont {O.}~\bibnamefont {Willi}},
  \bibinfo {author} {\bibfnamefont {M.}~\bibnamefont {Borghesi}}, \ and\
  \bibinfo {author} {\bibfnamefont {S.}~\bibnamefont {Kar}},\ }\href {\doibase
  10.1088/1367-2630/acaf99} {\bibfield  {journal} {\bibinfo  {journal} {New
  Journal of Physics}\ }\textbf {\bibinfo {volume} {25}},\ \bibinfo {pages}
  {013006} (\bibinfo {year} {2023})}\BibitemShut {NoStop}%
\bibitem [{\citenamefont {Liu}\ \emph {et~al.}(2018)\citenamefont {Liu},
  \citenamefont {Weng}, \citenamefont {Wang}, \citenamefont {Chen},
  \citenamefont {Zhao}, \citenamefont {Sheng}, \citenamefont {He},
  \citenamefont {Li},\ and\ \citenamefont {Zhang}}]{Liu.PoP.2018}%
  \BibitemOpen
  \bibfield  {author} {\bibinfo {author} {\bibfnamefont {M.}~\bibnamefont
  {Liu}}, \bibinfo {author} {\bibfnamefont {S.~M.}\ \bibnamefont {Weng}},
  \bibinfo {author} {\bibfnamefont {H.~C.}\ \bibnamefont {Wang}}, \bibinfo
  {author} {\bibfnamefont {M.}~\bibnamefont {Chen}}, \bibinfo {author}
  {\bibfnamefont {Q.}~\bibnamefont {Zhao}}, \bibinfo {author} {\bibfnamefont
  {Z.~M.}\ \bibnamefont {Sheng}}, \bibinfo {author} {\bibfnamefont {M.~Q.}\
  \bibnamefont {He}}, \bibinfo {author} {\bibfnamefont {Y.~T.}\ \bibnamefont
  {Li}}, \ and\ \bibinfo {author} {\bibfnamefont {J.}~\bibnamefont {Zhang}},\
  }\href {\doibase 10.1063/1.5033991} {\bibfield  {journal} {\bibinfo
  {journal} {Physics of Plasmas}\ }\textbf {\bibinfo {volume} {25}},\ \bibinfo
  {pages} {063103} (\bibinfo {year} {2018})}\BibitemShut {NoStop}%
\bibitem [{\citenamefont {An}\ \emph {et~al.}(2023)\citenamefont {An},
  \citenamefont {Wang}, \citenamefont {Xiong}, \citenamefont {Wang},
  \citenamefont {Pan}, \citenamefont {Ouyang}, \citenamefont {Jiang},
  \citenamefont {Xie}, \citenamefont {Wang}, \citenamefont {Yao}, \citenamefont
  {Hua}, \citenamefont {Wang}, \citenamefont {Jiang}, \citenamefont {Xiao},
  \citenamefont {Ding}, \citenamefont {Wan}, \citenamefont {Liu}, \citenamefont
  {Wang}, \citenamefont {Fang}, \citenamefont {Yang}, \citenamefont {Jiang},
  \citenamefont {Zhang}, \citenamefont {Zhu}, \citenamefont {Sun},
  \citenamefont {Qiao}, \citenamefont {Lei},\ and\ \citenamefont
  {Zhu}}]{An.HPLSE.2023}%
  \BibitemOpen
  \bibfield  {author} {\bibinfo {author} {\bibfnamefont {H.~H.}\ \bibnamefont
  {An}}, \bibinfo {author} {\bibfnamefont {W.}~\bibnamefont {Wang}}, \bibinfo
  {author} {\bibfnamefont {J.}~\bibnamefont {Xiong}}, \bibinfo {author}
  {\bibfnamefont {C.}~\bibnamefont {Wang}}, \bibinfo {author} {\bibfnamefont
  {X.}~\bibnamefont {Pan}}, \bibinfo {author} {\bibfnamefont {X.~P.}\
  \bibnamefont {Ouyang}}, \bibinfo {author} {\bibfnamefont {S.}~\bibnamefont
  {Jiang}}, \bibinfo {author} {\bibfnamefont {Z.~Y.}\ \bibnamefont {Xie}},
  \bibinfo {author} {\bibfnamefont {P.~P.}\ \bibnamefont {Wang}}, \bibinfo
  {author} {\bibfnamefont {Y.~L.}\ \bibnamefont {Yao}}, \bibinfo {author}
  {\bibfnamefont {N.}~\bibnamefont {Hua}}, \bibinfo {author} {\bibfnamefont
  {Y.}~\bibnamefont {Wang}}, \bibinfo {author} {\bibfnamefont {Z.~C.}\
  \bibnamefont {Jiang}}, \bibinfo {author} {\bibfnamefont {Q.}~\bibnamefont
  {Xiao}}, \bibinfo {author} {\bibfnamefont {F.~C.}\ \bibnamefont {Ding}},
  \bibinfo {author} {\bibfnamefont {Y.~T.}\ \bibnamefont {Wan}}, \bibinfo
  {author} {\bibfnamefont {X.}~\bibnamefont {Liu}}, \bibinfo {author}
  {\bibfnamefont {R.~R.}\ \bibnamefont {Wang}}, \bibinfo {author}
  {\bibfnamefont {Z.~H.}\ \bibnamefont {Fang}}, \bibinfo {author}
  {\bibfnamefont {P.~Q.}\ \bibnamefont {Yang}}, \bibinfo {author}
  {\bibfnamefont {Y.~E.}\ \bibnamefont {Jiang}}, \bibinfo {author}
  {\bibfnamefont {P.~Z.}\ \bibnamefont {Zhang}}, \bibinfo {author}
  {\bibfnamefont {B.~Q.}\ \bibnamefont {Zhu}}, \bibinfo {author} {\bibfnamefont
  {J.~R.}\ \bibnamefont {Sun}}, \bibinfo {author} {\bibfnamefont
  {B.}~\bibnamefont {Qiao}}, \bibinfo {author} {\bibfnamefont {A.~L.}\
  \bibnamefont {Lei}}, \ and\ \bibinfo {author} {\bibfnamefont {J.~Q.}\
  \bibnamefont {Zhu}},\ }\href {\doibase 10.1017/HPL.2023.54} {\bibfield
  {journal} {\bibinfo  {journal} {High Power Laser Science and Engineering}\
  }\textbf {\bibinfo {volume} {11}},\ \bibinfo {pages} {e63} (\bibinfo {year}
  {2023})}\BibitemShut {NoStop}%
\bibitem [{\citenamefont {Bychenkov}\ and\ \citenamefont
  {Dudnikova}(2007)}]{Bychenkov.PPR.2007}%
  \BibitemOpen
  \bibfield  {author} {\bibinfo {author} {\bibfnamefont {V.~Y.}\ \bibnamefont
  {Bychenkov}}\ and\ \bibinfo {author} {\bibfnamefont {G.~I.}\ \bibnamefont
  {Dudnikova}},\ }\href {\doibase 10.1134/S1063780X07080065} {\bibfield
  {journal} {\bibinfo  {journal} {Plasma Physics Reports}\ }\textbf {\bibinfo
  {volume} {33}},\ \bibinfo {pages} {655} (\bibinfo {year} {2007})}\BibitemShut
  {NoStop}%
\bibitem [{\citenamefont {Yogo}\ \emph {et~al.}(2017)\citenamefont {Yogo},
  \citenamefont {Mima}, \citenamefont {Iwata}, \citenamefont {Tosaki},
  \citenamefont {Morace}, \citenamefont {Arikawa}, \citenamefont {Fujioka},
  \citenamefont {Johzaki}, \citenamefont {Sentoku}, \citenamefont {Nishimura},
  \citenamefont {Sagisaka}, \citenamefont {Matsuo}, \citenamefont
  {Kamitsukasa}, \citenamefont {Kojima}, \citenamefont {Nagatomo},
  \citenamefont {Nakai}, \citenamefont {Shiraga}, \citenamefont {Murakami},
  \citenamefont {Tokita}, \citenamefont {Kawanaka}, \citenamefont {Miyanaga},
  \citenamefont {Yamanoi}, \citenamefont {Norimatsu}, \citenamefont {Sakagami},
  \citenamefont {Bulanov}, \citenamefont {Kondo},\ and\ \citenamefont
  {Azechi}}]{Yogo.SciRep.2017}%
  \BibitemOpen
  \bibfield  {author} {\bibinfo {author} {\bibfnamefont {A.}~\bibnamefont
  {Yogo}}, \bibinfo {author} {\bibfnamefont {K.}~\bibnamefont {Mima}}, \bibinfo
  {author} {\bibfnamefont {N.}~\bibnamefont {Iwata}}, \bibinfo {author}
  {\bibfnamefont {S.}~\bibnamefont {Tosaki}}, \bibinfo {author} {\bibfnamefont
  {A.}~\bibnamefont {Morace}}, \bibinfo {author} {\bibfnamefont
  {Y.}~\bibnamefont {Arikawa}}, \bibinfo {author} {\bibfnamefont
  {S.}~\bibnamefont {Fujioka}}, \bibinfo {author} {\bibfnamefont
  {T.}~\bibnamefont {Johzaki}}, \bibinfo {author} {\bibfnamefont
  {Y.}~\bibnamefont {Sentoku}}, \bibinfo {author} {\bibfnamefont
  {H.}~\bibnamefont {Nishimura}}, \bibinfo {author} {\bibfnamefont
  {A.}~\bibnamefont {Sagisaka}}, \bibinfo {author} {\bibfnamefont
  {K.}~\bibnamefont {Matsuo}}, \bibinfo {author} {\bibfnamefont
  {N.}~\bibnamefont {Kamitsukasa}}, \bibinfo {author} {\bibfnamefont
  {S.}~\bibnamefont {Kojima}}, \bibinfo {author} {\bibfnamefont
  {H.}~\bibnamefont {Nagatomo}}, \bibinfo {author} {\bibfnamefont
  {M.}~\bibnamefont {Nakai}}, \bibinfo {author} {\bibfnamefont
  {H.}~\bibnamefont {Shiraga}}, \bibinfo {author} {\bibfnamefont
  {M.}~\bibnamefont {Murakami}}, \bibinfo {author} {\bibfnamefont
  {S.}~\bibnamefont {Tokita}}, \bibinfo {author} {\bibfnamefont
  {J.}~\bibnamefont {Kawanaka}}, \bibinfo {author} {\bibfnamefont
  {N.}~\bibnamefont {Miyanaga}}, \bibinfo {author} {\bibfnamefont
  {K.}~\bibnamefont {Yamanoi}}, \bibinfo {author} {\bibfnamefont
  {T.}~\bibnamefont {Norimatsu}}, \bibinfo {author} {\bibfnamefont
  {H.}~\bibnamefont {Sakagami}}, \bibinfo {author} {\bibfnamefont {S.~V.}\
  \bibnamefont {Bulanov}}, \bibinfo {author} {\bibfnamefont {K.}~\bibnamefont
  {Kondo}}, \ and\ \bibinfo {author} {\bibfnamefont {H.}~\bibnamefont
  {Azechi}},\ }\href {\doibase 10.1038/srep42451} {\bibfield  {journal}
  {\bibinfo  {journal} {Scientific Reports}\ }\textbf {\bibinfo {volume} {7}},\
  \bibinfo {pages} {42451} (\bibinfo {year} {2017})}\BibitemShut {NoStop}%
\bibitem [{\citenamefont {Ferri}\ \emph {et~al.}(2019)\citenamefont {Ferri},
  \citenamefont {Siminos},\ and\ \citenamefont
  {Fülöp}}]{Ferri.CommPhys.2019}%
  \BibitemOpen
  \bibfield  {author} {\bibinfo {author} {\bibfnamefont {J.}~\bibnamefont
  {Ferri}}, \bibinfo {author} {\bibfnamefont {E.}~\bibnamefont {Siminos}}, \
  and\ \bibinfo {author} {\bibfnamefont {T.}~\bibnamefont {Fülöp}},\ }\href
  {\doibase 10.1038/s42005-019-0140-x} {\bibfield  {journal} {\bibinfo
  {journal} {Communications Physics}\ }\textbf {\bibinfo {volume} {2}},\
  \bibinfo {pages} {1} (\bibinfo {year} {2019})},\ \bibinfo {note} {publisher:
  Nature Publishing Group}\BibitemShut {NoStop}%
\bibitem [{\citenamefont {Kim}\ \emph {et~al.}(2022)\citenamefont {Kim},
  \citenamefont {Wilks}, \citenamefont {Kemp}, \citenamefont {Sherlock},
  \citenamefont {Ma}, \citenamefont {Beg},\ and\ \citenamefont
  {Mariscal}}]{Kim.PRR.2022}%
  \BibitemOpen
  \bibfield  {author} {\bibinfo {author} {\bibfnamefont {J.}~\bibnamefont
  {Kim}}, \bibinfo {author} {\bibfnamefont {S.}~\bibnamefont {Wilks}}, \bibinfo
  {author} {\bibfnamefont {A.}~\bibnamefont {Kemp}}, \bibinfo {author}
  {\bibfnamefont {M.}~\bibnamefont {Sherlock}}, \bibinfo {author}
  {\bibfnamefont {T.}~\bibnamefont {Ma}}, \bibinfo {author} {\bibfnamefont
  {F.}~\bibnamefont {Beg}}, \ and\ \bibinfo {author} {\bibfnamefont
  {D.}~\bibnamefont {Mariscal}},\ }\href {\doibase
  10.1103/PhysRevResearch.4.L032003} {\bibfield  {journal} {\bibinfo  {journal}
  {Physical Review Research}\ }\textbf {\bibinfo {volume} {4}},\ \bibinfo
  {pages} {L032003} (\bibinfo {year} {2022})}\BibitemShut {NoStop}%
\bibitem [{\citenamefont {Wang}\ \emph {et~al.}(2017)\citenamefont {Wang},
  \citenamefont {Weng}, \citenamefont {Murakami}, \citenamefont {Sheng},
  \citenamefont {Chen}, \citenamefont {Zhao},\ and\ \citenamefont
  {Zhang}}]{Wang.PoP.2017}%
  \BibitemOpen
  \bibfield  {author} {\bibinfo {author} {\bibfnamefont {H.~C.}\ \bibnamefont
  {Wang}}, \bibinfo {author} {\bibfnamefont {S.~M.}\ \bibnamefont {Weng}},
  \bibinfo {author} {\bibfnamefont {M.}~\bibnamefont {Murakami}}, \bibinfo
  {author} {\bibfnamefont {Z.~M.}\ \bibnamefont {Sheng}}, \bibinfo {author}
  {\bibfnamefont {M.}~\bibnamefont {Chen}}, \bibinfo {author} {\bibfnamefont
  {Q.}~\bibnamefont {Zhao}}, \ and\ \bibinfo {author} {\bibfnamefont
  {J.}~\bibnamefont {Zhang}},\ }\href {\doibase 10.1063/1.5000104} {\bibfield
  {journal} {\bibinfo  {journal} {Physics of Plasmas}\ }\textbf {\bibinfo
  {volume} {24}} (\bibinfo {year} {2017}),\ 10.1063/1.5000104}\BibitemShut
  {NoStop}%
\bibitem [{\citenamefont {Ting}\ \emph {et~al.}(2017)\citenamefont {Ting},
  \citenamefont {Hafizi}, \citenamefont {Helle}, \citenamefont {Chen},
  \citenamefont {Gordon}, \citenamefont {Kaganovich}, \citenamefont
  {Polyanskiy}, \citenamefont {Pogorelsky}, \citenamefont {Babzien},
  \citenamefont {Miao}, \citenamefont {Dover}, \citenamefont {Najmudin},\ and\
  \citenamefont {Ettlinger}}]{Ting.AAC.2017}%
  \BibitemOpen
  \bibfield  {author} {\bibinfo {author} {\bibfnamefont {A.}~\bibnamefont
  {Ting}}, \bibinfo {author} {\bibfnamefont {B.}~\bibnamefont {Hafizi}},
  \bibinfo {author} {\bibfnamefont {M.}~\bibnamefont {Helle}}, \bibinfo
  {author} {\bibfnamefont {Y.~H.}\ \bibnamefont {Chen}}, \bibinfo {author}
  {\bibfnamefont {D.}~\bibnamefont {Gordon}}, \bibinfo {author} {\bibfnamefont
  {D.}~\bibnamefont {Kaganovich}}, \bibinfo {author} {\bibfnamefont
  {M.}~\bibnamefont {Polyanskiy}}, \bibinfo {author} {\bibfnamefont
  {I.}~\bibnamefont {Pogorelsky}}, \bibinfo {author} {\bibfnamefont
  {M.}~\bibnamefont {Babzien}}, \bibinfo {author} {\bibfnamefont
  {C.}~\bibnamefont {Miao}}, \bibinfo {author} {\bibfnamefont {N.}~\bibnamefont
  {Dover}}, \bibinfo {author} {\bibfnamefont {Z.}~\bibnamefont {Najmudin}}, \
  and\ \bibinfo {author} {\bibfnamefont {O.}~\bibnamefont {Ettlinger}}\
  }(\bibinfo  {publisher} {American Institute of Physics Inc.},\ \bibinfo
  {year} {2017})\BibitemShut {NoStop}%
\bibitem [{\citenamefont {Kim}\ \emph {et~al.}(2016)\citenamefont {Kim},
  \citenamefont {Göde},\ and\ \citenamefont {Glenzer}}]{Kim2016}%
  \BibitemOpen
  \bibfield  {author} {\bibinfo {author} {\bibfnamefont {J.~B.}\ \bibnamefont
  {Kim}}, \bibinfo {author} {\bibfnamefont {S.}~\bibnamefont {Göde}}, \ and\
  \bibinfo {author} {\bibfnamefont {S.~H.}\ \bibnamefont {Glenzer}},\ }\href
  {\doibase 10.1063/1.4961089} {\bibfield  {journal} {\bibinfo  {journal}
  {Review of Scientific Instruments}\ }\textbf {\bibinfo {volume} {87}},\
  \bibinfo {pages} {11E328} (\bibinfo {year} {2016})},\ \Eprint
  {http://arxiv.org/abs/https://pubs.aip.org/aip/rsi/article-pdf/doi/10.1063/1.4961089/16054498/11e328\_1\_online.pdf}
  {https://pubs.aip.org/aip/rsi/article-pdf/doi/10.1063/1.4961089/16054498/11e328\_1\_online.pdf}
  \BibitemShut {NoStop}%
\bibitem [{\citenamefont {Gong}\ \emph {et~al.}(2022)\citenamefont {Gong},
  \citenamefont {Bulanov}, \citenamefont {Toncian},\ and\ \citenamefont
  {Arefiev}}]{gong.prr.2022}%
  \BibitemOpen
  \bibfield  {author} {\bibinfo {author} {\bibfnamefont {Z.}~\bibnamefont
  {Gong}}, \bibinfo {author} {\bibfnamefont {S.~S.}\ \bibnamefont {Bulanov}},
  \bibinfo {author} {\bibfnamefont {T.}~\bibnamefont {Toncian}}, \ and\
  \bibinfo {author} {\bibfnamefont {A.}~\bibnamefont {Arefiev}},\ }\href
  {\doibase 10.1103/PhysRevResearch.4.L042031} {\bibfield  {journal} {\bibinfo
  {journal} {Phys. Rev. Res.}\ }\textbf {\bibinfo {volume} {4}},\ \bibinfo
  {pages} {L042031} (\bibinfo {year} {2022})}\BibitemShut {NoStop}%
\bibitem [{\citenamefont {Lezhnin}\ and\ \citenamefont
  {Bulanov}(2022)}]{Lezhnin.PRR.2022}%
  \BibitemOpen
  \bibfield  {author} {\bibinfo {author} {\bibfnamefont {K.~V.}\ \bibnamefont
  {Lezhnin}}\ and\ \bibinfo {author} {\bibfnamefont {S.~V.}\ \bibnamefont
  {Bulanov}},\ }\href {\doibase 10.1103/PhysRevResearch.4.033248} {\bibfield
  {journal} {\bibinfo  {journal} {Phys. Rev. Res.}\ }\textbf {\bibinfo {volume}
  {4}},\ \bibinfo {pages} {033248} (\bibinfo {year} {2022})}\BibitemShut
  {NoStop}%
\bibitem [{\citenamefont {Zhu}\ \emph {et~al.}(2022)\citenamefont {Zhu},
  \citenamefont {Liu}, \citenamefont {Chen}, \citenamefont {Weng},
  \citenamefont {McKenna}, \citenamefont {Sheng},\ and\ \citenamefont
  {Zhang}}]{zhu.prappl.2022}%
  \BibitemOpen
  \bibfield  {author} {\bibinfo {author} {\bibfnamefont {X.-L.}\ \bibnamefont
  {Zhu}}, \bibinfo {author} {\bibfnamefont {W.-Y.}\ \bibnamefont {Liu}},
  \bibinfo {author} {\bibfnamefont {M.}~\bibnamefont {Chen}}, \bibinfo {author}
  {\bibfnamefont {S.-M.}\ \bibnamefont {Weng}}, \bibinfo {author}
  {\bibfnamefont {P.}~\bibnamefont {McKenna}}, \bibinfo {author} {\bibfnamefont
  {Z.-M.}\ \bibnamefont {Sheng}}, \ and\ \bibinfo {author} {\bibfnamefont
  {J.}~\bibnamefont {Zhang}},\ }\href@noop {} {\bibfield  {journal} {\bibinfo
  {journal} {Physical Review Applied}\ }\textbf {\bibinfo {volume} {18}},\
  \bibinfo {pages} {044051} (\bibinfo {year} {2022})}\BibitemShut {NoStop}%
\bibitem [{\citenamefont {Tazes}\ \emph {et~al.}(2022)\citenamefont {Tazes},
  \citenamefont {Passalidis}, \citenamefont {Kaselouris}, \citenamefont
  {Fitilis}, \citenamefont {Bakarezos}, \citenamefont {Papadogiannis},
  \citenamefont {Tatarakis},\ and\ \citenamefont
  {Dimitriou}}]{Tazes.HPLSE.2022}%
  \BibitemOpen
  \bibfield  {author} {\bibinfo {author} {\bibfnamefont {I.}~\bibnamefont
  {Tazes}}, \bibinfo {author} {\bibfnamefont {S.}~\bibnamefont {Passalidis}},
  \bibinfo {author} {\bibfnamefont {E.}~\bibnamefont {Kaselouris}}, \bibinfo
  {author} {\bibfnamefont {I.}~\bibnamefont {Fitilis}}, \bibinfo {author}
  {\bibfnamefont {M.}~\bibnamefont {Bakarezos}}, \bibinfo {author}
  {\bibfnamefont {N.~A.}\ \bibnamefont {Papadogiannis}}, \bibinfo {author}
  {\bibfnamefont {M.}~\bibnamefont {Tatarakis}}, \ and\ \bibinfo {author}
  {\bibfnamefont {V.}~\bibnamefont {Dimitriou}},\ }\href {\doibase
  10.1017/hpl.2022.16} {\bibfield  {journal} {\bibinfo  {journal} {High Power
  Laser Science and Engineering}\ }\textbf {\bibinfo {volume} {10}} (\bibinfo
  {year} {2022}),\ 10.1017/hpl.2022.16}\BibitemShut {NoStop}%
\bibitem [{\citenamefont {Bulanov}\ \emph {et~al.}(2015)\citenamefont
  {Bulanov}, \citenamefont {Esarey}, \citenamefont {Schroeder}, \citenamefont
  {Leemans}, \citenamefont {Bulanov}, \citenamefont {Margarone}, \citenamefont
  {Korn},\ and\ \citenamefont {Haberer}}]{bulanov.prab.2015}%
  \BibitemOpen
  \bibfield  {author} {\bibinfo {author} {\bibfnamefont {S.~S.}\ \bibnamefont
  {Bulanov}}, \bibinfo {author} {\bibfnamefont {E.}~\bibnamefont {Esarey}},
  \bibinfo {author} {\bibfnamefont {C.~B.}\ \bibnamefont {Schroeder}}, \bibinfo
  {author} {\bibfnamefont {W.~P.}\ \bibnamefont {Leemans}}, \bibinfo {author}
  {\bibfnamefont {S.~V.}\ \bibnamefont {Bulanov}}, \bibinfo {author}
  {\bibfnamefont {D.}~\bibnamefont {Margarone}}, \bibinfo {author}
  {\bibfnamefont {G.}~\bibnamefont {Korn}}, \ and\ \bibinfo {author}
  {\bibfnamefont {T.}~\bibnamefont {Haberer}},\ }\href {\doibase
  10.1103/PhysRevSTAB.18.061302} {\bibfield  {journal} {\bibinfo  {journal}
  {Phys. Rev. ST Accel. Beams}\ }\textbf {\bibinfo {volume} {18}},\ \bibinfo
  {pages} {061302} (\bibinfo {year} {2015})}\BibitemShut {NoStop}%
\bibitem [{\citenamefont {Fedeli}\ \emph {et~al.}(2022)\citenamefont {Fedeli},
  \citenamefont {Huebl}, \citenamefont {Boillod-Cerneux}, \citenamefont
  {Clark}, \citenamefont {Gott}, \citenamefont {Hillairet}, \citenamefont
  {Jaure}, \citenamefont {Leblanc}, \citenamefont {Lehe}, \citenamefont
  {Myers}, \citenamefont {Piechurski}, \citenamefont {Sato}, \citenamefont
  {Zaim}, \citenamefont {Zhang}, \citenamefont {Vay},\ and\ \citenamefont
  {Vincenti}}]{FedeliHuebl2022}%
  \BibitemOpen
  \bibfield  {author} {\bibinfo {author} {\bibfnamefont {L.}~\bibnamefont
  {Fedeli}}, \bibinfo {author} {\bibfnamefont {A.}~\bibnamefont {Huebl}},
  \bibinfo {author} {\bibfnamefont {F.}~\bibnamefont {Boillod-Cerneux}},
  \bibinfo {author} {\bibfnamefont {T.}~\bibnamefont {Clark}}, \bibinfo
  {author} {\bibfnamefont {K.}~\bibnamefont {Gott}}, \bibinfo {author}
  {\bibfnamefont {C.}~\bibnamefont {Hillairet}}, \bibinfo {author}
  {\bibfnamefont {S.}~\bibnamefont {Jaure}}, \bibinfo {author} {\bibfnamefont
  {A.}~\bibnamefont {Leblanc}}, \bibinfo {author} {\bibfnamefont
  {R.}~\bibnamefont {Lehe}}, \bibinfo {author} {\bibfnamefont {A.}~\bibnamefont
  {Myers}}, \bibinfo {author} {\bibfnamefont {C.}~\bibnamefont {Piechurski}},
  \bibinfo {author} {\bibfnamefont {M.}~\bibnamefont {Sato}}, \bibinfo {author}
  {\bibfnamefont {N.}~\bibnamefont {Zaim}}, \bibinfo {author} {\bibfnamefont
  {W.}~\bibnamefont {Zhang}}, \bibinfo {author} {\bibfnamefont {J.-L.}\
  \bibnamefont {Vay}}, \ and\ \bibinfo {author} {\bibfnamefont
  {H.}~\bibnamefont {Vincenti}},\ }in\ \href {\doibase
  10.1109/SC41404.2022.00008} {\emph {\bibinfo {booktitle} {SC22: International
  Conference for High Performance Computing, Networking, Storage and
  Analysis}}}\ (\bibinfo {year} {2022})\ pp.\ \bibinfo {pages}
  {1--12}\BibitemShut {NoStop}%
\bibitem [{\citenamefont {Vay}\ \emph {et~al.}(2018)\citenamefont {Vay},
  \citenamefont {Almgren}, \citenamefont {Amorim}, \citenamefont {Andriyash},
  \citenamefont {Belkin}, \citenamefont {Bizzozero}, \citenamefont {Blelly},
  \citenamefont {Clark}, \citenamefont {Fedeli}, \citenamefont {Garten},
  \citenamefont {Ge}, \citenamefont {Gott}, \citenamefont {Harrison},
  \citenamefont {Huebl}, \citenamefont {Giacomel}, \citenamefont {Groenewald},
  \citenamefont {Grote}, \citenamefont {Gu}, \citenamefont {Jambunathan},
  \citenamefont {Klion}, \citenamefont {Kumar}, \citenamefont {Thevenet},
  \citenamefont {Richardson}, \citenamefont {Shapoval}, \citenamefont {Lehe},
  \citenamefont {Loring}, \citenamefont {Miller}, \citenamefont {Myers},
  \citenamefont {Rheaume}, \citenamefont {Rowan}, \citenamefont {Sandberg},
  \citenamefont {Scherpelz}, \citenamefont {Yang}, \citenamefont {Zhang},
  \citenamefont {Zhao}, \citenamefont {Zhu}, \citenamefont {Zoni},\ and\
  \citenamefont {Zaim}}]{WarpX}%
  \BibitemOpen
  \bibfield  {author} {\bibinfo {author} {\bibfnamefont {J.-L.}\ \bibnamefont
  {Vay}}, \bibinfo {author} {\bibfnamefont {A.}~\bibnamefont {Almgren}},
  \bibinfo {author} {\bibfnamefont {L.~D.}\ \bibnamefont {Amorim}}, \bibinfo
  {author} {\bibfnamefont {I.}~\bibnamefont {Andriyash}}, \bibinfo {author}
  {\bibfnamefont {D.}~\bibnamefont {Belkin}}, \bibinfo {author} {\bibfnamefont
  {D.}~\bibnamefont {Bizzozero}}, \bibinfo {author} {\bibfnamefont
  {A.}~\bibnamefont {Blelly}}, \bibinfo {author} {\bibfnamefont {S.~E.}\
  \bibnamefont {Clark}}, \bibinfo {author} {\bibfnamefont {L.}~\bibnamefont
  {Fedeli}}, \bibinfo {author} {\bibfnamefont {M.}~\bibnamefont {Garten}},
  \bibinfo {author} {\bibfnamefont {L.}~\bibnamefont {Ge}}, \bibinfo {author}
  {\bibfnamefont {K.}~\bibnamefont {Gott}}, \bibinfo {author} {\bibfnamefont
  {C.}~\bibnamefont {Harrison}}, \bibinfo {author} {\bibfnamefont
  {A.}~\bibnamefont {Huebl}}, \bibinfo {author} {\bibfnamefont
  {L.}~\bibnamefont {Giacomel}}, \bibinfo {author} {\bibfnamefont {R.~E.}\
  \bibnamefont {Groenewald}}, \bibinfo {author} {\bibfnamefont
  {D.}~\bibnamefont {Grote}}, \bibinfo {author} {\bibfnamefont
  {J.}~\bibnamefont {Gu}}, \bibinfo {author} {\bibfnamefont {R.}~\bibnamefont
  {Jambunathan}}, \bibinfo {author} {\bibfnamefont {H.}~\bibnamefont {Klion}},
  \bibinfo {author} {\bibfnamefont {P.}~\bibnamefont {Kumar}}, \bibinfo
  {author} {\bibfnamefont {M.}~\bibnamefont {Thevenet}}, \bibinfo {author}
  {\bibfnamefont {G.}~\bibnamefont {Richardson}}, \bibinfo {author}
  {\bibfnamefont {O.}~\bibnamefont {Shapoval}}, \bibinfo {author}
  {\bibfnamefont {R.}~\bibnamefont {Lehe}}, \bibinfo {author} {\bibfnamefont
  {B.}~\bibnamefont {Loring}}, \bibinfo {author} {\bibfnamefont
  {P.}~\bibnamefont {Miller}}, \bibinfo {author} {\bibfnamefont
  {A.}~\bibnamefont {Myers}}, \bibinfo {author} {\bibfnamefont
  {E.}~\bibnamefont {Rheaume}}, \bibinfo {author} {\bibfnamefont {M.~E.}\
  \bibnamefont {Rowan}}, \bibinfo {author} {\bibfnamefont {R.~T.}\ \bibnamefont
  {Sandberg}}, \bibinfo {author} {\bibfnamefont {P.}~\bibnamefont {Scherpelz}},
  \bibinfo {author} {\bibfnamefont {E.}~\bibnamefont {Yang}}, \bibinfo {author}
  {\bibfnamefont {W.}~\bibnamefont {Zhang}}, \bibinfo {author} {\bibfnamefont
  {Y.}~\bibnamefont {Zhao}}, \bibinfo {author} {\bibfnamefont {K.~Z.}\
  \bibnamefont {Zhu}}, \bibinfo {author} {\bibfnamefont {E.}~\bibnamefont
  {Zoni}}, \ and\ \bibinfo {author} {\bibfnamefont {N.}~\bibnamefont {Zaim}},\
  }\href {\doibase 10.5281/zenodo.4571577} {\enquote {\bibinfo {title}
  {{WarpX}},}\ } (\bibinfo {year} {2018}),\ \bibinfo {note}
  {https://ecp-warpx.github.io
  https://doi.org/10.5281/zenodo.4571577}\BibitemShut {NoStop}%
\bibitem [{\citenamefont {Obst-Huebl}\ \emph {et~al.}(2023)\citenamefont
  {Obst-Huebl}, \citenamefont {Nakamura}, \citenamefont {Hakimi}, \citenamefont
  {Chant}, \citenamefont {Jewell}, \citenamefont {Stassel}, \citenamefont
  {Snijders}, \citenamefont {Gonsalves}, \citenamefont {van Tilborg},
  \citenamefont {Eisentraut}, \citenamefont {Harvey}, \citenamefont
  {Willingale}, \citenamefont {Toth}, \citenamefont {Schroeder}, \citenamefont
  {Geddes},\ and\ \citenamefont {Esarey}}]{ObstHueblSPIE2023}%
  \BibitemOpen
  \bibfield  {author} {\bibinfo {author} {\bibfnamefont {L.}~\bibnamefont
  {Obst-Huebl}}, \bibinfo {author} {\bibfnamefont {K.}~\bibnamefont
  {Nakamura}}, \bibinfo {author} {\bibfnamefont {S.}~\bibnamefont {Hakimi}},
  \bibinfo {author} {\bibfnamefont {J.~T.~D.}\ \bibnamefont {Chant}}, \bibinfo
  {author} {\bibfnamefont {A.}~\bibnamefont {Jewell}}, \bibinfo {author}
  {\bibfnamefont {B.}~\bibnamefont {Stassel}}, \bibinfo {author} {\bibfnamefont
  {A.~M.}\ \bibnamefont {Snijders}}, \bibinfo {author} {\bibfnamefont {A.~J.}\
  \bibnamefont {Gonsalves}}, \bibinfo {author} {\bibfnamefont {J.}~\bibnamefont
  {van Tilborg}}, \bibinfo {author} {\bibfnamefont {Z.}~\bibnamefont
  {Eisentraut}}, \bibinfo {author} {\bibfnamefont {Z.}~\bibnamefont {Harvey}},
  \bibinfo {author} {\bibfnamefont {L.}~\bibnamefont {Willingale}}, \bibinfo
  {author} {\bibfnamefont {C.}~\bibnamefont {Toth}}, \bibinfo {author}
  {\bibfnamefont {C.~B.}\ \bibnamefont {Schroeder}}, \bibinfo {author}
  {\bibfnamefont {C.~G.~R.}\ \bibnamefont {Geddes}}, \ and\ \bibinfo {author}
  {\bibfnamefont {E.}~\bibnamefont {Esarey}},\ }in\ \href {\doibase
  10.1117/12.2669162} {\emph {\bibinfo {booktitle} {Applying Laser-driven
  Particle Acceleration III: Using Distinctive Energetic Particle and Photon
  Sources}}},\ Vol.\ \bibinfo {volume} {12583},\ \bibinfo {editor} {edited by\
  \bibinfo {editor} {\bibfnamefont {J.}~\bibnamefont {Schreiber}}\ and\
  \bibinfo {editor} {\bibfnamefont {P.~R.}\ \bibnamefont {Bolton}}},\ \bibinfo
  {organization} {International Society for Optics and Photonics}\ (\bibinfo
  {publisher} {SPIE},\ \bibinfo {year} {2023})\ p.\ \bibinfo {pages}
  {1258305}\BibitemShut {NoStop}%
\bibitem [{\citenamefont {Kapchinskij}\ and\ \citenamefont
  {Vladimirskij}(1959)}]{kvdist.cern.1959}%
  \BibitemOpen
  \bibfield  {author} {\bibinfo {author} {\bibfnamefont {I.}~\bibnamefont
  {Kapchinskij}}\ and\ \bibinfo {author} {\bibfnamefont {V.}~\bibnamefont
  {Vladimirskij}},\ }in\ \href@noop {} {\emph {\bibinfo {booktitle}
  {Proceedings of the International Conference on High Energy Accelerators and
  Instrumentation}}}\ (\bibinfo {organization} {CERN Scientific Information
  Service Geneva},\ \bibinfo {year} {1959})\ p.\ \bibinfo {pages}
  {274}\BibitemShut {NoStop}%
\bibitem [{\citenamefont {Lindstr\o{}m}\ and\ \citenamefont
  {Adli}(2016)}]{Lindstroem2016}%
  \BibitemOpen
  \bibfield  {author} {\bibinfo {author} {\bibfnamefont {C.~A.}\ \bibnamefont
  {Lindstr\o{}m}}\ and\ \bibinfo {author} {\bibfnamefont {E.}~\bibnamefont
  {Adli}},\ }\href {\doibase 10.1103/PhysRevAccelBeams.19.071002} {\bibfield
  {journal} {\bibinfo  {journal} {Phys. Rev. Accel. Beams}\ }\textbf {\bibinfo
  {volume} {19}},\ \bibinfo {pages} {071002} (\bibinfo {year}
  {2016})}\BibitemShut {NoStop}%
\bibitem [{\citenamefont {Brack}\ \emph {et~al.}(2020)\citenamefont {Brack},
  \citenamefont {Kroll}, \citenamefont {Gaus}, \citenamefont {Bernert},
  \citenamefont {Beyreuther}, \citenamefont {Cowan}, \citenamefont {Karsch},
  \citenamefont {Kraft}, \citenamefont {Kunz-Schughart}, \citenamefont
  {Lessmann}, \citenamefont {Metzkes-Ng}, \citenamefont {Obst-Huebl},
  \citenamefont {Pawelke}, \citenamefont {Rehwald}, \citenamefont
  {Schlenvoigt}, \citenamefont {Schramm}, \citenamefont {Sobiella},
  \citenamefont {Szabó}, \citenamefont {Ziegler},\ and\ \citenamefont
  {Zeil}}]{Brack2020}%
  \BibitemOpen
  \bibfield  {author} {\bibinfo {author} {\bibfnamefont {F.-E.}\ \bibnamefont
  {Brack}}, \bibinfo {author} {\bibfnamefont {F.}~\bibnamefont {Kroll}},
  \bibinfo {author} {\bibfnamefont {L.}~\bibnamefont {Gaus}}, \bibinfo {author}
  {\bibfnamefont {C.}~\bibnamefont {Bernert}}, \bibinfo {author} {\bibfnamefont
  {E.}~\bibnamefont {Beyreuther}}, \bibinfo {author} {\bibfnamefont {T.~E.}\
  \bibnamefont {Cowan}}, \bibinfo {author} {\bibfnamefont {L.}~\bibnamefont
  {Karsch}}, \bibinfo {author} {\bibfnamefont {S.}~\bibnamefont {Kraft}},
  \bibinfo {author} {\bibfnamefont {L.~A.}\ \bibnamefont {Kunz-Schughart}},
  \bibinfo {author} {\bibfnamefont {E.}~\bibnamefont {Lessmann}}, \bibinfo
  {author} {\bibfnamefont {J.}~\bibnamefont {Metzkes-Ng}}, \bibinfo {author}
  {\bibfnamefont {L.}~\bibnamefont {Obst-Huebl}}, \bibinfo {author}
  {\bibfnamefont {J.}~\bibnamefont {Pawelke}}, \bibinfo {author} {\bibfnamefont
  {M.}~\bibnamefont {Rehwald}}, \bibinfo {author} {\bibfnamefont {H.-P.}\
  \bibnamefont {Schlenvoigt}}, \bibinfo {author} {\bibfnamefont
  {U.}~\bibnamefont {Schramm}}, \bibinfo {author} {\bibfnamefont
  {M.}~\bibnamefont {Sobiella}}, \bibinfo {author} {\bibfnamefont {E.~R.}\
  \bibnamefont {Szabó}}, \bibinfo {author} {\bibfnamefont {T.}~\bibnamefont
  {Ziegler}}, \ and\ \bibinfo {author} {\bibfnamefont {K.}~\bibnamefont
  {Zeil}},\ }\href {\doibase 10.1038/s41598-020-65775-7} {\bibfield  {journal}
  {\bibinfo  {journal} {Scientific Reports}\ }\textbf {\bibinfo {volume}
  {10}},\ \bibinfo {pages} {9118} (\bibinfo {year} {2020})}\BibitemShut
  {NoStop}%
\bibitem [{\citenamefont {Wang}\ \emph {et~al.}(2020)\citenamefont {Wang},
  \citenamefont {Zhu}, \citenamefont {Easton}, \citenamefont {Li},
  \citenamefont {Lin},\ and\ \citenamefont {Yan}}]{Wang2020}%
  \BibitemOpen
  \bibfield  {author} {\bibinfo {author} {\bibfnamefont {K.~D.}\ \bibnamefont
  {Wang}}, \bibinfo {author} {\bibfnamefont {K.}~\bibnamefont {Zhu}}, \bibinfo
  {author} {\bibfnamefont {M.~J.}\ \bibnamefont {Easton}}, \bibinfo {author}
  {\bibfnamefont {Y.~J.}\ \bibnamefont {Li}}, \bibinfo {author} {\bibfnamefont
  {C.}~\bibnamefont {Lin}}, \ and\ \bibinfo {author} {\bibfnamefont {X.~Q.}\
  \bibnamefont {Yan}},\ }\href {\doibase 10.1103/PhysRevAccelBeams.23.111302}
  {\bibfield  {journal} {\bibinfo  {journal} {Phys. Rev. Accel. Beams}\
  }\textbf {\bibinfo {volume} {23}},\ \bibinfo {pages} {111302} (\bibinfo
  {year} {2020})}\BibitemShut {NoStop}%
\bibitem [{\citenamefont {Busold}\ \emph {et~al.}(2014)\citenamefont {Busold},
  \citenamefont {Schumacher}, \citenamefont {Brabetz}, \citenamefont {Deppert},
  \citenamefont {Kroll}, \citenamefont {Blazevic}, \citenamefont {Bagnoud},\
  and\ \citenamefont {Roth}}]{Busold2014}%
  \BibitemOpen
  \bibfield  {author} {\bibinfo {author} {\bibfnamefont {S.}~\bibnamefont
  {Busold}}, \bibinfo {author} {\bibfnamefont {D.}~\bibnamefont {Schumacher}},
  \bibinfo {author} {\bibfnamefont {C.}~\bibnamefont {Brabetz}}, \bibinfo
  {author} {\bibfnamefont {O.}~\bibnamefont {Deppert}}, \bibinfo {author}
  {\bibfnamefont {F.}~\bibnamefont {Kroll}}, \bibinfo {author} {\bibfnamefont
  {A.}~\bibnamefont {Blazevic}}, \bibinfo {author} {\bibfnamefont
  {V.}~\bibnamefont {Bagnoud}}, \ and\ \bibinfo {author} {\bibfnamefont
  {M.}~\bibnamefont {Roth}},\ }\href {\doibase
  10.18429/JACoW-IPAC2014-TUPME030} {\bibfield  {journal} {\bibinfo  {journal}
  {IPAC 2014}\ } (\bibinfo {year} {2014}),\
  10.18429/JACoW-IPAC2014-TUPME030}\BibitemShut {NoStop}%
\bibitem [{\citenamefont {Courant}\ and\ \citenamefont
  {Snyder}(1958)}]{CourantSnyder.AoP.1958}%
  \BibitemOpen
  \bibfield  {author} {\bibinfo {author} {\bibfnamefont {E.~D.}\ \bibnamefont
  {Courant}}\ and\ \bibinfo {author} {\bibfnamefont {H.~S.}\ \bibnamefont
  {Snyder}},\ }\href {\doibase 10.1016/0003-4916(58)90012-5} {\bibfield
  {journal} {\bibinfo  {journal} {Annals of Physics}\ }\textbf {\bibinfo
  {volume} {3}},\ \bibinfo {pages} {1} (\bibinfo {year} {1958})}\BibitemShut
  {NoStop}%
\bibitem [{\citenamefont {Piazza}\ \emph {et~al.}(2022)\citenamefont {Piazza},
  \citenamefont {Willingale},\ and\ \citenamefont
  {Zuegel}}]{dipiazza2022multipetawatt}%
  \BibitemOpen
  \bibfield  {author} {\bibinfo {author} {\bibfnamefont {A.~D.}\ \bibnamefont
  {Piazza}}, \bibinfo {author} {\bibfnamefont {L.}~\bibnamefont {Willingale}},
  \ and\ \bibinfo {author} {\bibfnamefont {J.~D.}\ \bibnamefont {Zuegel}},\
  }\href@noop {} {\enquote {\bibinfo {title} {Multi-petawatt physics
  prioritization (mp3) workshop report},}\ } (\bibinfo {year} {2022}),\ \Eprint
  {http://arxiv.org/abs/2211.13187} {arXiv:2211.13187} \BibitemShut {NoStop}%
\bibitem [{\citenamefont {Garten}\ \emph {et~al.}(2023)\citenamefont {Garten},
  \citenamefont {Bulanov}, \citenamefont {Hakimi}, \citenamefont {Obst-Huebl},
  \citenamefont {Schroeder}, \citenamefont {Esarey}, \citenamefont {Geddes},
  \citenamefont {Vay},\ and\ \citenamefont {Huebl}}]{supplementary_material}%
  \BibitemOpen
  \bibfield  {author} {\bibinfo {author} {\bibfnamefont {M.}~\bibnamefont
  {Garten}}, \bibinfo {author} {\bibfnamefont {S.~S.}\ \bibnamefont {Bulanov}},
  \bibinfo {author} {\bibfnamefont {S.}~\bibnamefont {Hakimi}}, \bibinfo
  {author} {\bibfnamefont {C.~E.}\ \bibnamefont {Obst-Huebl}, \bibfnamefont
  {Lieselotte~Mitchell}}, \bibinfo {author} {\bibfnamefont {C.}~\bibnamefont
  {Schroeder}}, \bibinfo {author} {\bibfnamefont {E.}~\bibnamefont {Esarey}},
  \bibinfo {author} {\bibfnamefont {C.~G.~R.}\ \bibnamefont {Geddes}}, \bibinfo
  {author} {\bibfnamefont {J.-L.}\ \bibnamefont {Vay}}, \ and\ \bibinfo
  {author} {\bibfnamefont {A.}~\bibnamefont {Huebl}},\ }\href {\doibase
  10.5281/zenodo.8210206} {\enquote {\bibinfo {title} {{Data Archive}},}\ }
  (\bibinfo {year} {2023})\BibitemShut {NoStop}%
\end{thebibliography}%

\end{document}